\documentclass[twocolumn,showpacs,preprintnumbers,amsmath,amssymb,nofootinbib]{revtex4}
\usepackage{graphicx}% Include figure files
\usepackage{dcolumn}% Align table columns on decimal point
\usepackage{bm}% bold math
\usepackage{relsize}
\usepackage{mdwlist}
\usepackage{soul}
\usepackage{hyperref}
\usepackage[all]{hypcap}
\def\cesrta{{C{\smaller[2]ESR}TA}}

\begin{document}

%\preprint{APS/123-QED}

\title{Measurement and Modeling of Electron Cloud in a Field Free Environment Using Retarding Field Analyzers}

\author{J.~R.~Calvey}
\author{G. Dugan}
\author{W.~Hartung}
\author{J.~A.~Livezey\footnote{Present address: Department of Physics, University of California, Berkeley, CA}}
\author{J.~Makita\footnote{Present address: Department of Physics, Old Dominion University, Norfolk, VA}}
\author{M.~A.~Palmer\footnote{Present address: Fermi National Accelerator Laboratory, P.O. Box 500, MS 221, Batavia, IL 60510-5011, USA.}}
\affiliation{Cornell Laboratory for Accelerator-based Sciences and
  Education, Cornell University, Ithaca, New York, USA}

\date{\today}% It is always \today, today,
             %  but any date may be explicitly specified

\begin{abstract}
       As part of the \cesrta~ program at Cornell, diagnostic devices to measure and quantify the electron cloud effect have been installed throughout the CESR ring.  One such device is the Retarding Field Analyzer (RFA), which provides information on the local electron cloud density and energy distribution.  In a magnetic field free environment, RFA measurements can be directly compared with simulation to study the growth and dynamics of the cloud on a quantitative level.  In particular, the photoemission and secondary emission characteristics of the instrumented chambers can be determined simultaneously.
\end{abstract}

% PACS, the Physics and Astronomy
                             % Classification Scheme.
%\keywords{Suggested keywords}%Use showkeys class option if keyword
                              %display desired
\maketitle

\section{\label{sec:intro} Introduction}

The electron cloud effect is a well known phenomenon in particle accelerators (see, for example,~\cite{ECLOUD12:Miguel}), in which a high density of low energy electrons builds up inside the vacuum chamber.  These electrons can cause a wide variety of undesirable effects, including emittance growth and beam instabilities~\cite{PhysRevSTAB.7.124801}.  Electron cloud has been observed in many facilities ~\cite{PRSTAB7:024402,PRSTAB14:071001,NIMA556:399to409,PRSTAB6:034402,PAC09:WE4GRC02,PhysRevSTAB.6.014203,PRSTAB16:011003}, and is expected to be a major limiting factor in next generation positron and proton storage rings.  In lepton machines, the cloud is usually seeded by photoelectrons generated by synchrotron radiation.  The collision of these electrons with the beam pipe can then produce one or more secondary electrons, depending on the secondary electron yield (SEY) of the material.  If the average SEY is greater than unity, the cloud density will grow exponentially, until a saturation is reached.

In 2008, the Cornell Electron Storage Ring (CESR) was reconfigured to study issues related to the design of the International Linear Collider (ILC~\cite{ILCREP2007:001}) damping ring, including electron cloud.  A significant component of this program, called CESR Test Accelerator (\cesrta), was the installation of several retarding field analyzers (RFAs) throughout the ring.  These detectors, which provide information on the local electron cloud density, energy, and transverse distributions, have been used to directly compare different electron cloud mitigation techniques~\cite{ARXIV:1402.1904}.  Quantitative analysis of RFA data requires detailed computer simulations, and is the subject of this paper.  More specifically, we will give a brief overview of the \cesrta ~electron cloud experimental program (Section~\ref{sec:hardware}), describe the use of computer simulations to model the cloud (Section~\ref{sec:sims}), detail our efforts at incorporating a model of an RFA into the simulation (Section~\ref{sec:modeling}), and explain how the comparison of data and simulation yields a holistic and self-consistent description of cloud generation and dynamics in a field free environment (Section~\ref{sec:comparison}).  As a result of this procedure, we obtain information on the primary and secondary emission properties of the instrumented chambers.

This technique has several additional advantages.  First, we are able to study the chambers in an actual accelerator environment, after processing with a lepton beam.  Also, by comparing data and simulation on a detailed level, we substantially validate the electron emission model embodied in the simulation codes, and therefore reinforce our confidence in their applicability in other situations, in particular to hadron storage rings.  Finally, we have been able to study several different mitigation techniques, and evaluate their effectiveness in preventing electron cloud build-up.

\subsection{Retarding Field Analyzers}

\begin{figure}
	\centering
	\includegraphics[width=.45\textwidth]{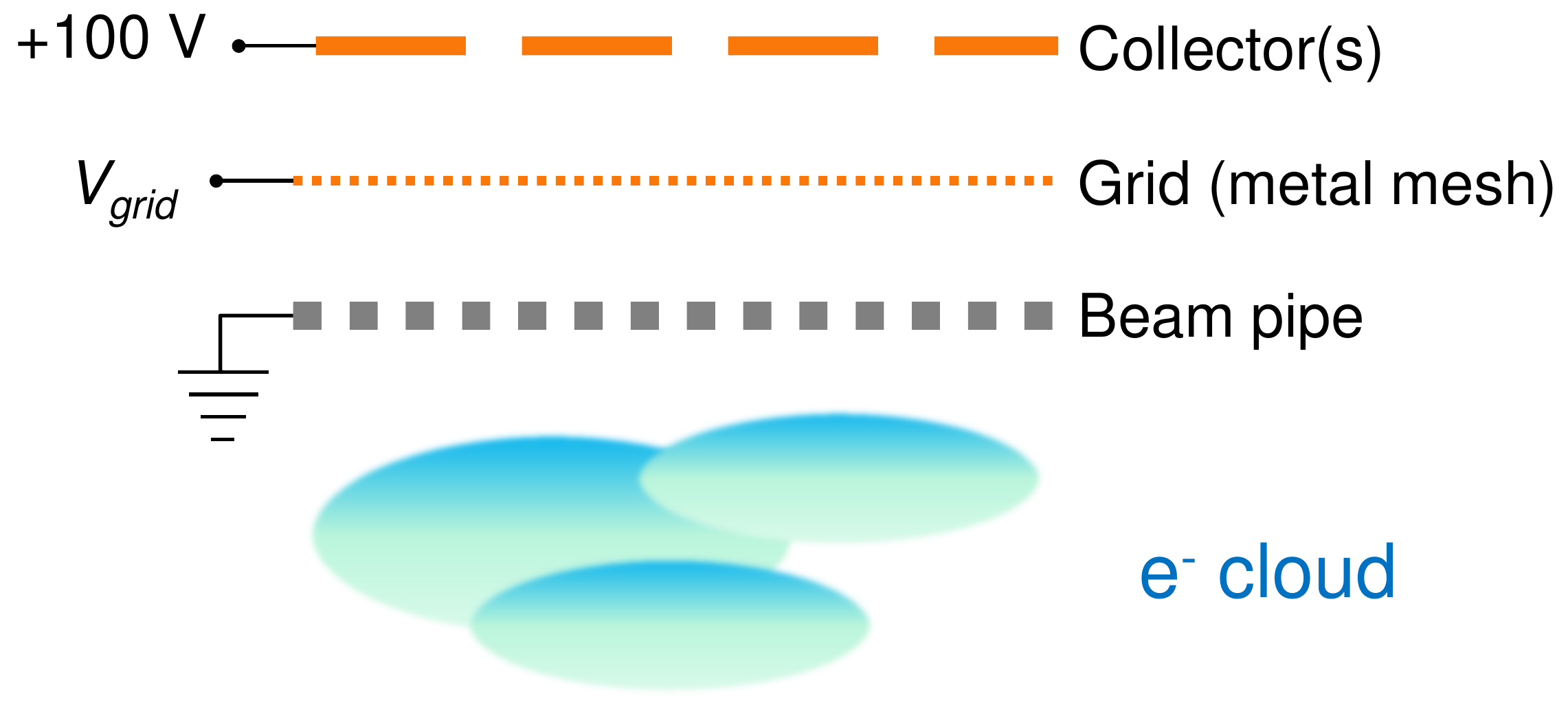}      %{rfa_diagram.png}
	\caption{Idealized diagram of a retarding field analyzer.}	
    \label{fig:rfa_diagram}
\end{figure}

A retarding field analyzer consists of three main components~\cite{NIMA453:507to513}: holes drilled in the beam pipe to allow electrons to enter the device; a ``retarding grid," to which a voltage can be applied, rejecting electrons with less than a certain energy; and a positively biased collector, to capture any electrons which make it past the grid (Fig.~\ref{fig:rfa_diagram}).  If space permits, additional (grounded) grids can be added to allow for a more ideal retarding field.  In addition, the collectors of most RFAs used in \cesrta~are segmented to allow characterization of the spatial structure of the cloud build-up.  Thus a single RFA measurement provides information on the local cloud density, energy, and transverse distribution.  Most of the data presented here are ``voltage scans," in which the retarding voltage is varied (typically from +100 to $-250$~V or $-400$~V) while beam conditions are held constant.  The collector was set to +100~V for all of our measurements.

\begin{figure}
   \centering
   \includegraphics*[width=0.45\textwidth]{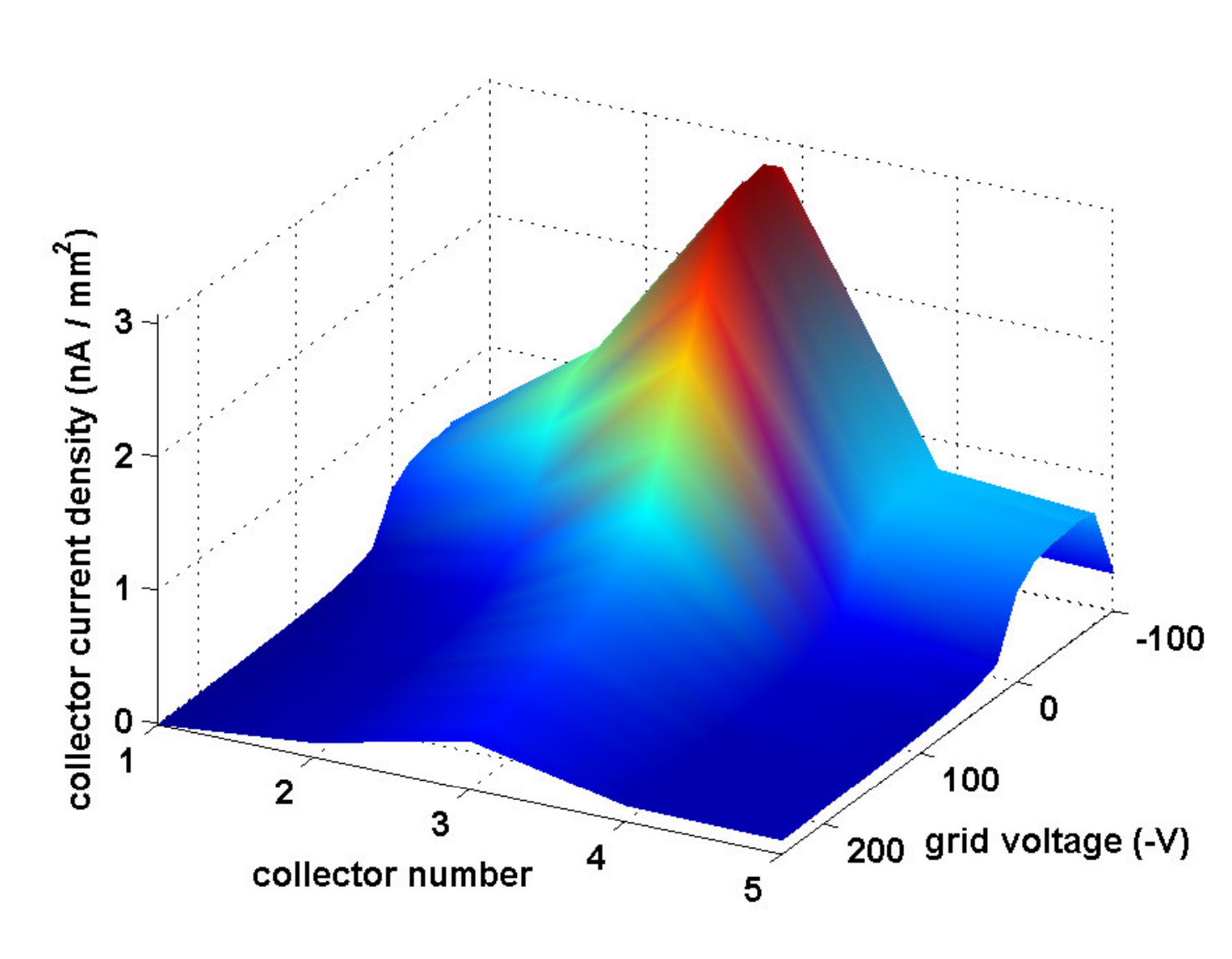}
   \caption[RFA voltage scan with an insertable segmented drift RFA]{\label{fig:seg_example} RFA voltage scan in a Cu chamber, 1x45x1.25 mA e$^+$, 14~ns, 5.3 GeV.}
\end{figure}

An example voltage scan is given in Fig.~\ref{fig:seg_example}.  The RFA response is plotted as a function of collector number and retarding voltage.  Roughly speaking, this is a description of the transverse and energy distribution of the cloud.  Collector  1 is closest to the outside of the chamber (where direct synchrotron radiation hits); the central collector (3 in this case) is aligned with the beam.  The sign convention for retarding voltage is chosen so that a positive value on this axis corresponds to a negative physical voltage on the grid (and thus a rejection of lower energy electrons).  The beam conditions are given as ``1x45x1.25~mA e$^+$, 14~ns, 5.3~GeV."  This notation indicates one train of 45 bunches, with 1.25~mA/bunch (1~mA = $1.6\times10^{10}$ particles), with positrons, 14~ns spacing, and at beam energy 5.3~GeV.

We have used RFAs to probe the local behavior of the cloud at multiple locations in CESR, under many different beam conditions, and in the presence of different mitigation schemes.  The primary method of reducing electron cloud density in a field free region is the use of beam pipe coatings, which reduce the primary and/or secondary emission yield of the chamber.  Coatings tested at \cesrta~include titanium nitride (TiN)~\cite{NIMA551:187to199}, amorphous carbon (aC)~\cite{PRSTAB14:071001}, diamond-like carbon (DLC)~\cite{ECLOUD10:MIT00}, and Ti-Zr-V non-evaporable getter (NEG)~\cite{NIMA554:92to113}.  Direct comparisons of RFA data taken in the various chambers showed that all the coatings are effective at reducing electron cloud, relative to uncoated aluminum or copper~\cite{ARXIV:1402.1904}.

\section{\label{sec:hardware} Experimental Program}

Detailed descriptions of the \cesrta~electron cloud experimental program, design of the drift RFAs, and data acquisition system can be found elsewhere~\cite{ARXIV:1402.1904}; here we provide only a brief summary.

There are five main electron cloud experimental sections of CESR instrumented with drift RFAs (Table~\ref{tab:rfa_master_list}).  These include long sections at Q14E and Q14W (the names refer to their proximity to the 14E and 14W quadrupoles, respectively), shorter sections at Q15E and Q15W, and a long straight section at L3.  The vacuum chambers at Q15E/W are approximately elliptical and made of aluminum (as is most of CESR), while the chambers at Q14E/W are rectangular and made of copper, and the pipe is circular stainless steel at L3.

%Fig.~\ref{fig:cesr_configuration} shows the locations of these experimental sections in the CESR ring.

\begin{table}
  \centering
  \footnotesize
    \caption{\label{tab:rfa_master_list} List of drift RFA locations.  ``Material" refers to the base material; some locations have tested one or more coatings.  The vacuum chambers at all locations are 5~cm in height by 9~cm in width, with the exception of the circular chambers, which are 4.5~cm in radius.}
  \begin{tabular}{ccccc}
    \hline \hline
  Location  &   RFA Type    &   Material    &   Coatings  &     Shape \\
  \hline
  14W   &   Ins. I  &   Cu  &  TiN   &   Rectangular \\
  15W   &   Thin, Ins. II   &   Al  &   TiN, aC &   Elliptical  \\
  L3    &   APS   &   SST &   NEG   &   Circular    \\
  15E   &   Thin, Ins. II   &   Al  &   TiN, aC, DLC    &   Elliptical  \\
  14E   &   APS, Ins. I  &   Cu  &  TiN   &   Rectangular \\
  \hline \hline
  \end{tabular}

\end{table}

% \subsection{\label{ssec:styles} RFA Styles}

Because of these differences in beam pipe geometry, several different styles of RFA were deployed throughout drift sections in CESR.  Table~\ref{tab:rfa_styles} summarizes the key parameters of each style.

\begin{table}
   \centering
       \setlength{\tabcolsep}{6pt}
   \caption{\label{tab:rfa_styles} Drift RFA styles deployed in CESR.  Each RFA has one retarding grid.  For RFAs with multiple grids, the additional grids are grounded.}
   \begin{tabular}{cccc}
       \hline \hline
       Type &   Grids   &   Collectors  &   Grid Transparency \\
       \hline
           APS  &   2   &   1   &    ~46\% \\
           Insertable I   &   2   &   5  &  ~40\%   \\
           Insertable II  &   3   &   11   &   ~90\%   \\
           Thin  &   1   &   9   &  ~90\%   \\
        \hline \hline
       %\bottomrule
   \end{tabular}
\end{table}

\subsection{CESR Parameters}

The primary advantage of CESR as a test accelerator is its flexibility.  At \cesrta, we have been able to study the behavior of the electron cloud as a function of several different beam parameters, varying the number of bunches, bunch current, bunch spacing, beam energy, and species.  As will be described in Section~\ref{sec:comparison}, this is very helpful for independently determining the photoelectron and secondary electron properties of the instrumented chambers.  Table~\ref{tab:mitigation_beam_condx} gives some of the basic parameters of CESR, and lists some of the beam parameters used for electron cloud mitigation studies with RFAs.  

%A more complete description of the full operating range of CESR can be found in~\cite{ECLOUD10:OPR06}.

\begin{table}
   \centering
   \caption{\label{tab:mitigation_beam_condx} CESR parameters and typical beam conditions for electron cloud mitigation studies}
   \begin{tabular}{ccc}
       \hline \hline
       Parameter &   Value(s)  &   Units \\
       \hline
            \textbf{General Parameters} &    &   \\
            Circumference   &   768 &   m   \\
            Revolution Period   &   2.56    &   $\mu$s \\
            Harmonic number     &   1281    &   -   \\
            Number of bunches  &   9, 20, 30, 45   &   -  \\
            %Bunch spacing  &   4, 8, 14, 16, 28, 56, 112, 280   &   ns   \\
            Bunch spacing  &   4 - 280   &   ns   \\
            Beam energy &   2.1, 4, 5.3  &   GeV  \\
       \hline
            \textbf{2.1 GeV Parameters} &    &   \\
            RMS Horizontal Emittance    &   2.6     &   nm  \\
            RMS Vertical Emittance  &   0.02    &   nm   \\
            RMS Bunch Length    &   12.2  &   mm  \\
            Bunch current  &   0 - 5   &   mA\footnote{1~mA = $1.6\times10^{10}$ particles}   \\
            Beam species &   e$^+$, e$^-$  &   - \\
       \hline
            \textbf{4 GeV Parameters} &    &   \\
            RMS Horizontal Emittance    &   23     &   nm  \\
            RMS Vertical Emittance  &   0.23    &   nm   \\
            RMS Bunch Length    &   9  &   mm  \\
            Bunch current  &   0 - 6   &   mA   \\
            Beam species &   e$^+$  &   - \\
       \hline
            \textbf{5.3 GeV Parameters} &    &   \\
            RMS Horizontal Emittance    &   144     &   nm  \\
            RMS Vertical Emittance  &   1.3    &   nm   \\
            RMS Bunch Length    &   20.1  &   mm  \\
            Bunch current  &   0 - 10   &   mA   \\
            Beam species &   e$^+$, e$^-$  &   - \\
        \hline \hline
   \end{tabular}
\end{table}

\section{\label{sec:sims} Cloud Buildup Simulations}

As the behavior of the electron cloud can be very complicated and depends on many parameters, it is best understood on a quantitative level through the use of computer simulations.  The results presented here were obtained with the particle tracking code POSINST~\cite{MBI97:170,LHC:ProjRep:180,PRSTAB5:124404}.  In this code, the electrons are dynamical (and represented by macroparticles), while the beam is not (and is instead represented by a prescribed function of time and space).  As such, it is useful for modeling buildup of the cloud, but not the effect of the cloud on the beam.

In POSINST, a simulated photoelectron is generated on the chamber surface and tracked under the action of the beam.  Secondary electrons are generated via a probabilistic process.  Space charge and image charge are also included.  Electron motion is fully 3D, but the space and image charge forces are only calculated in two dimensions (effectively this assumes periodic boundary conditions).  POSINST has been used to study cloud buildup in a number of different contexts (e.g.~\cite{IPAC12:TUPPR063,IPAC12:WEPPR088,PAC11:WEP108,MBI97:170,PRSTAB11:091002,EPAC08:MOPP063,PAC07:THPMN118,HB06:TUAX05,LHC:ProjRep:180,PRSTAB7:024402,PAC01:FOAB004}).

\subsection{\label{ssec:parameters} Simulation Parameters}

\begin{figure}
\centering
\includegraphics[width=.45\textwidth]{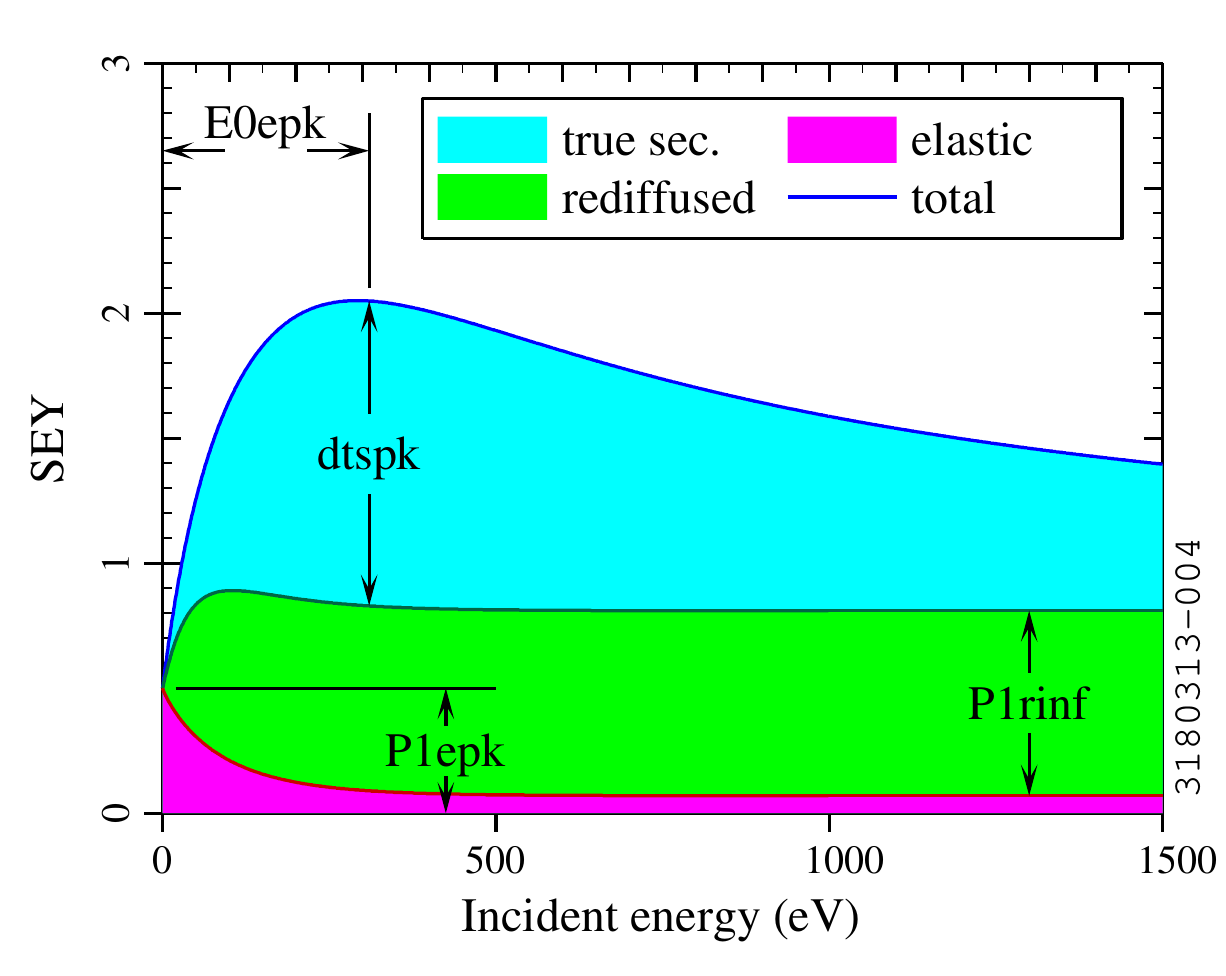}       %{sey_diagram_big.pdf}   %{3D_Model_RFA_section_white_big.eps} %{opera_rfa_model.eps} \\
\caption{\label{fig:sey_diagram} Secondary electron yield as a function of energy, with important POSINST parameters indicated.}
\end{figure}

There are many parameters related to primary and secondary electron emission that are relevant to this analysis.  The secondary electron yield model in POSINST contains three components: ``true" secondaries, which are emitted at low ($\leq$20 eV) energy regardless of the incident particle energy; ``elastic" secondaries, which are emitted at the same energy as the incident particle; and ``rediffused" secondaries, which are emitted with a uniform energy spectrum, ranging between 0 and the incident particle energy.  The peak true secondary yield (characterized by the parameter \texttt{dtspk} in POSINST) occurs for primary electrons with an incident energy (POSINST parameter \texttt{E0epk}) around 300~eV.  The peak elastic yield (POSINST parameter \texttt{P1epk}) occurs at low energy (we assume 0~eV), while the rediffused yield reaches a steady state value for high energy primaries (POSINST parameter \texttt{P1rinf}).  Another relevant secondary emission parameter is the ``shape parameter" \texttt{powts}, which determines the shape of the true secondary curve about its peak.  Fig.~\ref{fig:sey_diagram} shows a typical SEY curve, and indicates how each of the SEY components contributes to the total peak secondary yield (POSINST parameter \texttt{dtotpk}).

POSINST also makes use of several parameters that describe the properties of emitted secondary electrons.  The parameters that define the true-secondary emission energy distribution were chosen to give a peak emission energy of 1.5 eV~\cite{IPAC11:MOPS083}.  Secondaries are emitted with angular distribution $\frac{\partial N}{\partial \theta} \propto \sin(\theta) \cos(\theta)$, where $\theta$ is the angle relative to normal.

%, based on RFA measurements done in a dipole

The model for photoelectron emission in POSINST is simpler than the secondary model, but still involves several important parameters.  The most significant of these is the quantum efficiency (\texttt{queffp}).  Also, in order to explain the measurable RFA signal we see with an electron beam, there must be some photoelectrons with sufficiently high energy to overcome the repulsive force of the beam.  In the simulation, this is accomplished by using a Lorentzian photoelectron energy distribution (which has been observed in some measurements~\cite{PRSTAB2:063201}), with a low peak energy (5~eV).  By assumption, the width of the distribution is scaled with the average photon energy incident at the RFA position (this resulted in better agreement with data than using the same distribution for all energies).  For example, for an electron beam at Q15E, the width is 12~eV for a 2.1~GeV beam, and 150~eV for a 5.3~GeV beam.  The drift RFA data do not constrain the exact shape of the distribution; measurements with a shielded button detector~\cite{NIMA749:42to46} provide a method to probe these parameters in more detail.

Computational parameters in POSINST (Table~\ref{tab:computational_params}) were adjusted to give consistent results, without requiring prohibitively long run times.  Further increase of these parameters did not result in significant changes to the output of the simulation.

\begin{table}
   \centering
   \caption{\label{tab:computational_params} Computational parameters in POSINST}
      \footnotesize
   \begin{tabular}{ccc}
       \hline \hline
       Parameter &	 Description    &  Value \\
       \hline
            \texttt{nkicks}   &   Beam kicks per bunch passage &	 51 \\ %.01 \\
            \texttt{ngrexpx, ngrexpy}   &	Space charge grid parameter &   5 \footnote{Results in a 32x32 grid for space charge calculations} \\ %.005    \\
            \texttt{nsteps}  &   Steps between bunch passages     & 14     \\
            \texttt{macrophel}  &	Macroparticles generated per bunch    &  5000   \\
        \hline \hline
   \end{tabular}
%   \end{minipage}
\end{table}

\section{\label{sec:modeling} RFA Modeling}

%A retarding field analyzer consists of three main components: holes drilled in the beam pipe to allow electrons to enter the device; a ``retarding grid," to which a voltage can be applied, rejecting electrons with less than a certain energy; and a positively biased collector, to capture any electrons which make it past the grid.

\begin{figure*}[floatfix]
  % \begin{minipage}{.9\textwidth}
   \centering
   \includegraphics*[width=0.75\textwidth]{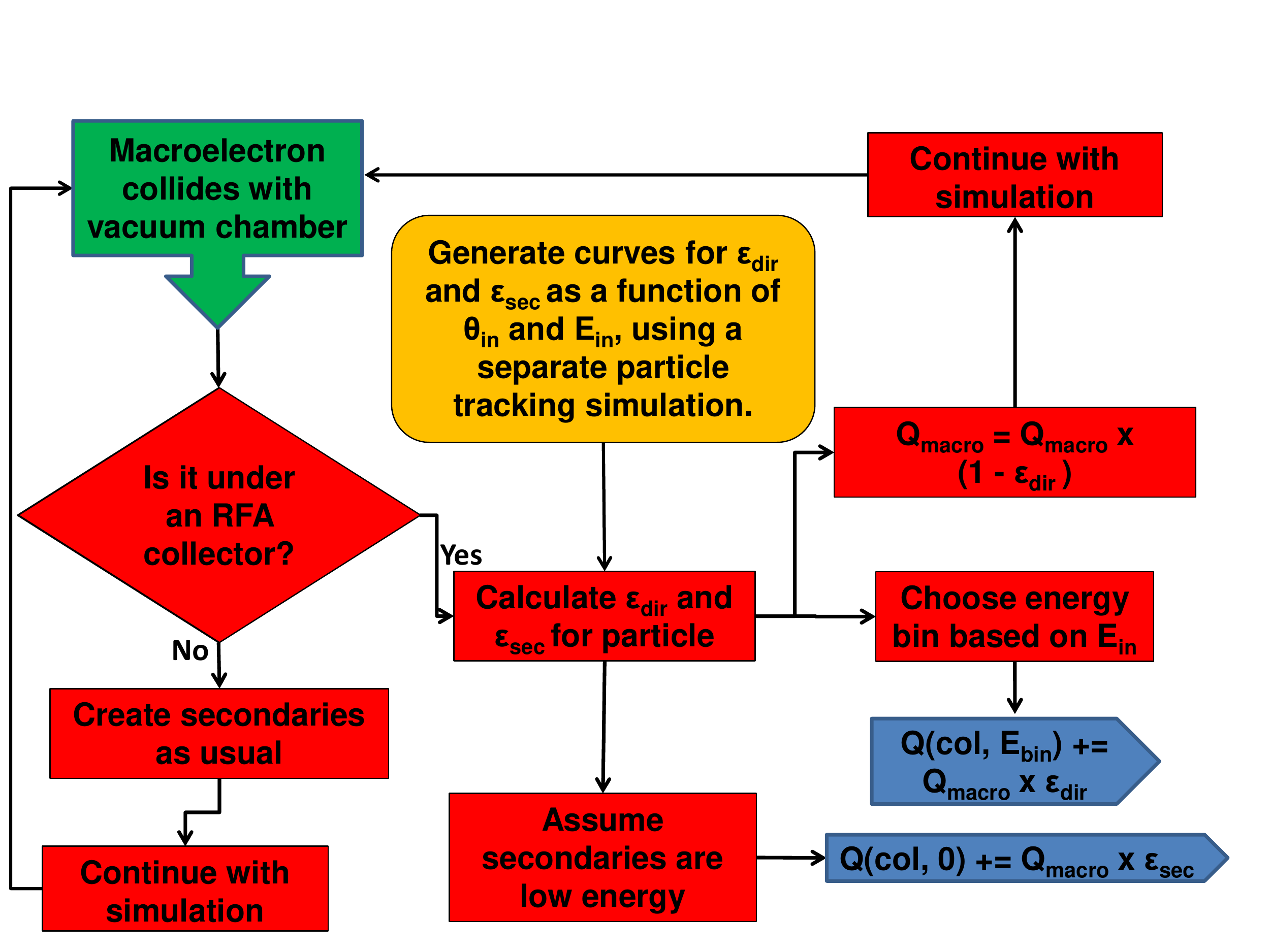}
   \caption{\label{fig:flowchart} Conceptual flowchart of the RFA model in POSINST.  The charge deposited in the collector is binned by energy and collector number ($Q(col, E_{bin})$).  The magnitude of this quantity depends on the macroelectron charge ($Q_{macro}$) and the efficiency of the RFA ($\varepsilon_{dir}$), which in turn depends on the incident particle energy ($E_{in}$) and angle ($\theta_{in}$). In addition, the macroelectron can generate low energy ``secondary" charge ($Q(col,0)$), depending on the secondary efficiency ($\varepsilon_{sec}$).  Secondaries which do not contribute to the collector signal are considered lost.  Charge that enters the RFA is removed from the macroelectron.}
  % \end{minipage}
\end{figure*}

To understand the RFA measurements on a more fundamental level, we need a way of translating an RFA measurement into physical quantities relating to the development of the electron cloud.  To bridge this gap, accurate models of both the cloud development and the RFA itself are required.  To this end, we have modified POSINST to include a model of the RFA, which automatically generates an output file containing the simulated RFA signals.  This integrated RFA model is implemented as a special function that is called when a macroelectron in the simulation collides with the vacuum chamber wall, immediately before the code section that simulates secondary emission.  First, this function checks whether the macroelectron is in the region covered by the RFA.  If so, a certain fraction of the particle's charge, which depends on the incident angle and energy (as well as the overall beam pipe transparency), is added to the collector signal.  The RFA acceptance as a function of angle and energy is calculated by a separate particle tracking code, described below.  The charge is binned by energy and transverse position, reproducing the energy and position resolution of the RFA.  The macroelectron then has its charge reduced by the amount that went into the detector, and the simulation continues as normal.  This process is shown diagrammatically in Fig.~\ref{fig:flowchart}.  In comparison to previous efforts at analyzing the RFA data, which relied on post-processing POSINST output files~\cite{PAC11:MOP214}, the integrated model is both faster and more self-consistent.

In order for this method to work, we need to know the RFA response to a particle with a given incident energy and angle.  To answer this question, we developed a specialized code which tracks electrons through a model of the RFA.  The model includes a detailed replica of the beam pipe, grid(s), and collector, as well as a realistic map of the electric fields inside the RFA, generated by the electrostatic calculation tool Opera 3D\footnote{http://operafea.com/}.  The tracking code also allows for the production of secondary electrons on both the beam pipe and grid(s).  The secondary emission model is a simplified version of the one used in POSINST, and includes both elastic and ``true" secondaries (see Section~\ref{ssec:parameters}).  The output of the simulation is a table which maps the incident particle energy and angle to both a ``direct" and (low energy) ``secondary" collector signal.  POSINST can then consult this table to determine the RFA response to a given macroelectron-wall collision.

\begin{figure}
\centering
\includegraphics[width=.45\textwidth]{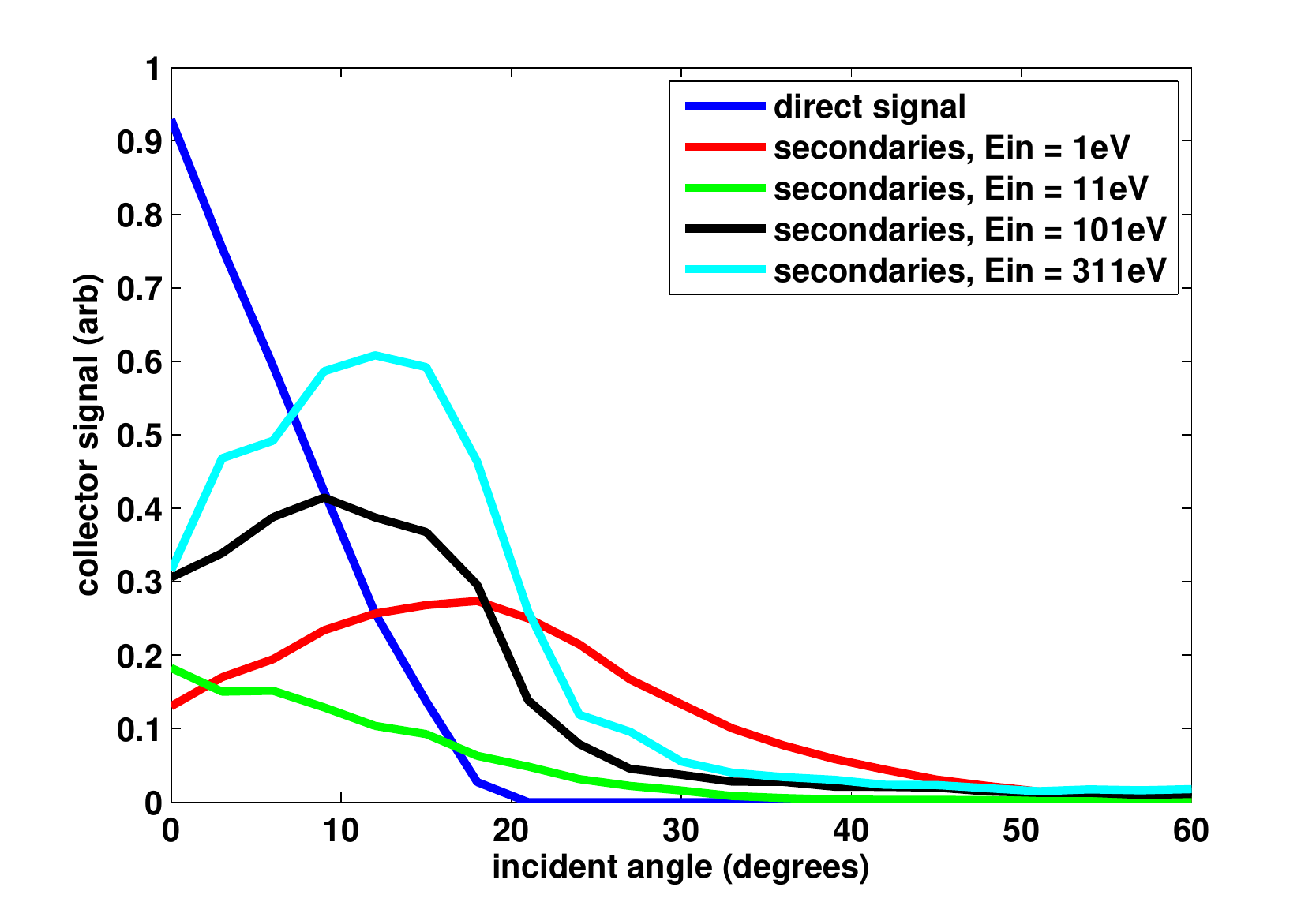}   %{3D_Model_RFA_section_white_big.eps} %{opera_rfa_model.eps} \\
\caption{\label{fig:beam_pipe_sec} Simulated collector current caused by a uniform beam of electrons incident on the thin RFA model, for various incident energies.  The direct signal is determined only by the angular acceptance of the beam pipe holes (and thus does not depend on energy).  The ``secondary" signal is caused by the production of (low energy) secondary electrons in the beam pipe holes and retarding grid, and depends on the energy of the incident electrons.  In general the secondary signal is highest for very low incident energies (due to the production of elastic secondaries), and around $E_{in} \approx$ \texttt{E0epk} (due to the production of ``true" secondaries).}
\end{figure}

The production of secondary electrons in the beam pipe holes and on the retarding grid is an especially important effect, and results in an enhanced low energy signal in most of our drift RFA measurements.  Fig.~\ref{fig:beam_pipe_sec} shows the simulated secondary signal in a thin style RFA, as a function of incident angle, for different incident electron energies.  The effect is particularly strong for electrons with high energy and moderate angle.

%Collector current due to the production of secondary electrons in the beam pipe holes and retarding grid, according to simulation.  The direct signal, which is determined only by the angular acceptance of the beam pipe holes, is also shown.

To aid in the development of our model, we constructed a bench experiment to study the response of a test RFA under controlled conditions.  Measurements with this system showed good agreement with our model~\cite{ARXIV:1402.1904}.

\section{\label{sec:comparison}  Comparison with Measurements}

%To this end, an RFA model has been incorporated into POSINST, enabling much faster analysis of the simulation.  This model includes most of the effects described in the previous section, including the enhancement of the collector signal at positive voltage due to secondaries from the beam pipe holes and retarding grid.

The large quantity of RFA data obtained during the \cesrta~program necessitates a systematic method for detailed analysis.  Our approach has been to take a large set of voltage scan data, and find a set of simulation parameters that bring data and simulation into as close agreement as possible.  Simultaneously fitting data taken under a wide variety of beam conditions gives us confidence that our model is producing a reasonable description of the growth and dynamics of the electron cloud.

More specifically, we want to minimize $\chi^2$, as defined in Eq. (\ref{eq:chi2def}).  Here $\mathbf{y_d}$ is a vector of data points, $\mathbf{y_s}$ is a vector of simulation points, $\mathbf{\beta_0}$ is the vector of nominal parameter values, and $\mathbf{\beta}$ is the vector of new parameter values.  $\mathbf{X}$ is the Jacobian matrix ($X_{i,j} \equiv \frac{\partial y_i}{\partial \beta_j}$), and  $\mathbf{W}$ is a diagonal matrix whose elements are $\frac{1}{\sigma_i^2}$, where $\sigma_i$ is the error on data point $i$.  Both the data and simulation can contribute to this error.  The $T$ superscript denotes the matrix transpose.  Note that $\mathbf{X}$ and $\mathbf{y_s}$ are both evaluated at $\mathbf{\beta_0}$.  Once a new set of parameter values is obtained, the process can be repeated with this new set as the ``nominal" values.  As this method uses a linear approximation for the dependence of $\mathbf{\varepsilon}$ on $\mathbf{\beta}$, it will need to be iterated a few times before it converges on the actual minimum value of $\chi^2$.
%The value of $\beta$ which will minimize $\chi^2$ is given in Eq.~\ref{eq:betadef}.

%\begin{eqnarray}
\begin{align}
\begin{split}
\label{eq:chi2def}
    \chi^2 & = \mathbf{\varepsilon}^T \: \mathbf{W} \: \mathbf{\varepsilon} \\
    \mathbf{\varepsilon} & \equiv \mathbf{y_d} - (\mathbf{y_s} +\mathbf{X}(\mathbf{\beta} - \mathbf{\beta_0}))
\end{split}
\end{align}

\subsection{\label{ssec:fit_param} Parameter Constraints}

We have generally found that up to three simulation parameters (Section~\ref{ssec:parameters}) can be robustly and independently determined by this fitting procedure.  In all cases, \texttt{dtspk} and \texttt{queffp} needed to be included in the fits to get good agreement with the RFA data.  The quantum efficiency was allowed to be different for different beam energies and species, since it will in general depend on photon energy~\cite{ECLOUD12:Laura}.  Other strong parameters include \texttt{P1epk}, \texttt{P1rinf}, and \texttt{powts}, but they are highly correlated with each other (i.e. have similar effects on the RFA simulation), so only one of the three is needed.  For the uncoated chambers (Al and Cu), we varied \texttt{P1epk}.  For the coated chambers (aC, TiN, DLC, NEG), we found this parameter usually tended towards 0 in the fits, so we assumed a low value (0.05), and varied \texttt{P1rinf} instead.

Additional parameters were determined with the help of other \cesrta~ measurements and simulations.  The photon flux and azimuthal distribution at the RFA are calculated by a 3 dimensional simulation of photon production and reflection~\cite{ECLOUD10:PST08}, which includes diffuse scattering and a realistic model of the CESR vacuum chamber geometry.  Other secondary emission parameters (\texttt{powts} and \texttt{E0epk}) were obtained from direct in-situ SEY measurements~\cite{ECLOUD10:PST12}.  In addition, the analysis uses one arbitrary parameter: a ``chamber hole SEY," which is an overall scaling of the effect of secondaries generated in the RFA on the low energy signal (Fig.~\ref{fig:beam_pipe_sec}).  The fitted values for this parameter are within the expected range; a typical number for the effective hole SEY is on the order of ~1.5.

%because the photon energy spectrum at any given location in CESR will be different for the two species.

Table~\ref{tab:parameter_summary} summarizes the POSINST parameters most relevant to our analysis and indicates whether the parameter was used in the fits.

%explains where the baseline values were derived from,

\begin{table}
%    \begin{minipage}{.9\textwidth}
   \centering
   \caption{\label{tab:parameter_summary} Summary of relevant POSINST parameters.  The last column indicates whether the parameter was used in fits always (A), in some cases (S), or never (N).}
   \begin{tabular}{ccc}
       \hline \hline
       Parameter &	 Description    &  Fit? \\
       \hline
            \texttt{dtspk}   &   True secondary yield &	 A \\ %.01 \\
            \texttt{P1epk}   &	Elastic yield &   S \\ %.005    \\
            \texttt{P1rinf}  &   Rediffused yield     & S     \\
            \texttt{dtotpk}  &	Total peak yield    &  N\footnote{Equal to the sum of the three SEY components at peak energy}   \\
            \texttt{E0epk}   &   Peak yield energy     & N  \\
            \texttt{powts}   &   Shape parameter    &    N   \\
            \texttt{queffp}  &	Quantum efficiency    &  A   \\

        \hline \hline
   \end{tabular}
%   \end{minipage}
\end{table}

\subsection{Fitting the Data}

%In performing the $\chi^2$ analysis, 

After choosing the parameters we want to fit, we must pick out a set of voltage scan data that determine these parameters as independently as possible.  For example, the true secondary yield (\texttt{dtspk}) is highest for $\sim$300~eV electrons, so it is best determined by data taken under beam conditions where a typical electron energy is on that order.  This tends to mean short bunch spacing and moderately high current.  The elastic yield (\texttt{P1epk}) mainly affects the decay of the cloud, when most of the cloud particles have low energy.  It is best derived from data where the cloud is repeatedly generated and allowed to decay, i.e. for large bunch spacing. The quantum efficiency (\texttt{queffp}) is most significant in regimes where secondary emission is less important, namely for low current data.  Table~\ref{tab:condx_list} gives a list of data sets used in one round of fitting, and indicates which parameter was best determined by each.

\begin{table*}
  %  \begin{minipage}{.9\textwidth}
   \centering
   \caption{\label{tab:condx_list} List of beam conditions used for one round of fitting (15W Al chamber, May 2010), and which parameter they most strongly determined}
   %\begin{tabular}{c c c c c c}
%   \begin{tabular}{p{.15\textwidth} p{.15\textwidth} p{.15\textwidth} p{.15\textwidth} p{.15\textwidth} p{.15\textwidth}}
    \setlength{\tabcolsep}{10pt}
    \begin{tabular}{c c c c c c}
       \hline \hline
       Index    &   Bunches &	Bunch current &   Bunch Spacing &    Beam Energy  &  Parameter  \\
       \hline
       1  &     45 e$^+$ & 0.75 mA & 14 ns & 5.3 GeV & \texttt{queffp} \\
       2  &     45 e$^+$ & 0.75 mA & 14 ns & 4 GeV & \texttt{queffp} \\
       3  &     45 e$^+$ & 0.75 mA & 14 ns & 2.1 GeV & \texttt{queffp} \\
       4  &     45 e$^+$ & 2.3 mA & 14 ns & 2.1 GeV & \texttt{dtspk} \\
       5  &     20 e$^+$ & 2.8 mA & 4 ns & 4 GeV & \texttt{dtspk} \\
       6  &     20 e$^+$ & 7.5 mA & 14 ns & 2.1 GeV & \texttt{dtspk} \\
       7  &     20 e$^+$ & 10.75 mA & 14 ns & 5.3 GeV & \texttt{P1epk} \\
       8  &     9 e$^+$ & 3.78 mA & 280 ns & 2.1 GeV & \texttt{P1epk} \\
       9  &     9 e$^+$ & 3.78 mA & 280 ns & 4 GeV & \texttt{P1epk} \\
       10 &     9 e$^+$ & 4.11 mA & 280 ns & 5.3 GeV & \texttt{P1epk} \\
       11 &     45 e$^-$ & 2.89 mA & 4 ns & 5.3 GeV & \texttt{dtspk} \\
       12 &     45 e$^-$ & 1.25 mA & 14 ns & 5.3 GeV & \texttt{queffp} \\
       13 &     45 e$^-$ & 2 mA & 14 ns & 2.1 GeV & \texttt{queffp} \\
       14 &     20 e$^-$ & 2.8 mA & 14 ns & 5.3 GeV & \texttt{dtspk} \\
       15 &     9 e$^-$ & 3.78 mA & 280 ns & 2.1 GeV & \texttt{P1epk} \\
        \hline \hline
       %\bottomrule
   \end{tabular}
 %  \end{minipage}
\end{table*}

%Several sources of error must be taken into account when constructing the error matrix ($\mathbf{W}$ in Eq.~\ref{eq:chi2def}).  They include:
Several sources of error can complicate the analysis, and must be added (in quadrature) when constructing the error matrix ($\mathbf{W}$ in Eq. (\ref{eq:chi2def})).  They are listed below.  For the purpose of comparison, a typical signal in the 15E/W RFAs is on the scale of 100's of nA.

\begin{itemize*}
\item Noise in the measurements (typically quite small, a few tenths of a nA)
\item Statistical errors in simulations.  This can be reduced by increasing the number of macroelectrons used in the simulation, at the cost of increased run time.  Typical values are on the order of a few nA.
\item We have observed a slow drift of “baseline” (zero current value) in measurements, on the order of $\sim 0.2\%$ of full scale.  This amounts to $\sim$20~nA on the lowest gain setting, and $\sim$0.02~nA on the highest one (2~nA for a typical case).
\item A general error of 10\% was added to account for systematic uncertainties in the data.  This value was chosen to reflect our confidence in the repeatability of the measurements.  Similarly, an extra 20\% error was added to the signal in the simulation caused by beam pipe hole secondaries, to account for the additional uncertainty in the modeling of this phenomenon.  We found that excluding these effects resulted in unrealistically small errors in the high current data, which lead to these data being over-emphasized in the fits.
\item Since the gradient for the Jacobian matrix ($\mathbf{X}$) is determined by simulation, it will also have an associated error.  This cannot be included in the $\mathbf{W}$ matrix, because it will be different for each parameter.  However, it can still be calculated, and its effect on the final parameter errors can be estimated.
\end{itemize*}

\subsection{Results}

Figs.~\ref{fig:rfa_results1}~-~\ref{fig:rfa_results3} in Appendix~\ref{app:plots} show the results of the $\chi^2$ analysis for an uncoated aluminum drift chamber.  The plots compare both the transverse and energy distribution of the data and fitted simulation (effectively these are cross sections of the full voltage scan shown in Fig.~\ref{fig:seg_example}).  The error bars shown reflect all of the uncertainties described above.  Overall the data and simulation are in good agreement for a wide variety of beam conditions, including different beam currents, train lengths, beam energies, bunch spacings, and species.  The biggest discrepancy occurs for high current electron beam data.  These are the conditions most likely to produce ion effects~\cite{PhysRevE.52.5487}, which are not included in our model, and may be leading to this disagreement.  

A sampling of results for the other chambers (TiN at Q15W, aC at Q15E, DLC at Q15E, Cu at Q14E, and NEG at L3) are shown in Figs.~\ref{fig:rfa_results_TiN}-\ref{fig:rfa_results_NEG}.  These fits also showed good agreement in general.

The covariance matrix for the parameters is $(\mathbf{X}^T \, \mathbf{W} \, \mathbf{X})^{-1}$.  The standard errors on each parameter are equal to the square root of the diagonal elements of this matrix.  These errors are one dimensional 68$\%$ confidence intervals for each parameter individually, without regard for the values of the other parameters.  The covariance matrix is multiplied by the ``$\chi^2$ per degree of freedom" ($\frac{\chi^2}{n - p}$, where $n$ is the number of data points and $p$ is the number of parameters fitted). Effectively this scales up the uncertainty on the data points, to include (in a somewhat ad hoc manner) any errors that have been left out of the analysis.  The error bars also include an estimate of the uncertainty introduced by errors in the Jacobian matrix, which is added in quadrature to the standard error.  The correlation coefficient of two parameters is defined as $\rho \equiv \frac{C_{i,j}}{\sqrt{C_{i,i} \times C_{j,j}}}$, where $C_{i,j}$ is the $i, j$th element of the covariance matrix.  In general the correlation between parameters is significant.  For example, in the fits shown in Figs.~\ref{fig:rfa_results1}~-~\ref{fig:rfa_results3}, $\rho$ = 0.42 for \texttt{dtspk} and \texttt{P1epk}, 0.22 for \texttt{dtspk} and \texttt{queffp}, and 0.31 for \texttt{P1epk} and \texttt{queffp}.

It should be noted that, with the number of parameters involved in the analysis, it is impossible to say whether we have arrived at the global minimum value of $\chi^2$ in parameter space.  Nonetheless, the ability of this method to achieve a good fit for data taken under a wide variety of beam conditions strongly suggests that the primary and secondary emission models used are reproducing reality to a reasonable degree.

The best fit values and 68$\%$ confidence intervals for the SEY parameters of each chamber are shown in Table \ref{tab:param_result}, and the best fit quantum efficiencies are listed in Table~\ref{tab:queff_list}.   Each of these results represents a fit using a series of voltage scans done during one \cesrta~machine studies run, typically within a few days of each other.  Several such fits were done for most of the chambers, and the results were usually found to be consistent, with a few exceptions.  In particular, some of the fits for aC showed a higher quantum efficiency, but somewhat lower rediffused yield. This may represent a different state of processing of the chamber.  In the results presented here, the fit with the lowest $\chi^2$ for each chamber was chosen.

The peak secondary yield (\texttt{dtopk}) for the uncoated Al chamber was found to be very high ($>$ 2).  This is consistent with values measured elsewhere~\cite{PRSTAB6:034402}. All of the coated chambers (aC, TiN, DLC, and NEG) had much lower values, corresponding in all cases to a peak SEY $\leq$ 0.9, and also consistent with direct measurements~\cite{Suetsugu2007470,NIMA554:92to113,ECLOUD10:MIT00,PRSTAB14:071001}.  The fitted values for TiN and DLC in particular are very low, implying a peak SEY on the order of 0.7.

\begin{figure}[h!]
\centering
\includegraphics[width=.45\textwidth]{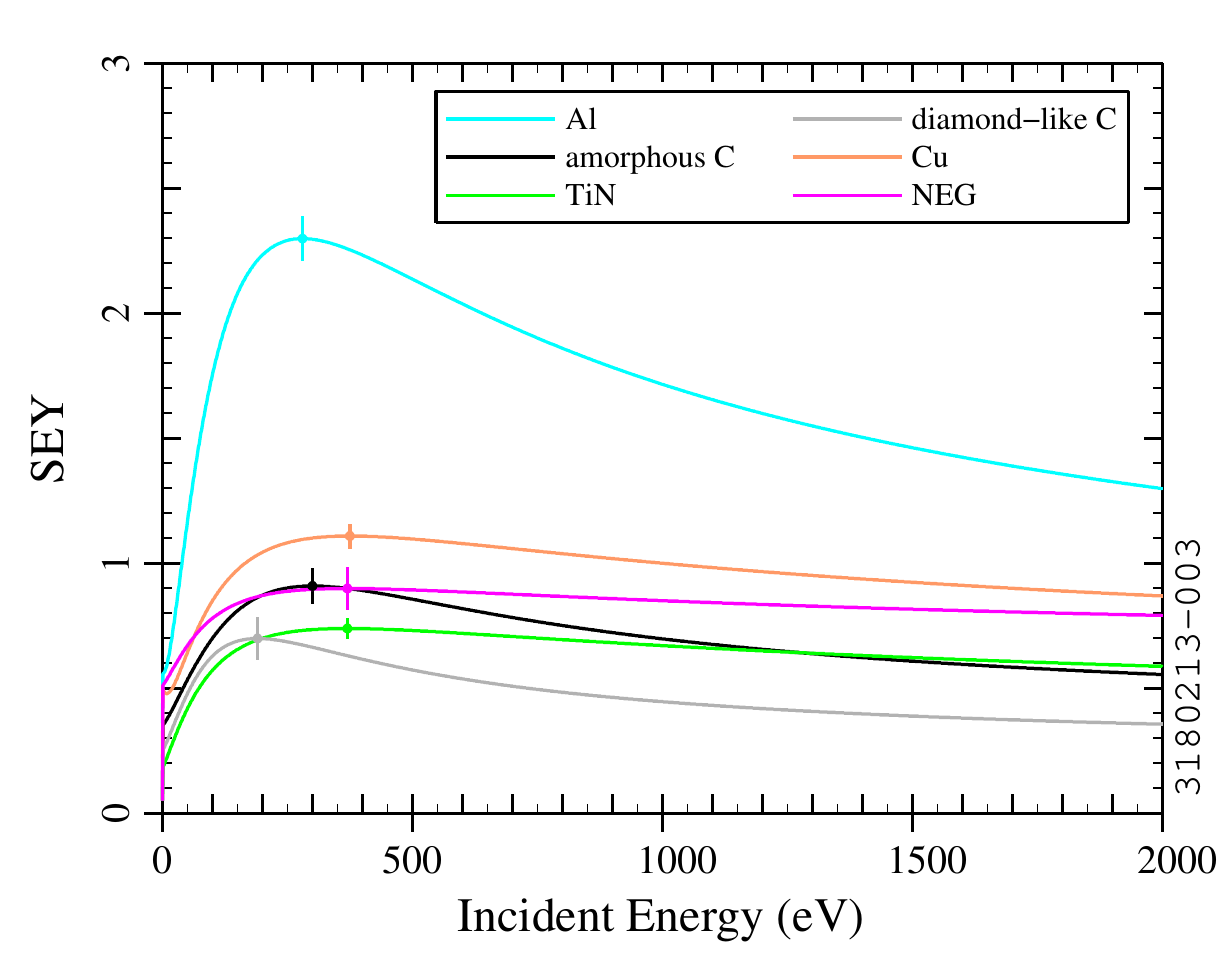}   %{3D_Model_RFA_section_white_big.eps} %{opera_rfa_model.eps} \\
\caption{\label{fig:best_fit_sey} Secondary electron yield curves (at normal incidence) generated by the best fit parameters for each chamber (Table~\ref{tab:param_result}).  Error bars are shown for the peak yield values.}
\end{figure}

The best fit value for the elastic yield (\texttt{P1epk}) was found to be low for both uncoated (Al and Cu) chambers.  As explained, in Section~\ref{ssec:fit_param}, the data in the coated chambers (TiN, aC, DLC, and NEG) was best fit  by assuming a low elastic yield, and varying the rediffused yield (\texttt{P1rinf}) instead.  Since we don't have a direct measurement of the SEY curve for NEG, the initial values for the parameters were (somewhat arbitrarily) taken from TiN.  The fitted values for NEG indicate a much higher rediffused yield than the other coated chambers.  The SEY curves generated by the best fit parameters for each chamber are shown in Fig.~\ref{fig:best_fit_sey}.

Notably, the DLC fit also required a very low value for the ``chamber hole SEY" parameter described in Section~\ref{ssec:fit_param}.   Bench measurements of the SEY of DLC indicate that the material can retain charge if bombarded with a sufficiently high electron flux, thus modifying the apparent SEY~\cite{ECLOUD10:PST12}.  This effect could result in charge around the beam pipe holes influencing the transmission of low energy electrons, reducing the apparent hole SEY~\cite{ARXIV:1402.1904}.

The best fit values for quantum efficiency (\texttt{queffp}) were also lower for the coated chambers.  Amorphous carbon consistently had very low values, less than 5\% for all cases.  In most cases, the quantum efficiency fit was significantly higher for 5.3~GeV than for 2.1~GeV.  Additional work is required to determine whether this is a real effect, or simply an artifact of our incomplete photoelectron model.  Generally speaking, quantum efficiencies on the order of 5~-~10\% are consistent with direct measurements of accelerator materials~\cite{PRSTAB2:063201}.

  %Table~\ref{tab:queff_list} gives all the best fit quantum efficiency values for each chamber.

\begin{table*}
   %  \begin{minipage}{.95\textwidth}
   \centering
%   \caption{\label{tab:param_result_NEG} Best fit parameters- L3 NEG-coated chamber, December 2010}
    \setlength{\tabcolsep}{8pt}
    \caption{\label{tab:param_result} Best fit SEY parameters.  Error bars are given for the parameters used in the fit.  Values without error bars were either assumed, or taken from in-situ SEY measurements (see Section~\ref{ssec:parameters}).}
   \begin{tabular}{ccccccc}
       \hline \hline
       Parameter  & Al    & Cu  &   TiN &   aC  &   DLC &   NEG \\
       \hline
            dtspk   &  	2.08 $\pm$ 0.09  &  0.81 $\pm$ 0.05  &	0.59 $\pm$ 0.03 &  0.59 $\pm$ 0.05    &  0.48 $\pm$ 0.06 &  0.42 $\pm$ 0.07 \\
            P1epk   &   0.36 $\pm$ 0.03   &  0.22  $\pm$ 0.07  & 0.05  &  0.05  & 0.05  & 0.05   \\
            P1rinf  &   0.2              &  0.28  &  0.30 $\pm$ 0.05   & 0.13 $\pm$ 0.03  &  0.20 $\pm$ 0.06    &  0.46 $\pm$ 0.05   \\
            \textbf{dtotpk}  &   \textbf{2.3 $\pm$ 0.1}    & \textbf{1.11 $\pm$ 0.09}  & \textbf{0.75  $\pm$ 0.04} &  \textbf{0.91 $\pm$ 0.07}  & \textbf{0.70 $\pm$ 0.08}  &  \textbf{0.90 $\pm$ 0.09} \\
            E0epk   &   280~eV           &   375~eV  &  370~eV &  370~eV   & 190~eV  & 370~eV \\
            powts   &   1.54            &   1.38 &  1.32 &  1.77    & 1.77  &  1.32  \\
        \hline \hline
   \end{tabular}
    %   \end{minipage}
\end{table*}

\begin{table*}
%   \begin{minipage}{.9\textwidth}
   \centering
   \setlength{\tabcolsep}{10pt}
   \caption{\label{tab:queff_list} Table of best fit quantum efficiencies (in percent).}
   \begin{tabular}{ccccccc}
       \hline \hline
        Beam            &   Al              &   Cu            &   TiN           &   aC              &   DLC             &   NEG             \\ 	
        \hline 
        2.1 GeV, e$^+$  &   11.3 $\pm$ 1.4  & 2.5 $\pm$ 0.8   & 4.9 $\pm$ 0.2   &   3.6 $\pm$ 0.5   &   4.5 $\pm$ 0.6   &   2.9 $\pm$ 0.9   \\
        2.1 GeV, e$^-$  &   8.0 $\pm$ 1.1   & 4.7 $\pm$ 0.7   & -               &   -               &   7.1 $\pm$ 0.6   &   -               \\
        4 GeV, e$^+$    &   10.0 $\pm$ 1.2  & 15.0 $\pm$ 2.0  & -               &   -               &   -               &   -               \\
        5.3 GeV, e$^+$  &   10.3 $\pm$ 1.2  & 15.3 $\pm$ 2.8  & 8.9 $\pm$ 0.7   &   4.6 $\pm$ 0.6   &   9.1 $\pm$ 1.1   &   14 $\pm$ 2      \\
        5.3 GeV, e$^-$  &   10.5 $\pm$ 1.4  & 12.1 $\pm$ 1.8  & 5.0 $\pm$ 0.4   &   4.9 $\pm$ 0.6   &   7.1 $\pm$ 0.6   &   -               \\
        Average         &   10.0            & 9.9             & 6.3             &   4.4             &   7.0             &   8.5             \\
        \hline \hline
       %\bottomrule
   \end{tabular}
    %   \end{minipage}
\end{table*}

\section{Conclusions}

Retarding field analyzers have been installed in drift regions around CESR, and a great deal of electron cloud data has been collected with them.  Detailed models of our RFAs have been developed, and integrated into the cloud simulation code POSINST, allowing for analysis on a more fundamental level.  This has enabled the calculation of best fit simulation parameters, which describe the primary and secondary electron emission characteristics of each material in situ.  The fits indicate that TiN and DLC have especially low secondary yields, while aC has the lowest quantum efficiency.

Electron emission properties of material surfaces, such as quantum efficiency and secondary emission yield, are traditionally measured employing dedicated, well-controlled laboratory devices applied to clean, smooth surfaces.  In contrast, our analysis determines several model parameters via a simultaneous, multi-parameter fit to data obtained with RFAs installed in the \cesrta~vacuum chamber. Thus, while none of the above-mentioned parameters is determined with great precision, our exercise amounts to a more global fit to the model, and yields reasonable values for the parameters. In combination with many other kinds of measurements and simulations within the \cesrta~program~\cite{NIMA749:42to46,ECLOUD10:PST08,ECLOUD10:PST12,PAC11:WEP108} , our results lend validity to the electron emission model embodied in the simulation code.

Our approach has the additional advantage that it allows the assessment of the performance of various chamber materials vis-\`{a}-vis the electron-cloud problem for actual chamber surfaces within a realistic storage ring environment. As such, our analysis takes intrinsic account of such issues as surface roughness, material composition, and beam conditioning. Given the ubiquitousness of the electron-cloud effect, our results are directly and immediately applicable to other high-energy or high-intensity storage rings, whether lepton or hadron.

%\section{Acknowledgements}

%\begin{acknowledgements}
\begin{acknowledgments}

This work was supported by the US National Science Foundation (PHY-0734867 and PHY-1002467) and Department of Energy (DE-FC02-08ER41538).

The results presented in this paper were made possible by the hard work of the \cesrta~collaboration, especially L. Bartnik, M.G. Billing, J.V. Conway, J.A. Crittenden, M. Forster, S. Greenwald, Y. Li, X. Liu, R.E. Meller, S. Roy, S. Santos, R.M. Schwartz, J. Sikora, K. Sonnad, and C.R. Strohman.  We are also grateful to S. Calatroni and G. Rumolo at CERN for providing us with the aC coated chambers, and S. Kato at KEK for the DLC chamber.

Finally, the authors would like to thank D. Rubin at Cornell for his advice and guidance, and M. Furman at LBNL for his support with the POSINST simulation code.

%\end{acknowledgements}
\end{acknowledgments}

\newpage %Just because of unusual number of tables stacked at end
\bibliographystyle{medium}
\bibliography{rfa_drift_prst}% Produces the bibliography via BibTeX.

\newpage

\begin{figure*}
\begin{minipage}{.98\textwidth}
\centering

\appendix
\section{\label{app:plots} Plots of Fit Results}

\vspace{10mm}

\begin{tabular}{cc}
\includegraphics[width=.32\linewidth]{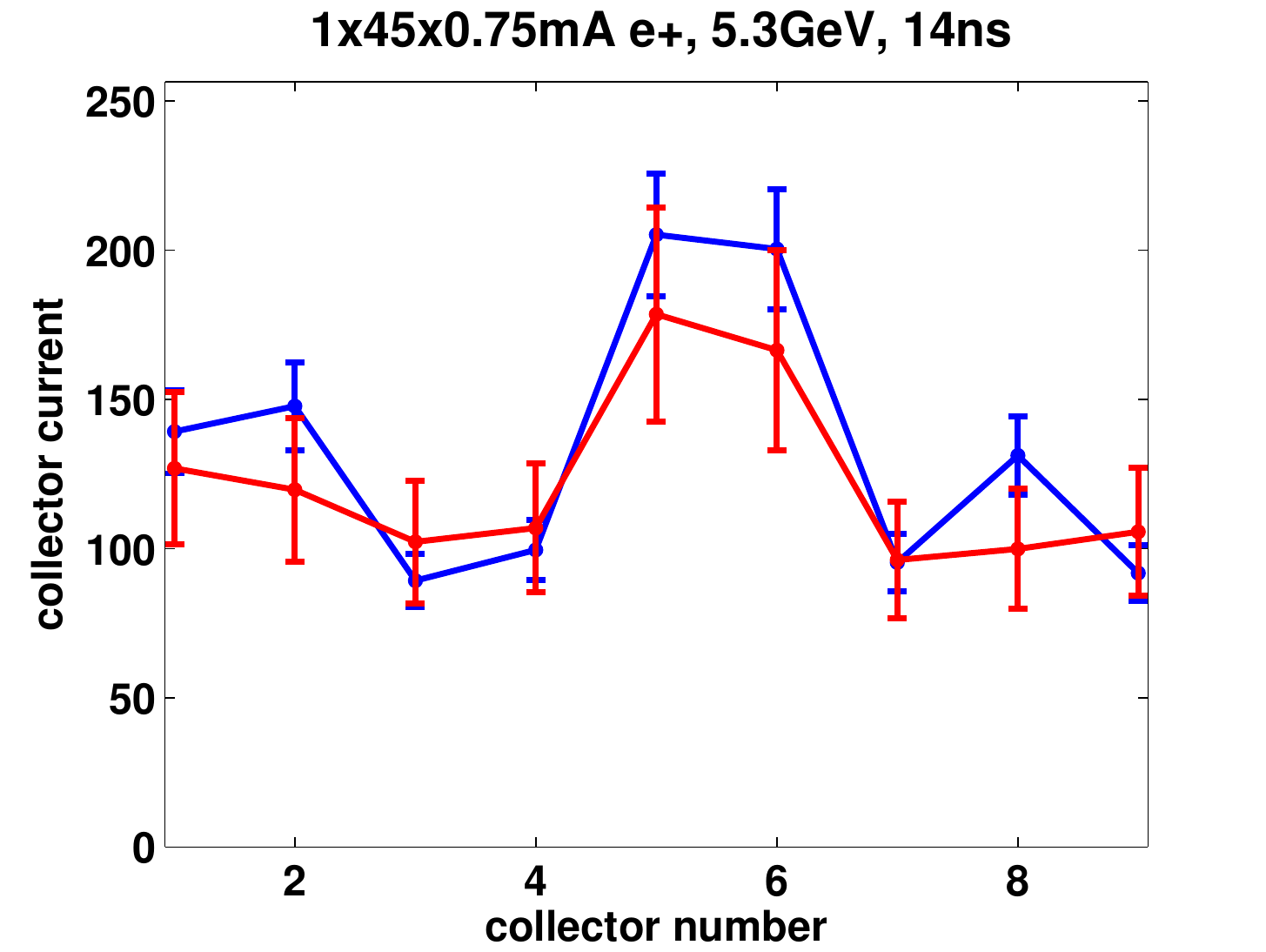}
\includegraphics[width=.32\linewidth]{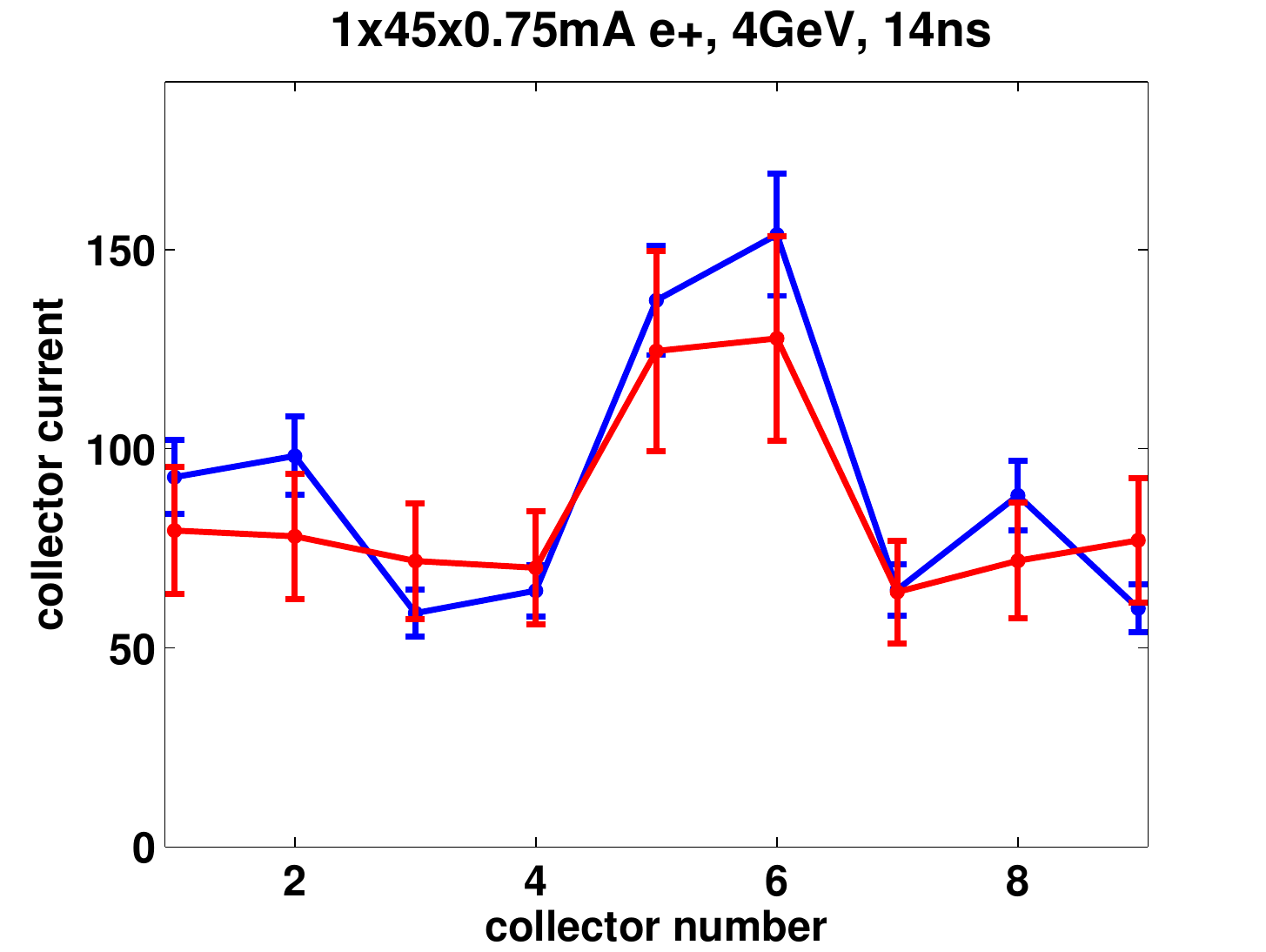}
\includegraphics[width=.32\linewidth]{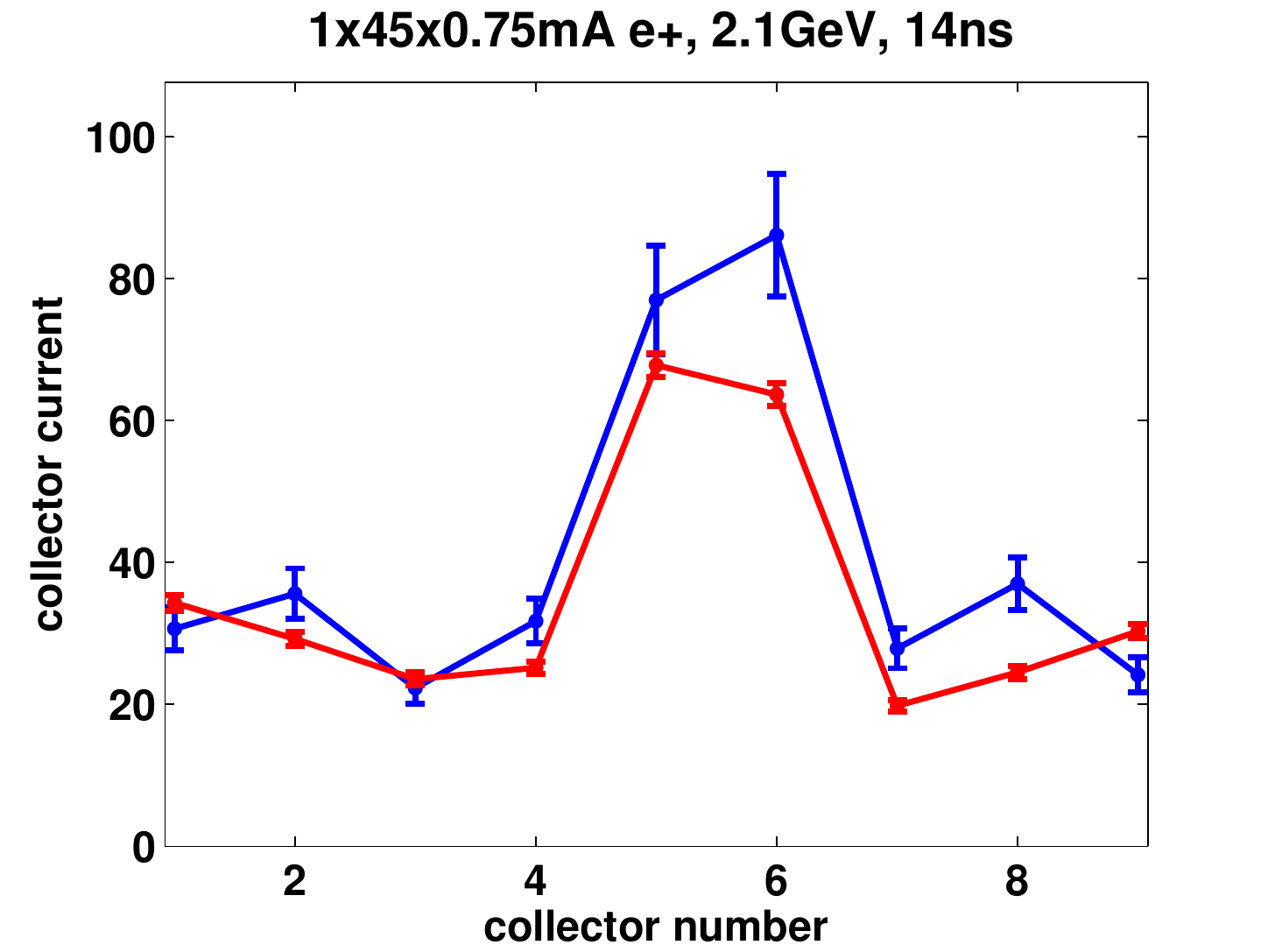} \\
\includegraphics[width=.32\linewidth]{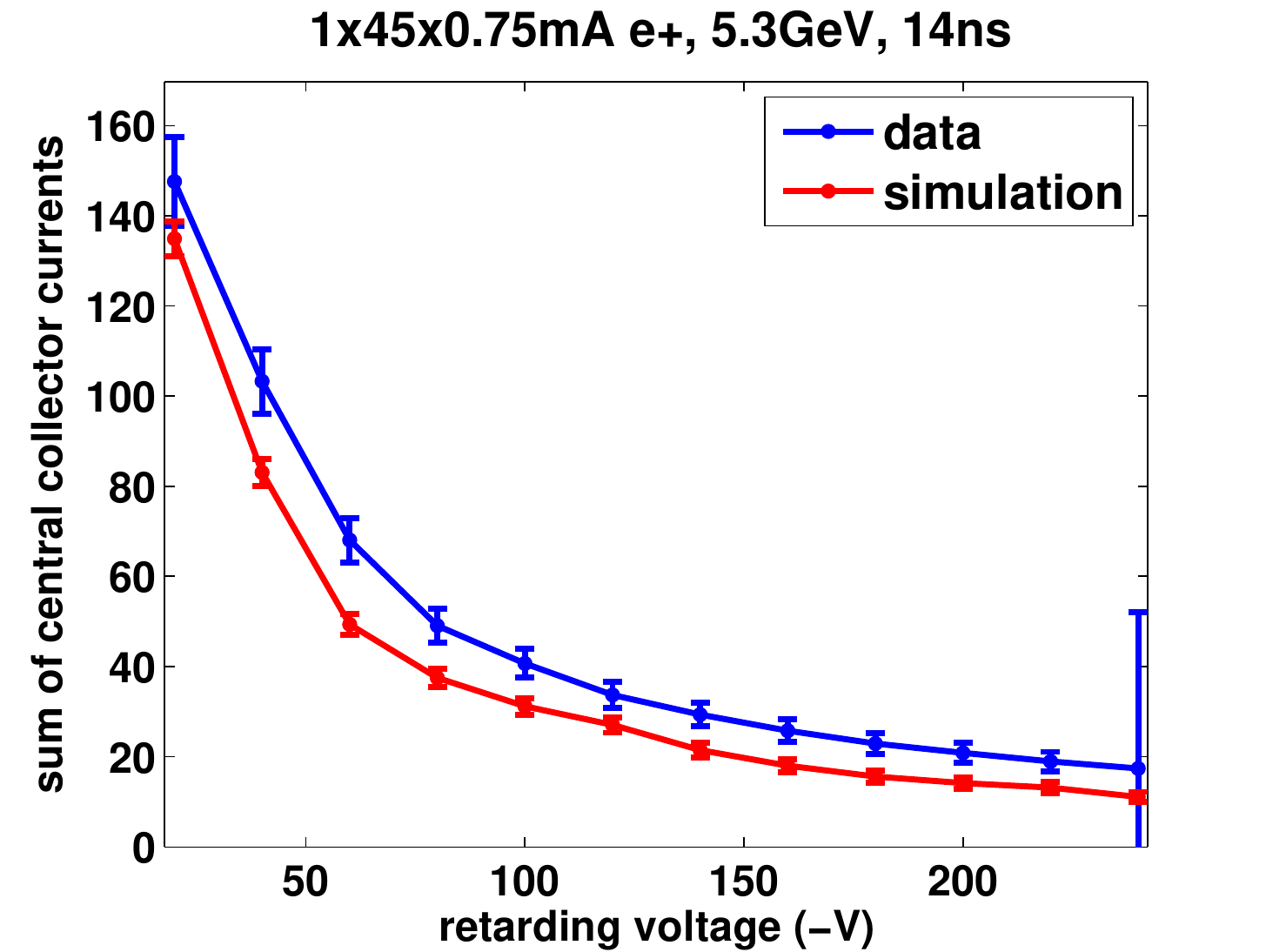}
\includegraphics[width=.32\linewidth]{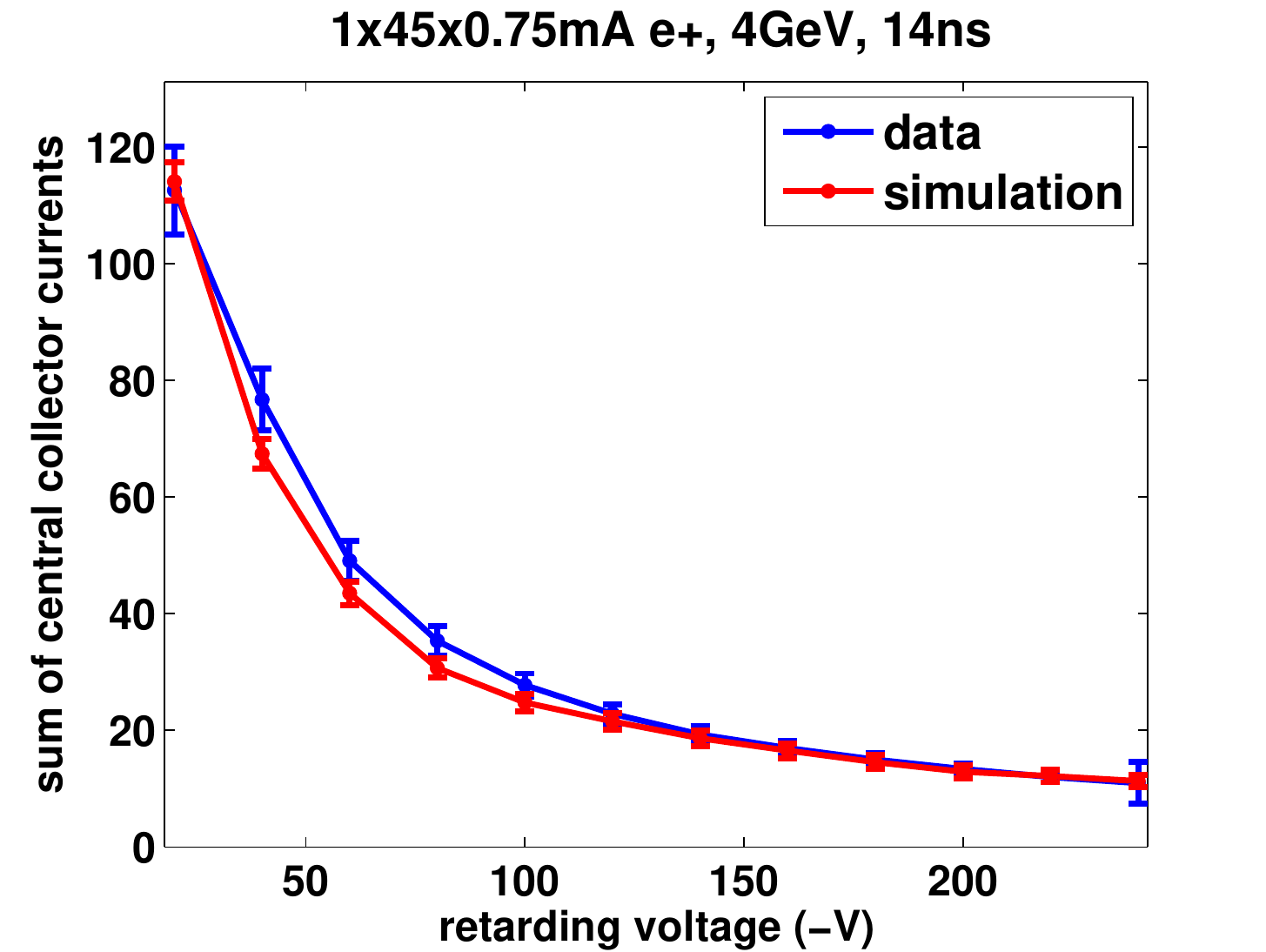}
\includegraphics[width=.32\linewidth]{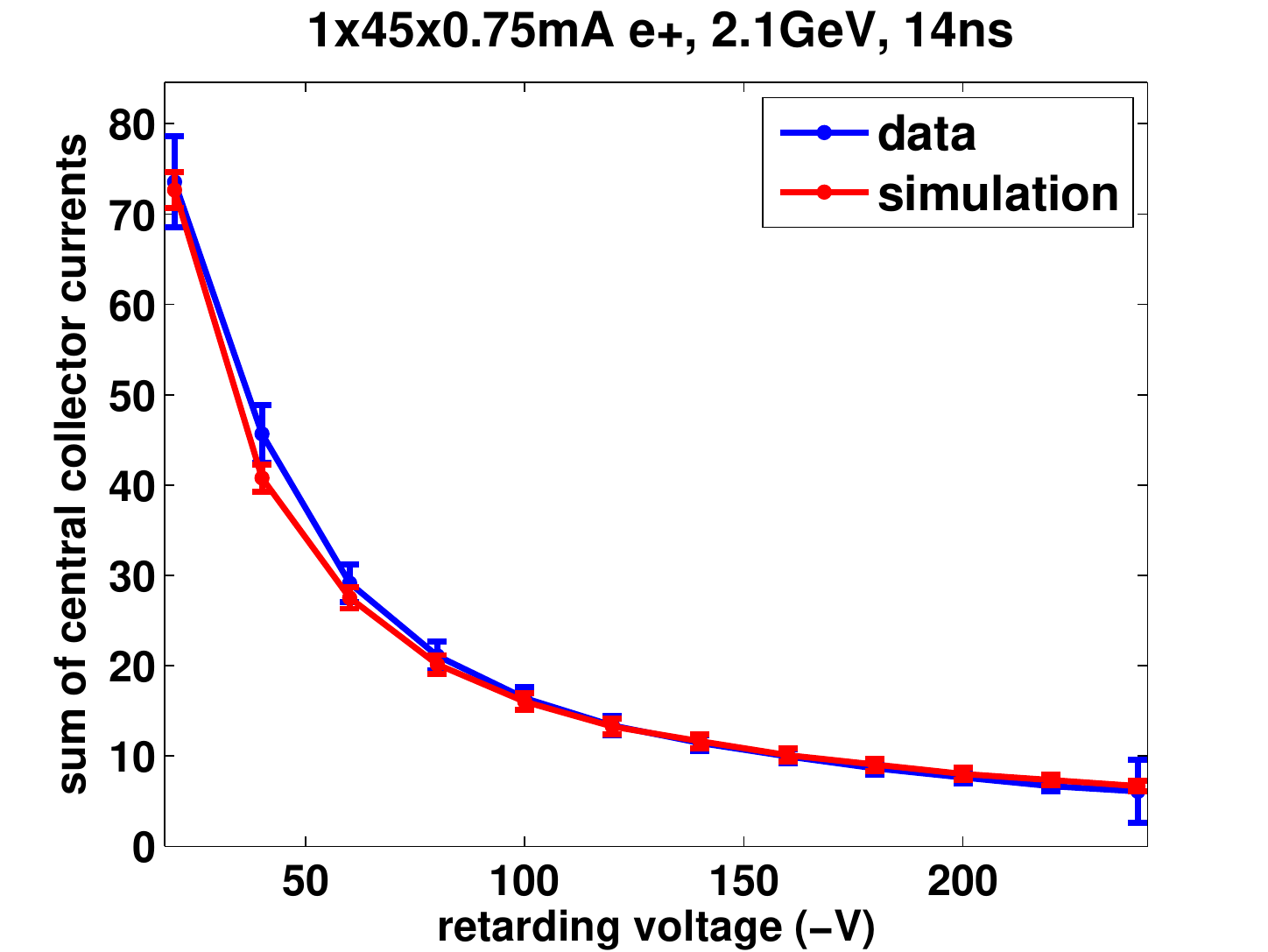} \\
\noalign{\vskip 10mm}

\includegraphics[width=.32\linewidth]{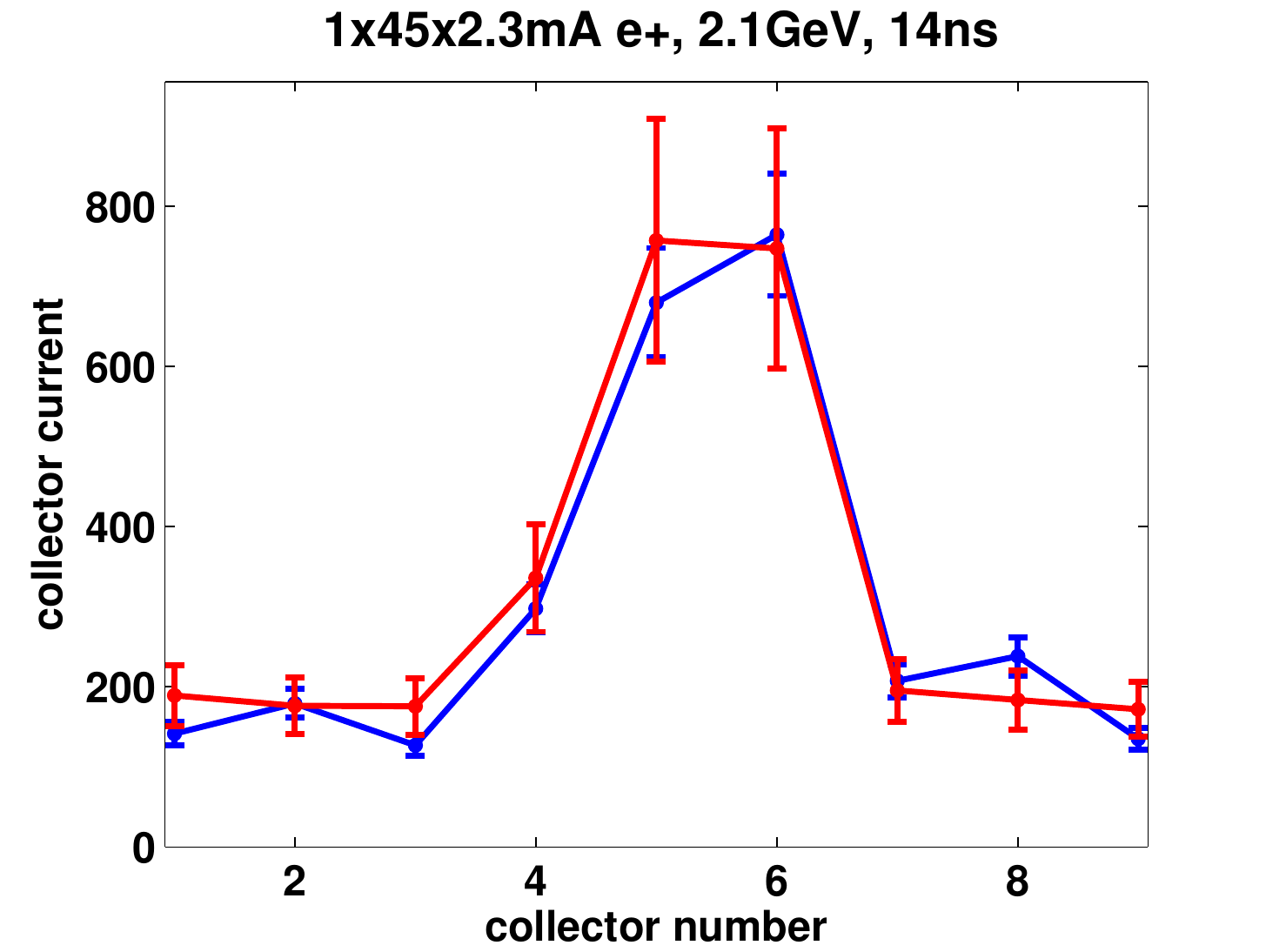}
\includegraphics[width=.32\linewidth]{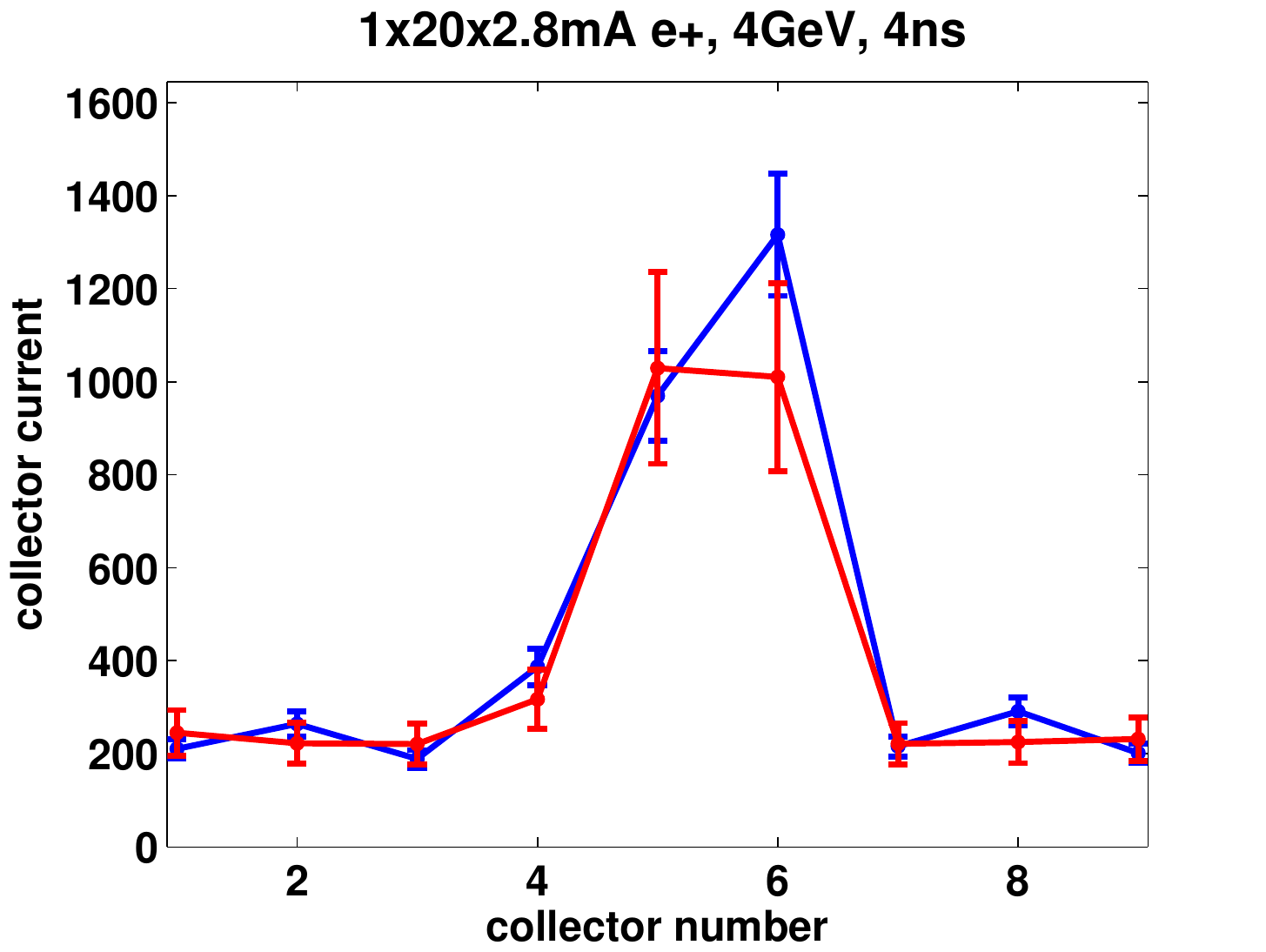}
\includegraphics[width=.32\linewidth]{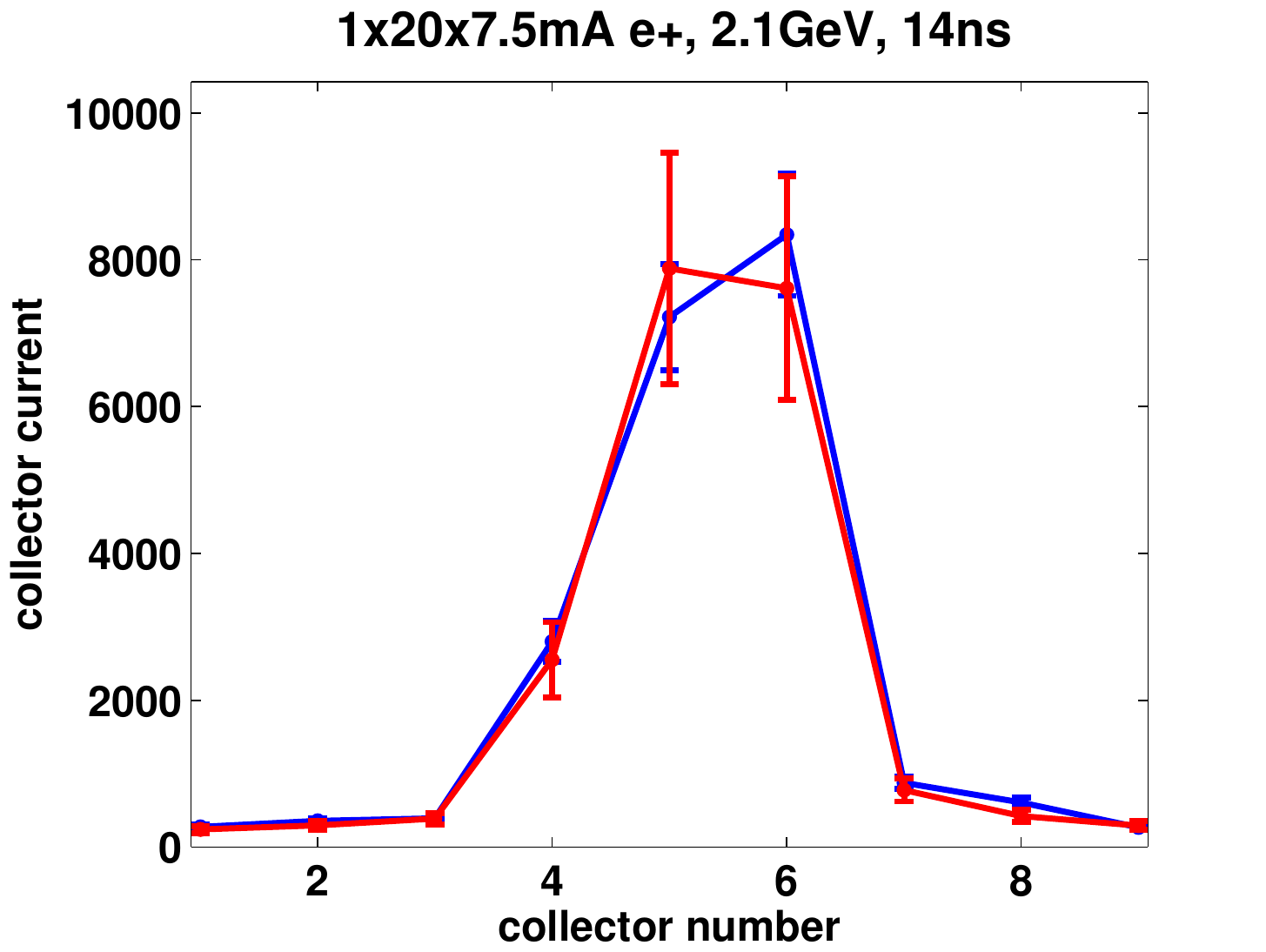} \\
\includegraphics[width=.32\linewidth]{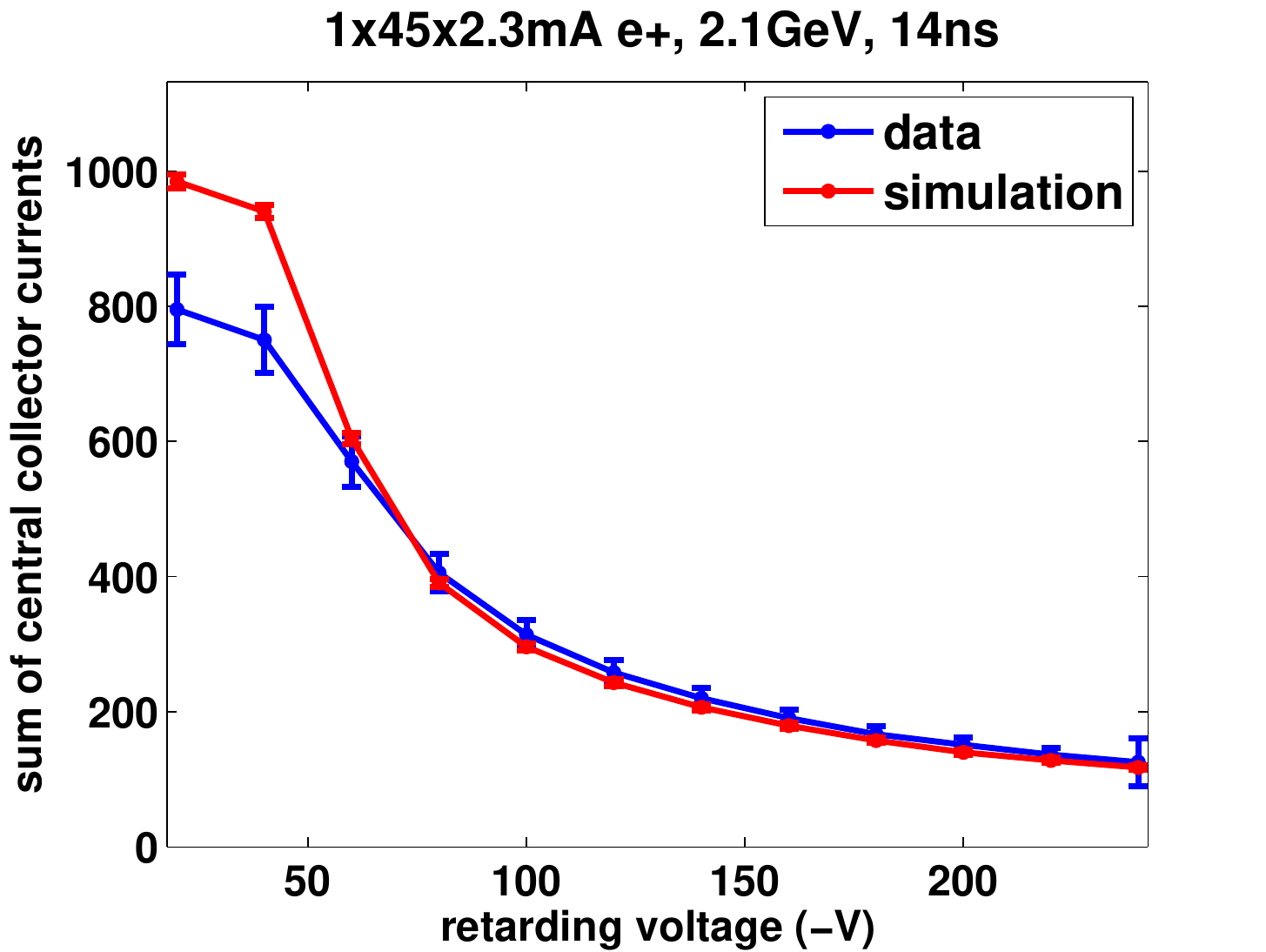}
\includegraphics[width=.32\linewidth]{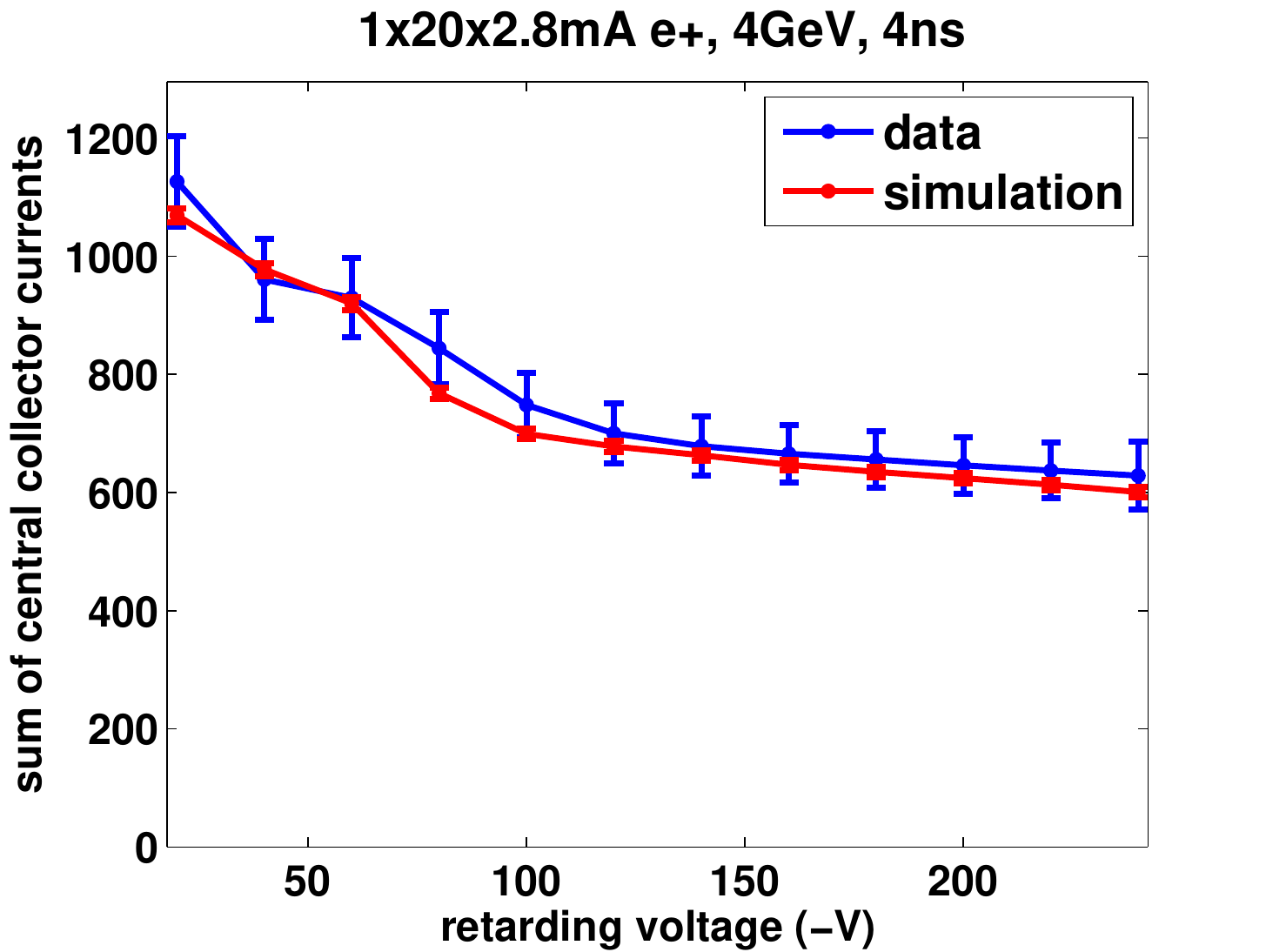}
\includegraphics[width=.32\linewidth]{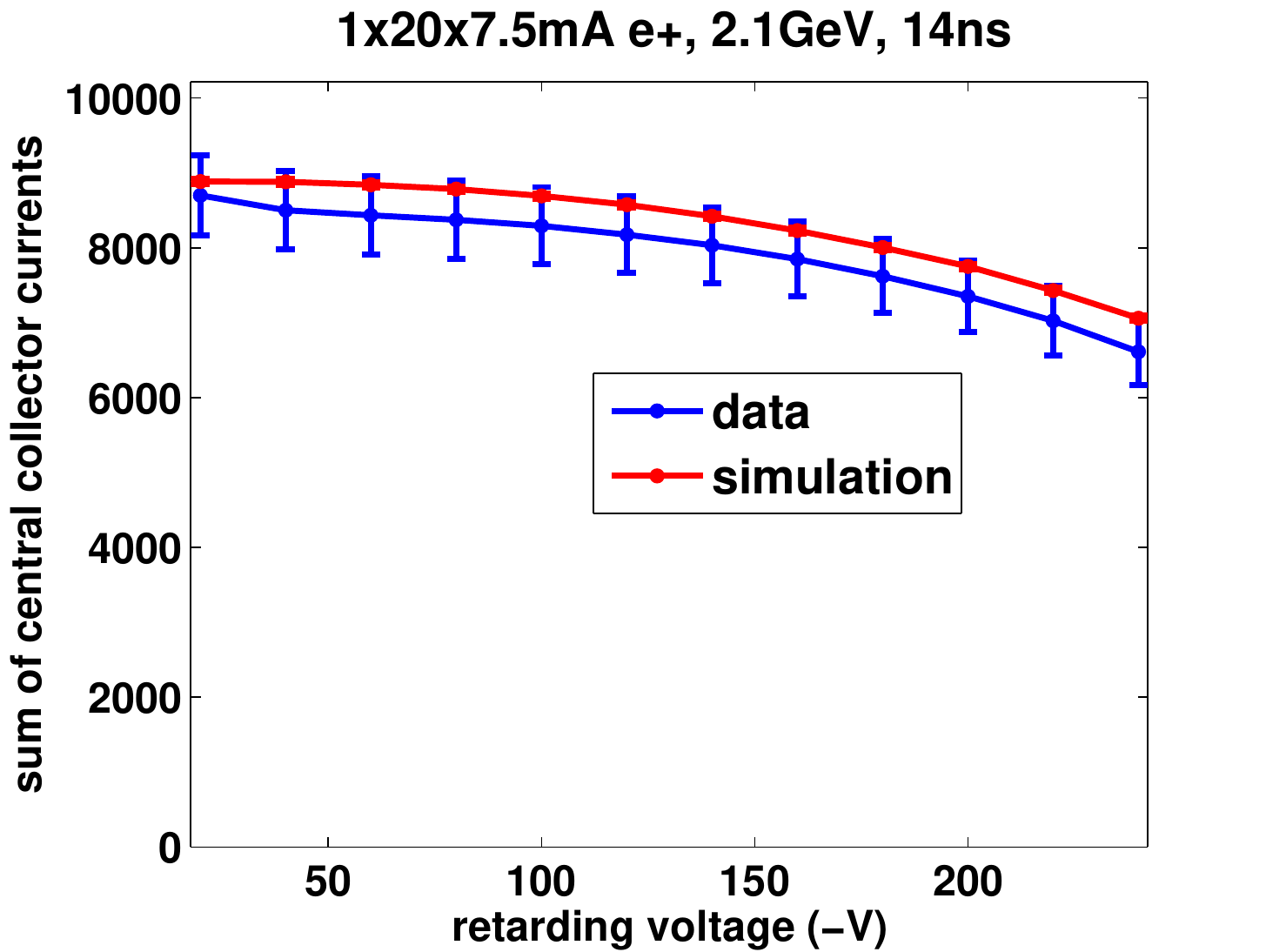} \\

\end{tabular}
\caption[Comparison of RFA data and simulation, using best fit parameters.]{\label{fig:rfa_results1} Comparison of Q15W Al RFA data and simulation, using best fit parameters (Table~\ref{tab:param_result}), conditions 1 - 6 (Table~\ref{tab:condx_list}).  The RFA is ``thin" style (Table~\ref{tab:rfa_styles}).  The top plots show the total signal across the 9 RFA collectors (with +50~V on the grid); the bottom plots show the signal in the central three collectors vs retarding voltage.}
%The plots show the signal across the 9 RFA collectors at three different retarding voltages.}
%\vskip 30mm
\end{minipage}
\end{figure*}

%\vspace{10cm}

\begin{figure*}
\begin{minipage}{.98\textwidth}
\centering
\begin{tabular}{cc}
\includegraphics[width=.32\linewidth]{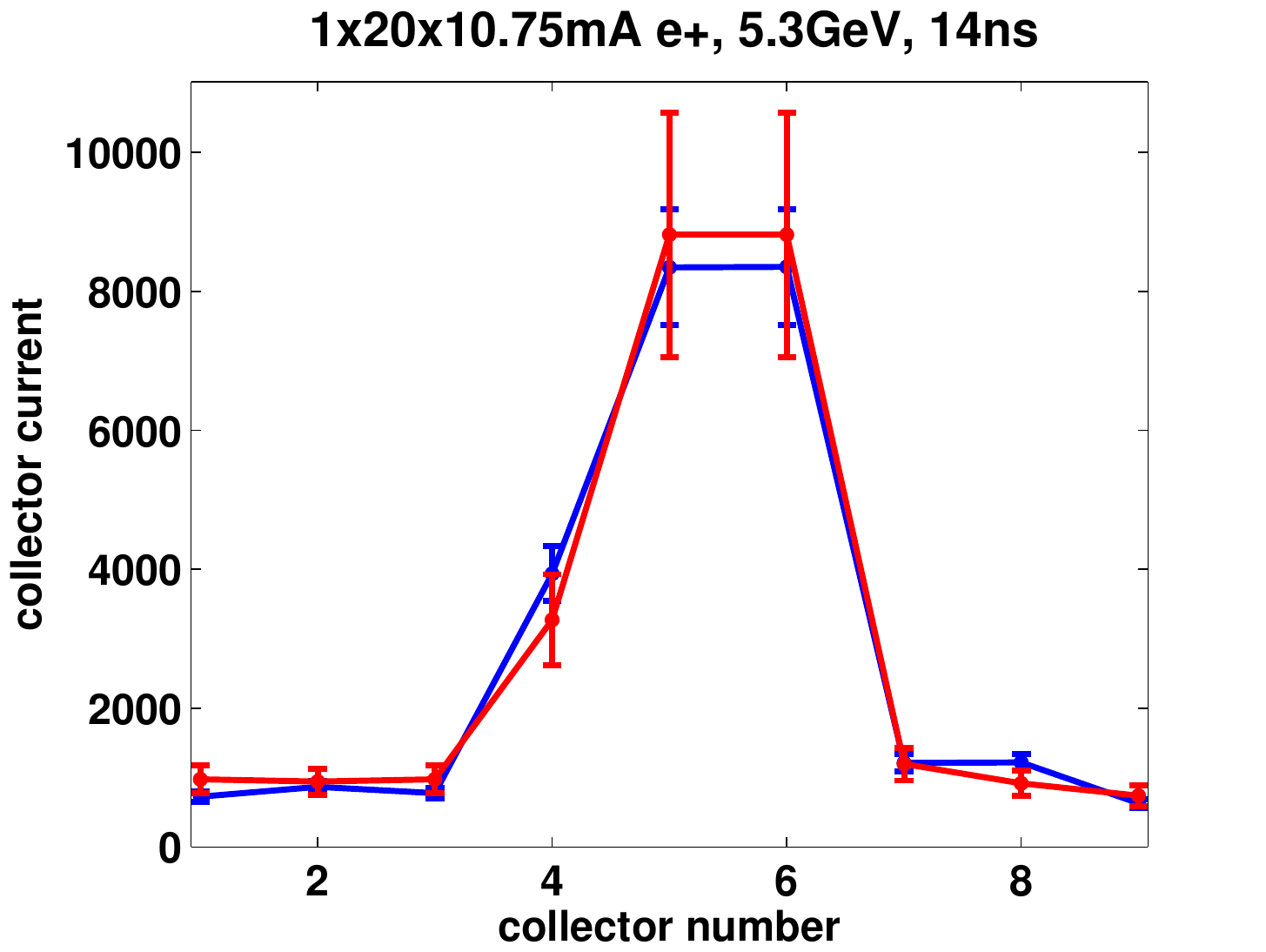}
\includegraphics[width=.32\linewidth]{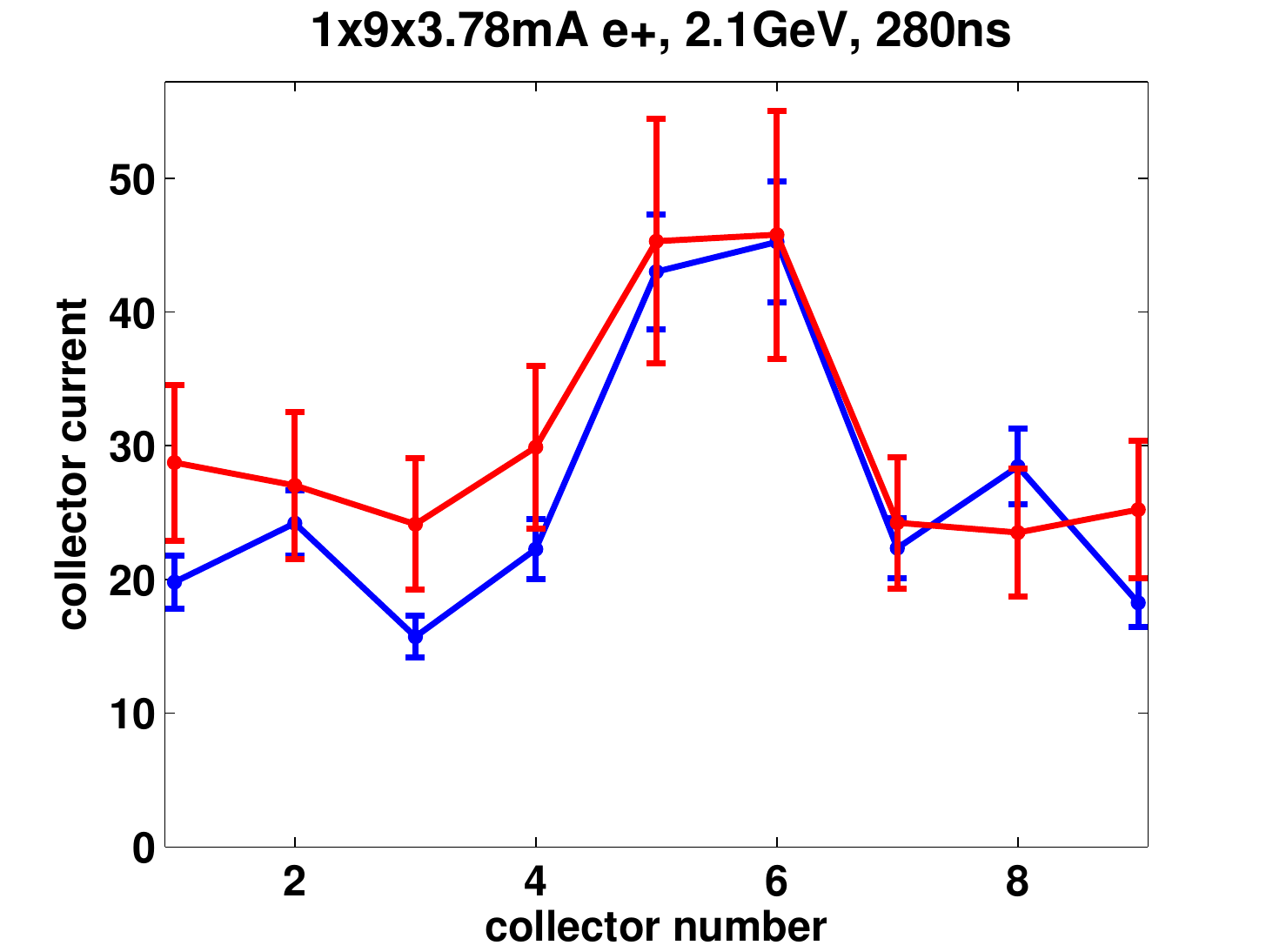}
\includegraphics[width=.32\linewidth]{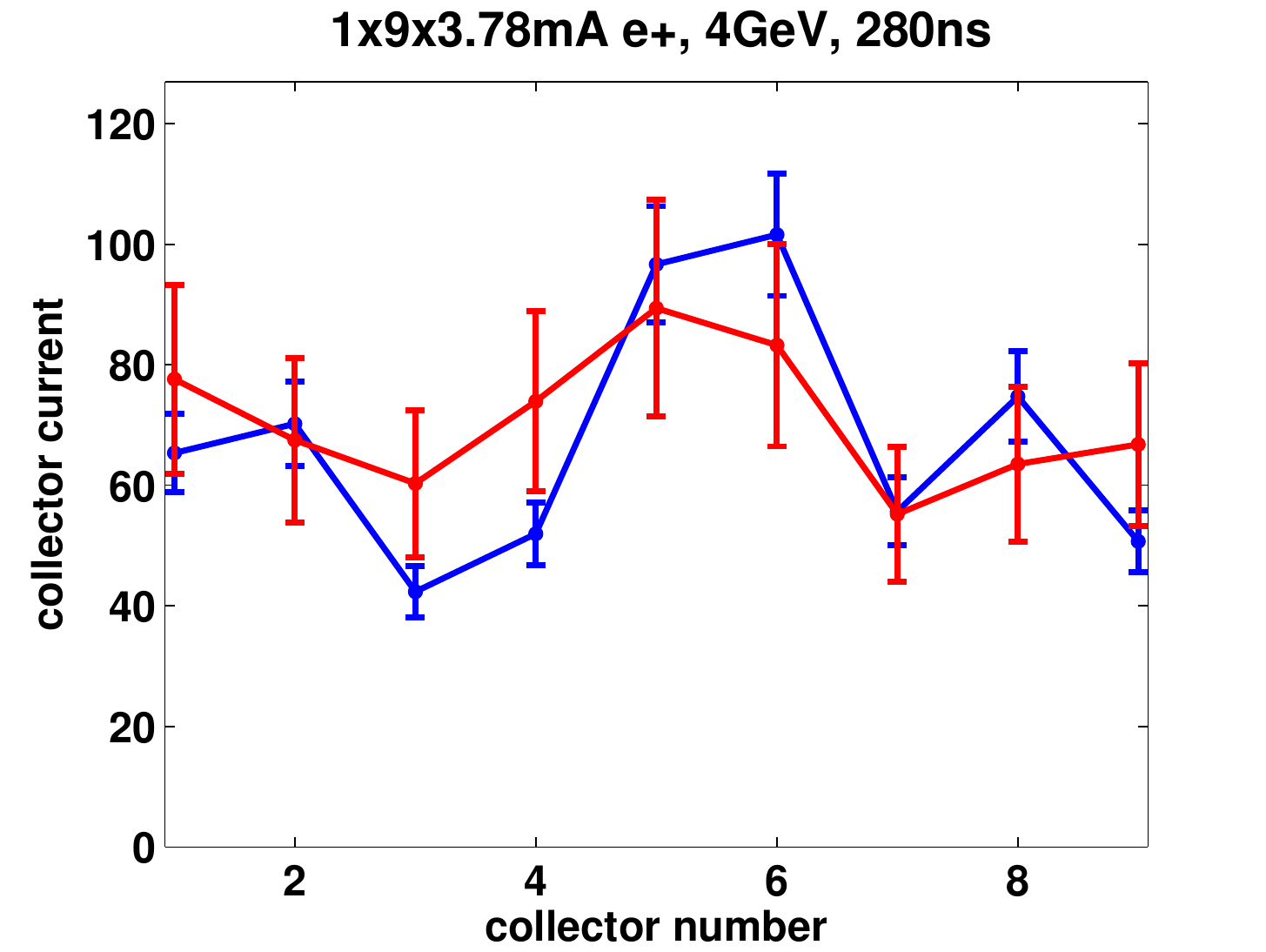} \\
\includegraphics[width=.32\linewidth]{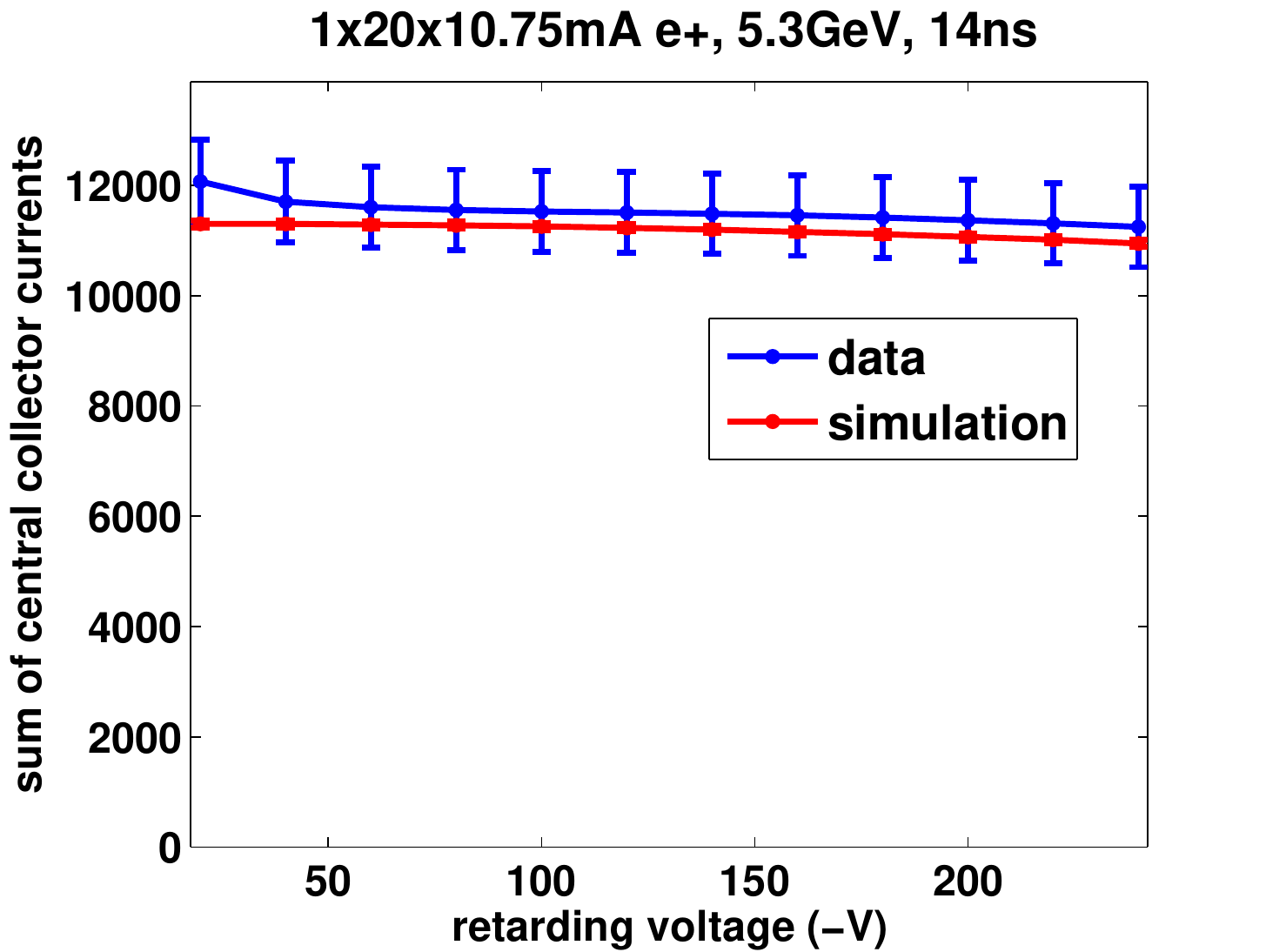}
\includegraphics[width=.32\linewidth]{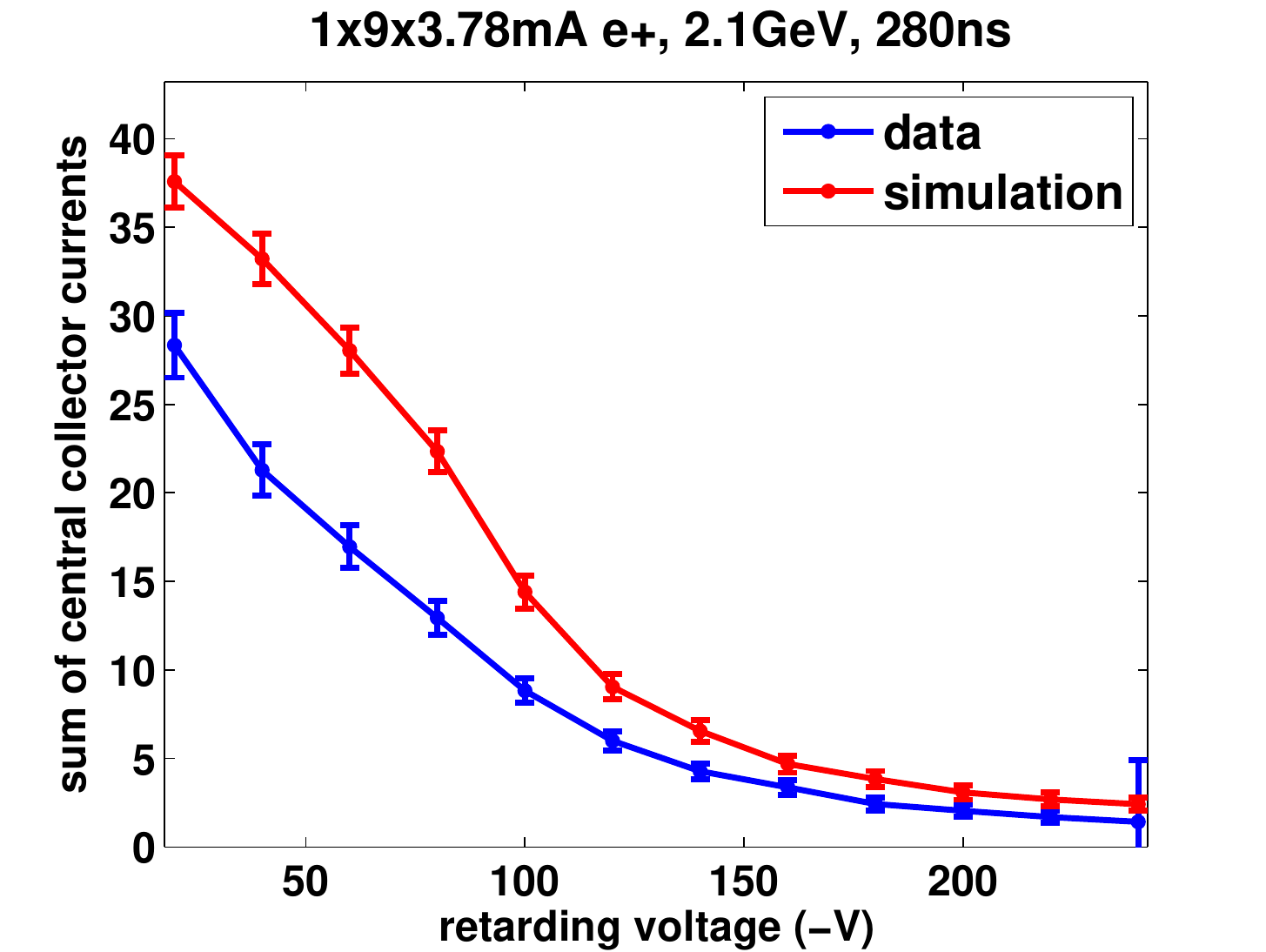}
\includegraphics[width=.32\linewidth]{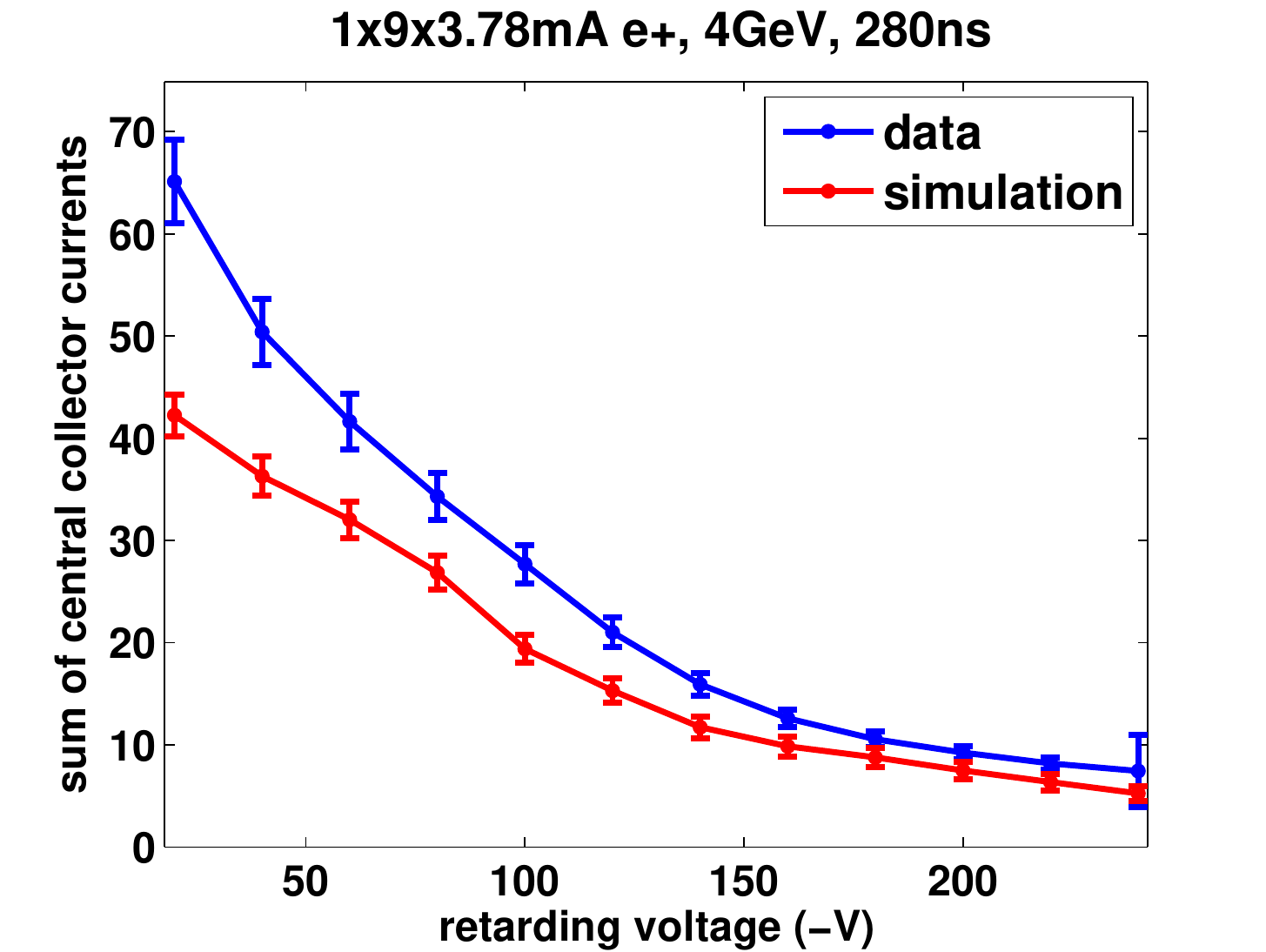} \\
\noalign{\vskip 10mm}
\includegraphics[width=.32\linewidth]{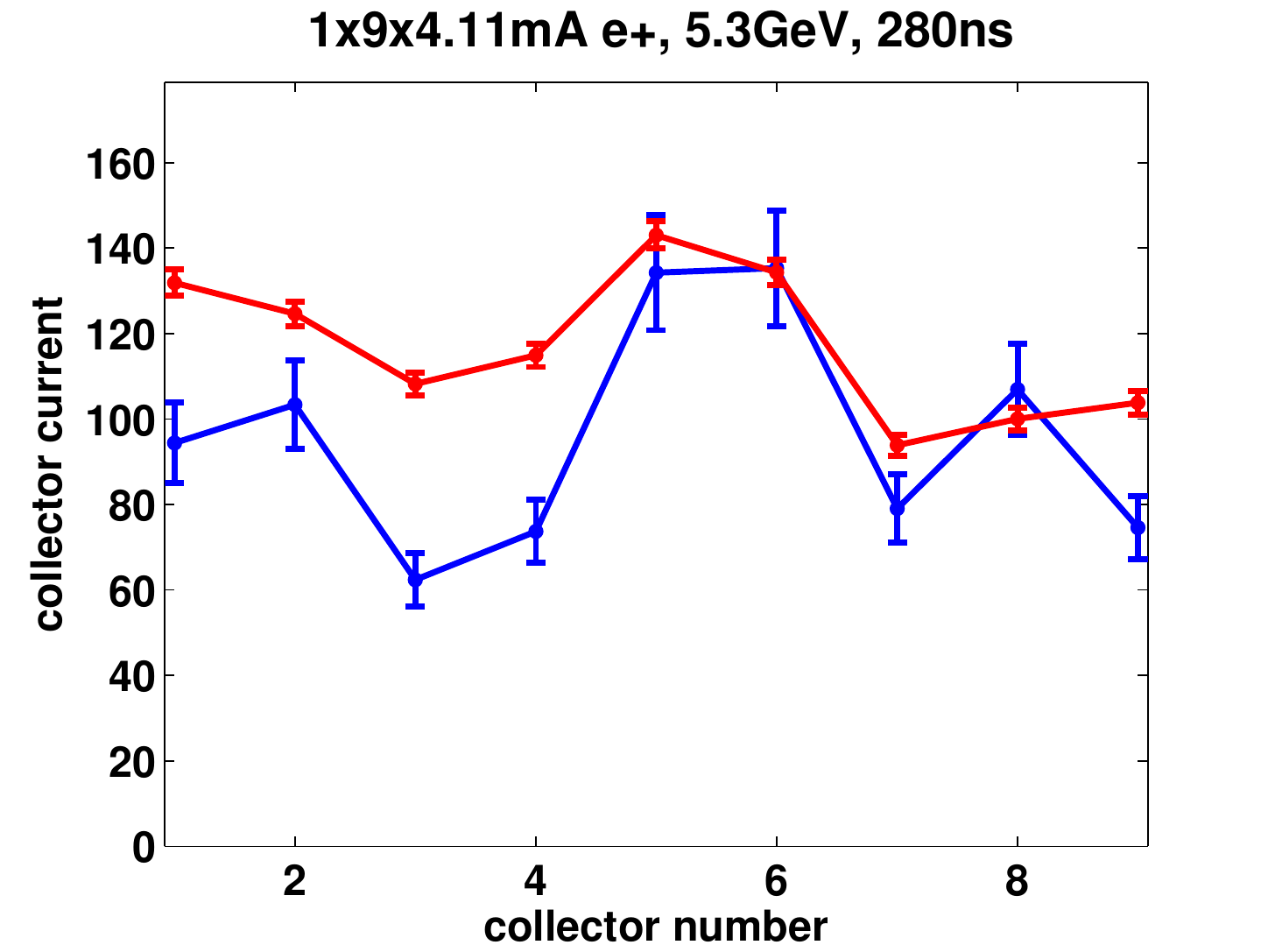}
\includegraphics[width=.32\linewidth]{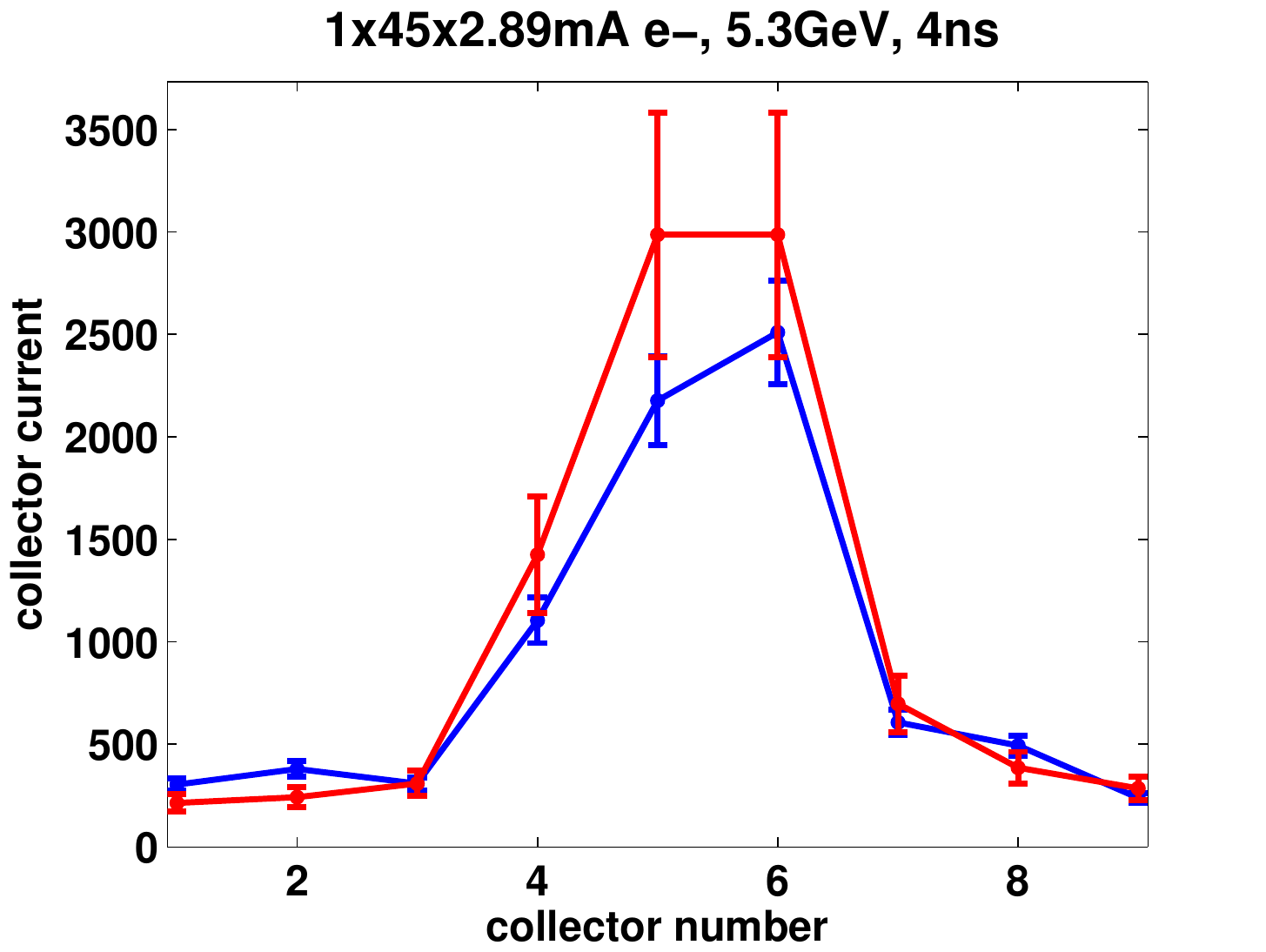}
\includegraphics[width=.32\linewidth]{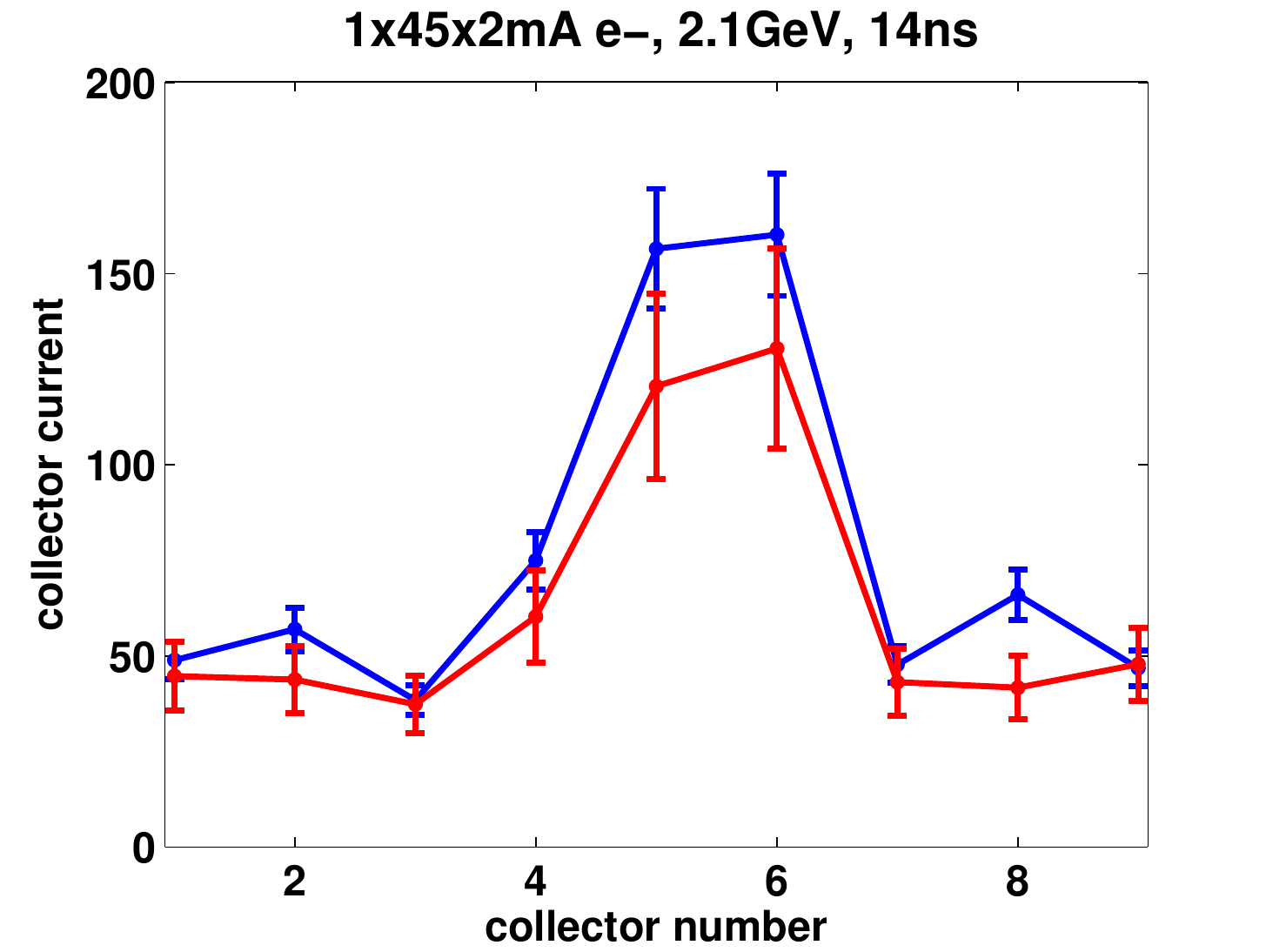} \\
\includegraphics[width=.32\linewidth]{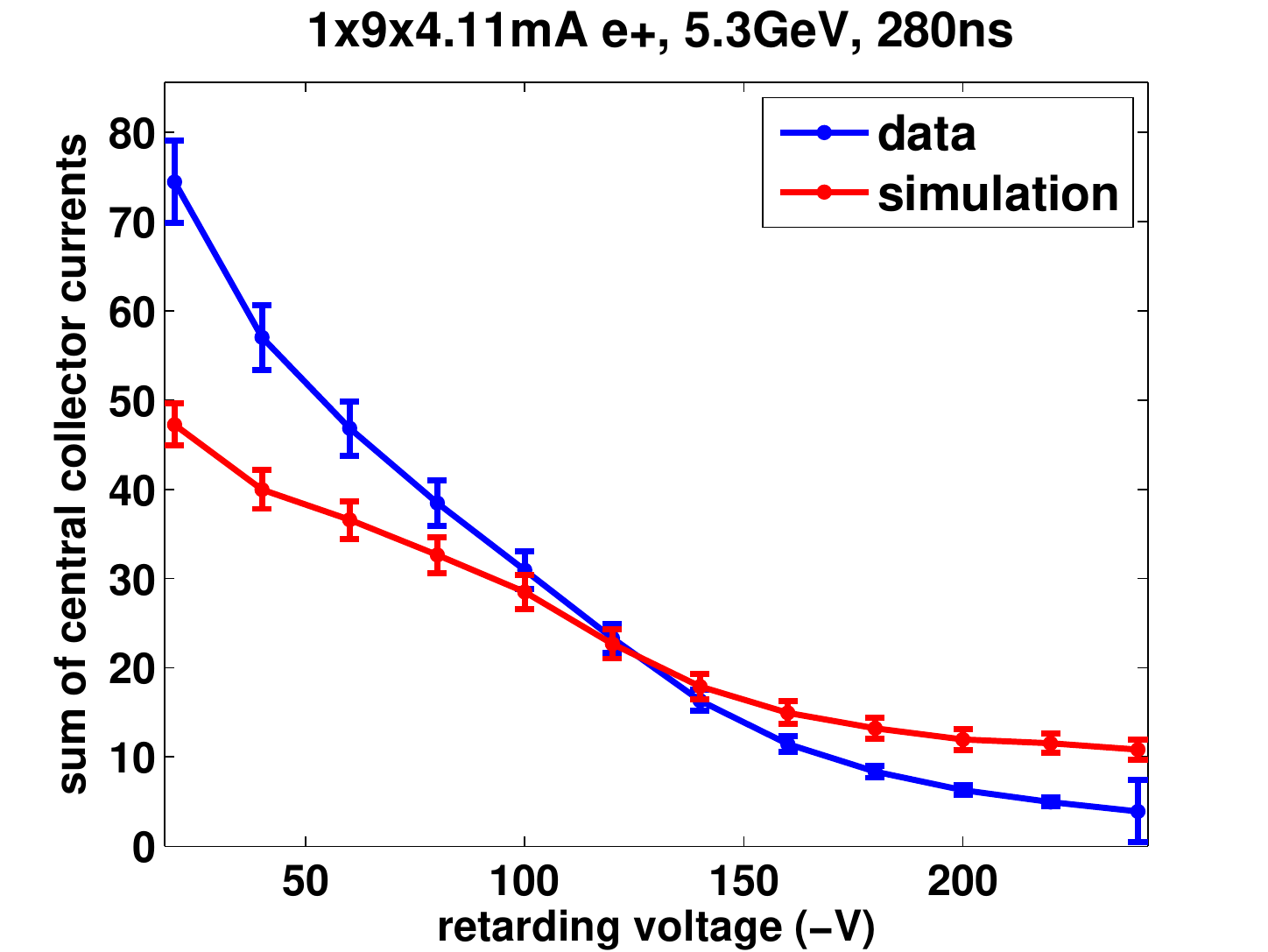}
\includegraphics[width=.32\linewidth]{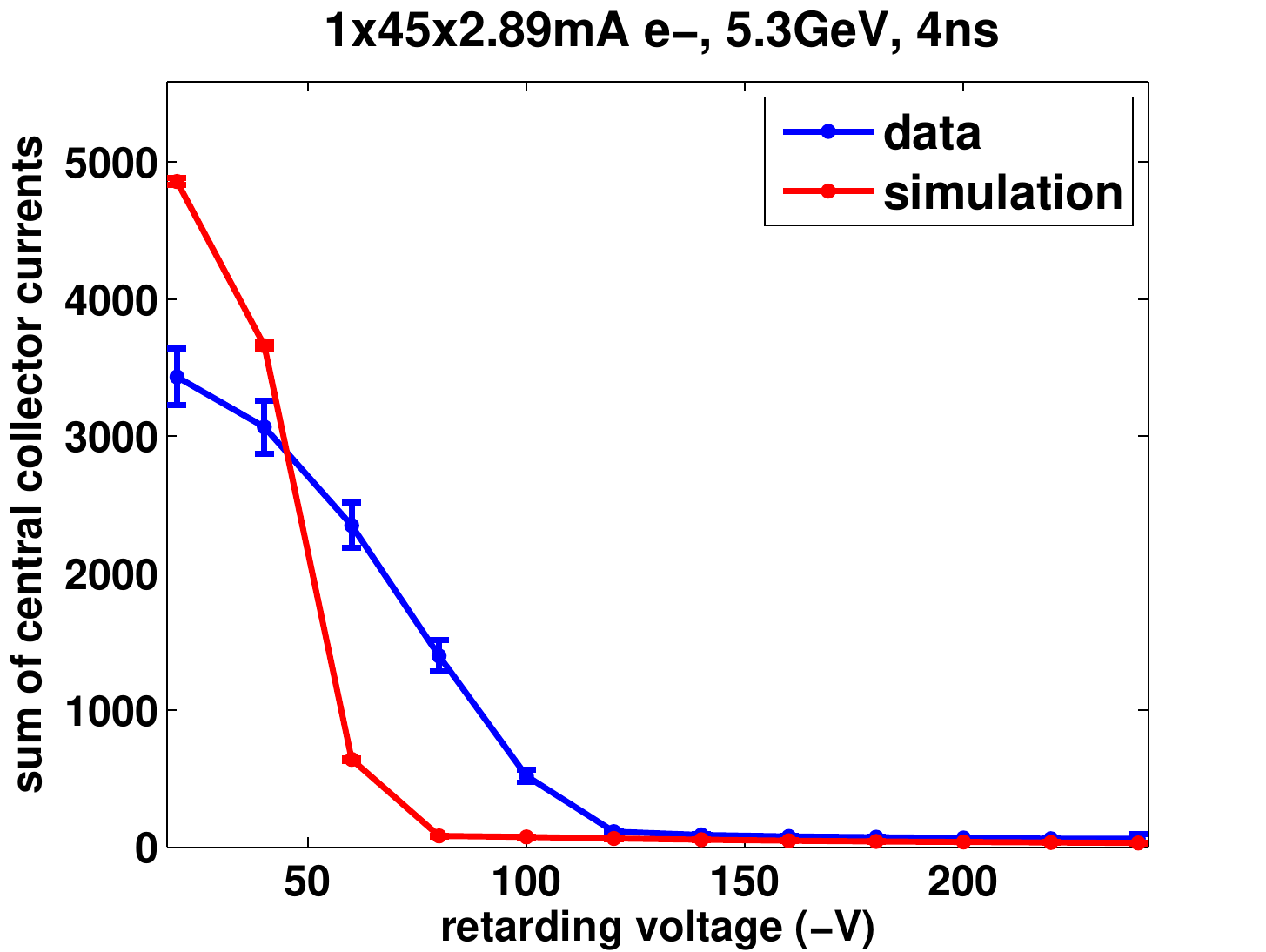}
\includegraphics[width=.32\linewidth]{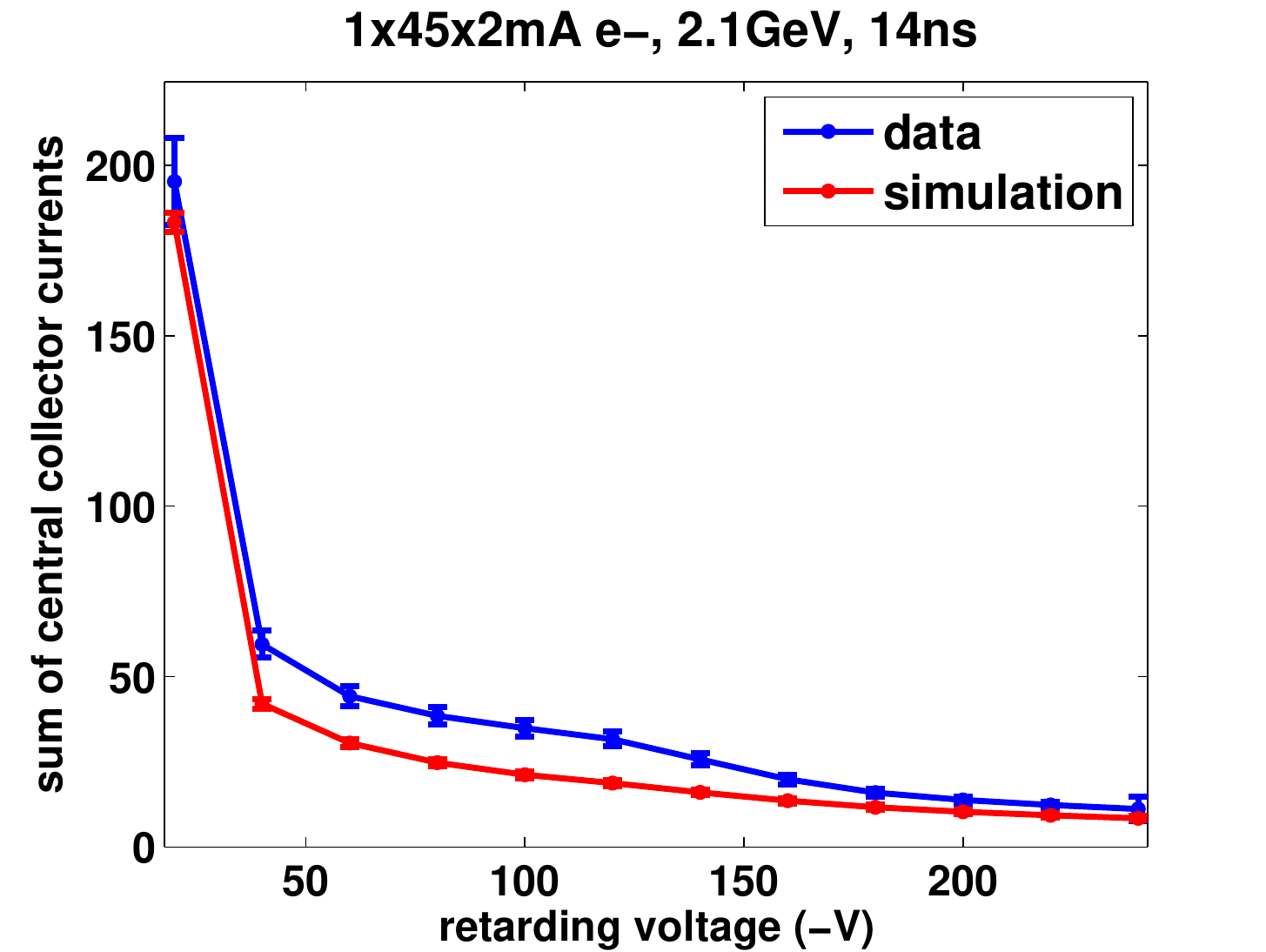} \\
%\hline \hline
\end{tabular}
\caption[Comparison of RFA data and simulation, using best fit parameters.]{\label{fig:rfa_results2} Comparison of Q15W Al RFA data and simulation, using best fit parameters (Table~\ref{tab:param_result}), conditions 7 - 12 (Table~\ref{tab:condx_list}).  The RFA is ``thin" style (Table~\ref{tab:rfa_styles}).  The top plots show the total signal across the 9 RFA collectors (with +50~V on the grid); the bottom plots show the signal in the central three collectors vs retarding voltage.}
%The plots show the signal across the 9 RFA collectors at three different retarding voltages.}
%\vskip 30mm
\end{minipage}
\end{figure*}

\begin{figure*}
\begin{minipage}{.98\textwidth}
\centering
\begin{tabular}{cc}
\includegraphics[width=.32\linewidth]{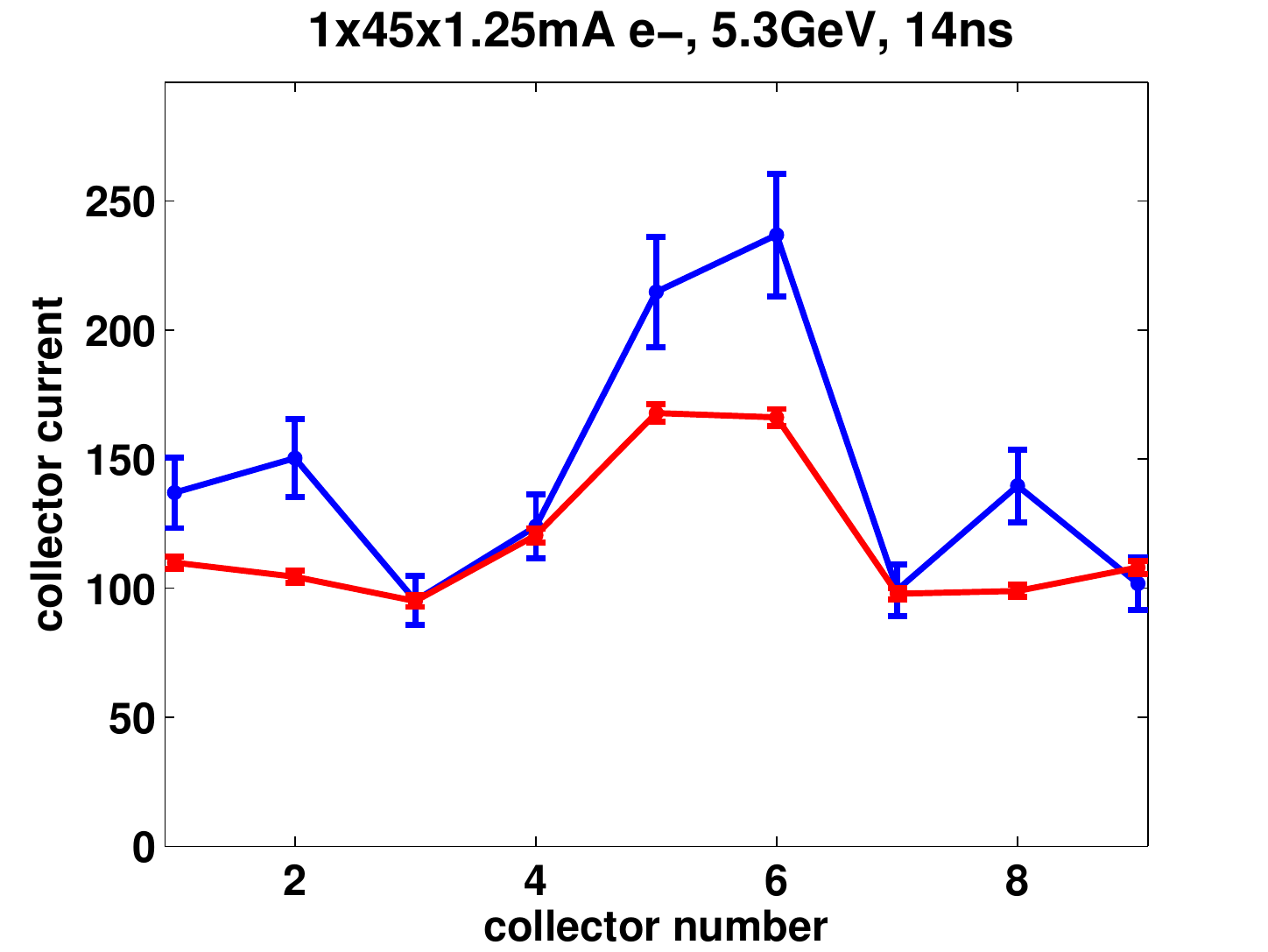}
\includegraphics[width=.32\linewidth]{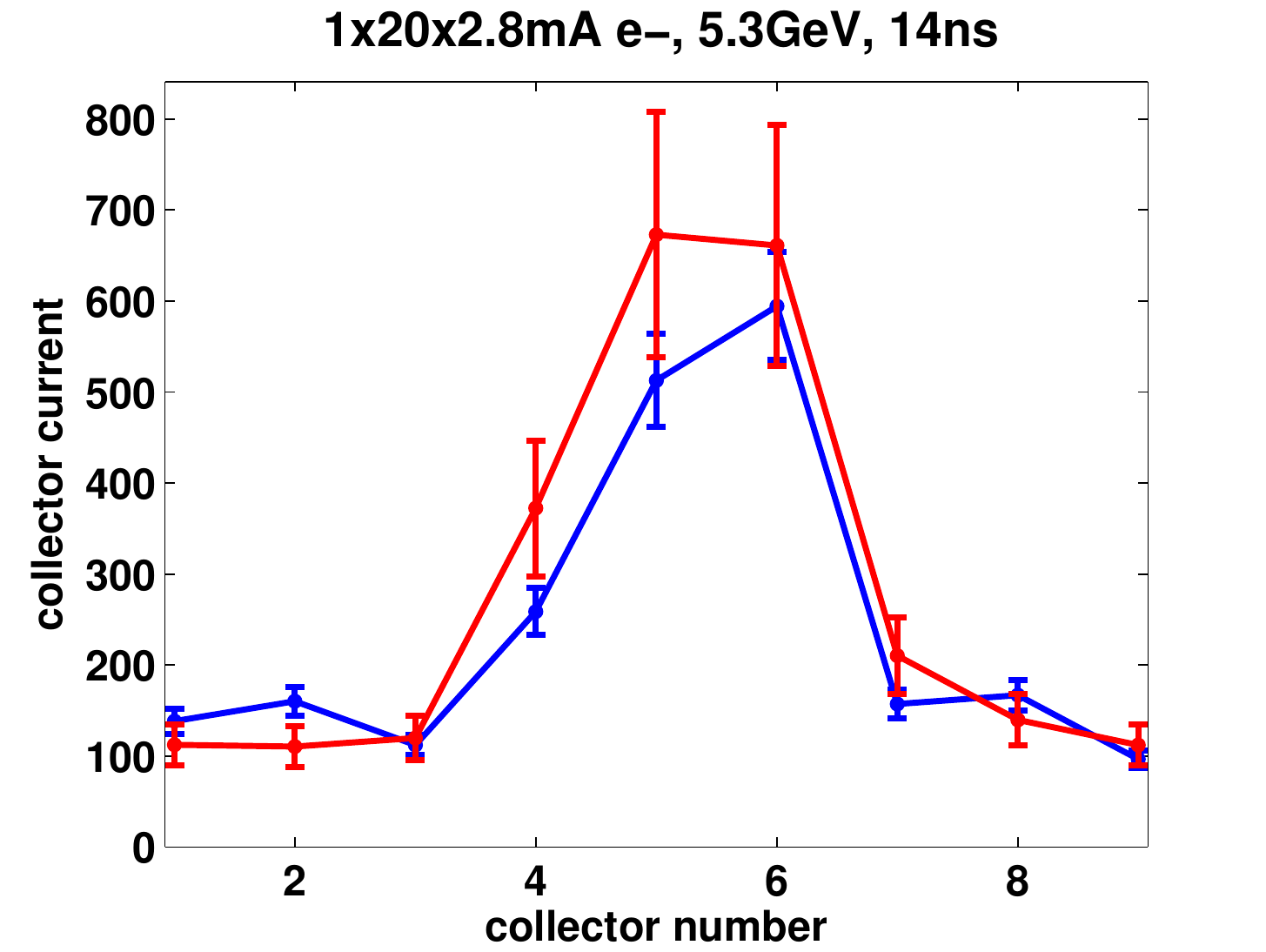}
\includegraphics[width=.32\linewidth]{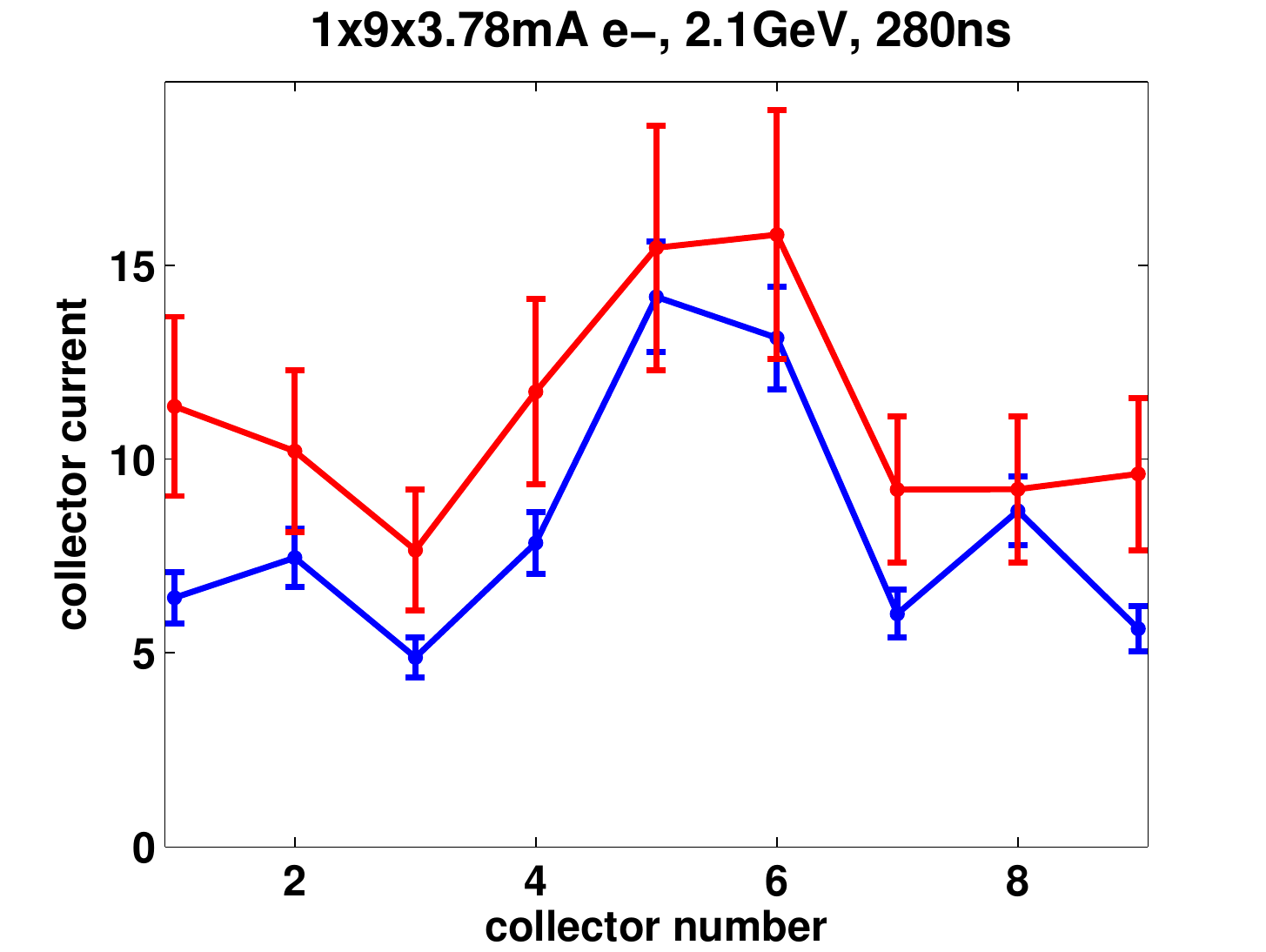} \\
\includegraphics[width=.32\linewidth]{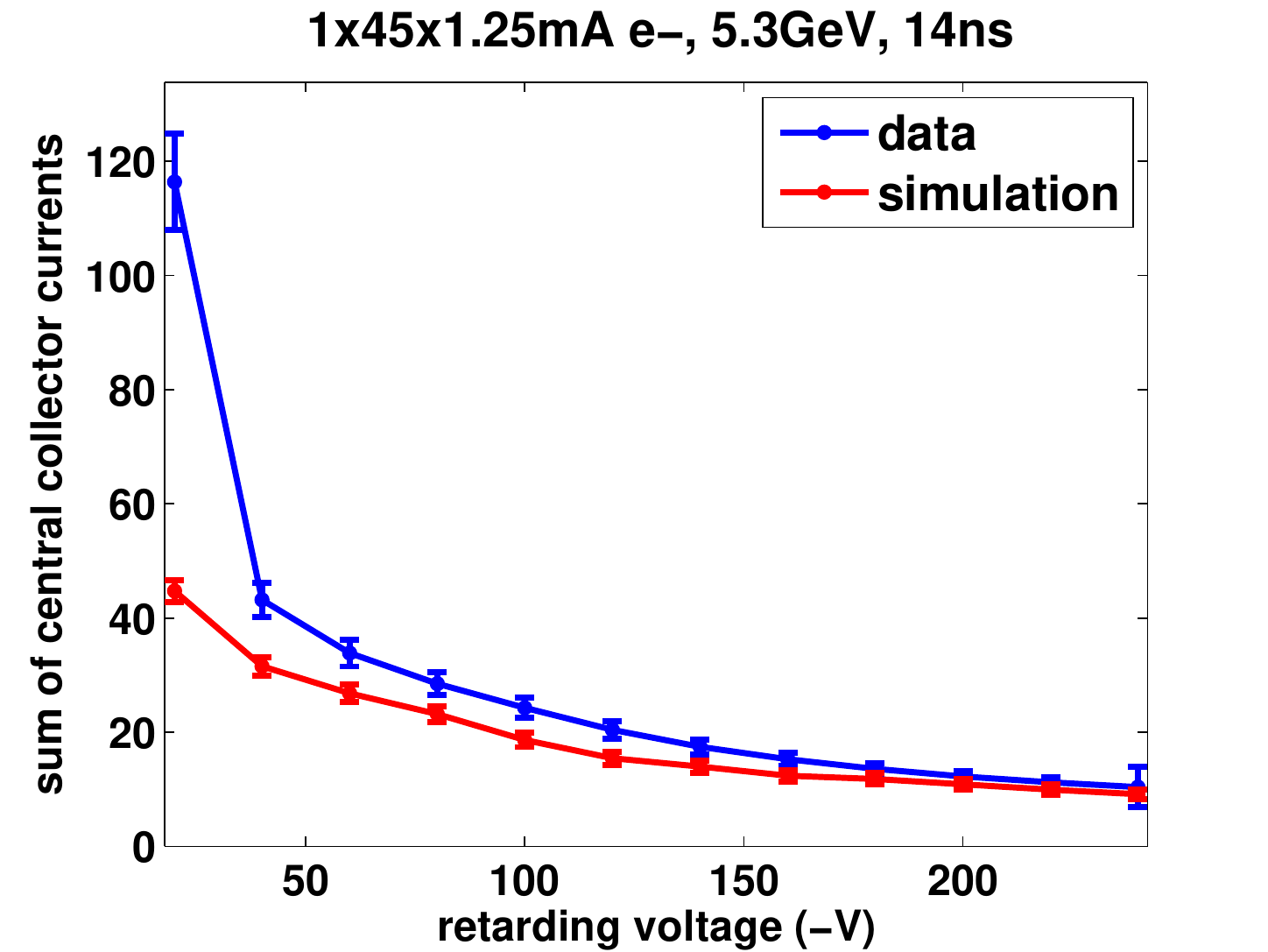}
\includegraphics[width=.32\linewidth]{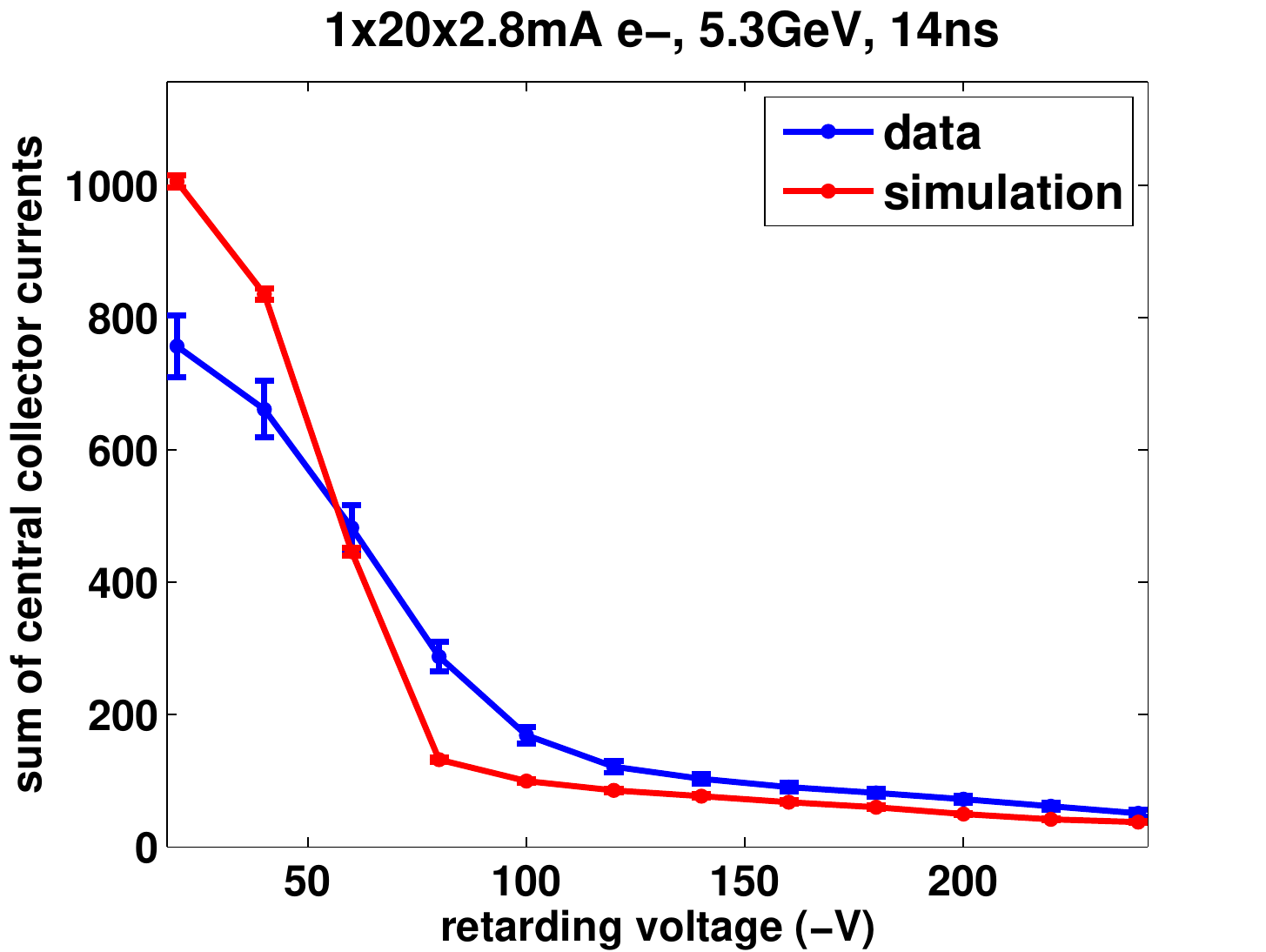}
\includegraphics[width=.32\linewidth]{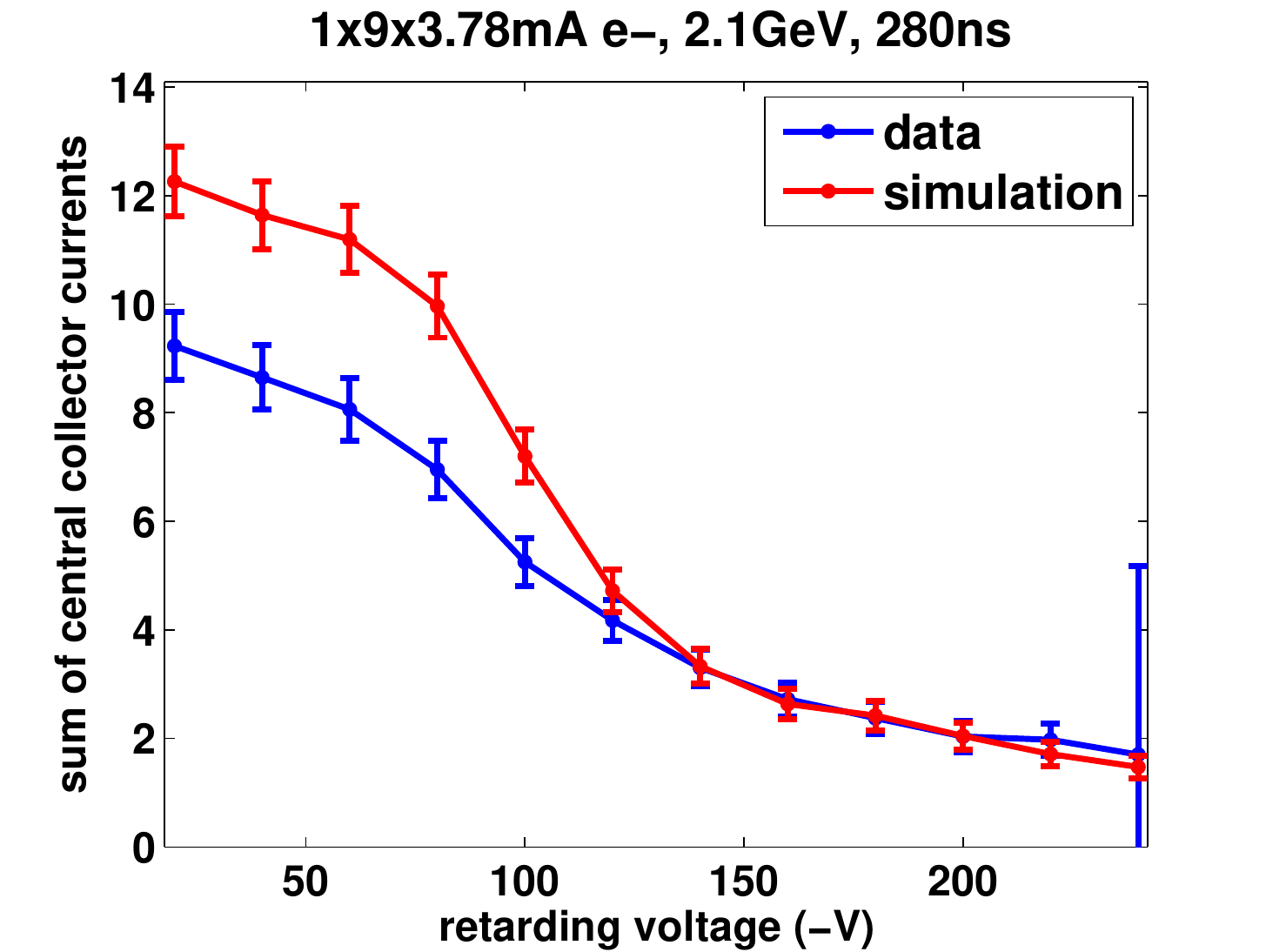} \\
%\hline \hline
\end{tabular}
\caption[Comparison of RFA data and simulation, using best fit parameters.]{\label{fig:rfa_results3} Comparison of Q15W Al RFA data and simulation, using best fit parameters (Table~\ref{tab:param_result}), conditions 13 - 15 (Table~\ref{tab:condx_list}).  The RFA is ``thin" style (Table~\ref{tab:rfa_styles}).  The top plots show the total signal across the 9 RFA collectors (with +50~V on the grid); the bottom plots show the signal in the central three collectors vs retarding voltage.}
%The plots show the signal across the 9 RFA collectors at three different retarding voltages.}
%\vskip 30mm
\end{minipage}
\end{figure*}

\begin{figure*}
\begin{minipage}{.98\textwidth}
\centering
\begin{tabular}{cc}
\includegraphics[width=.32\linewidth]{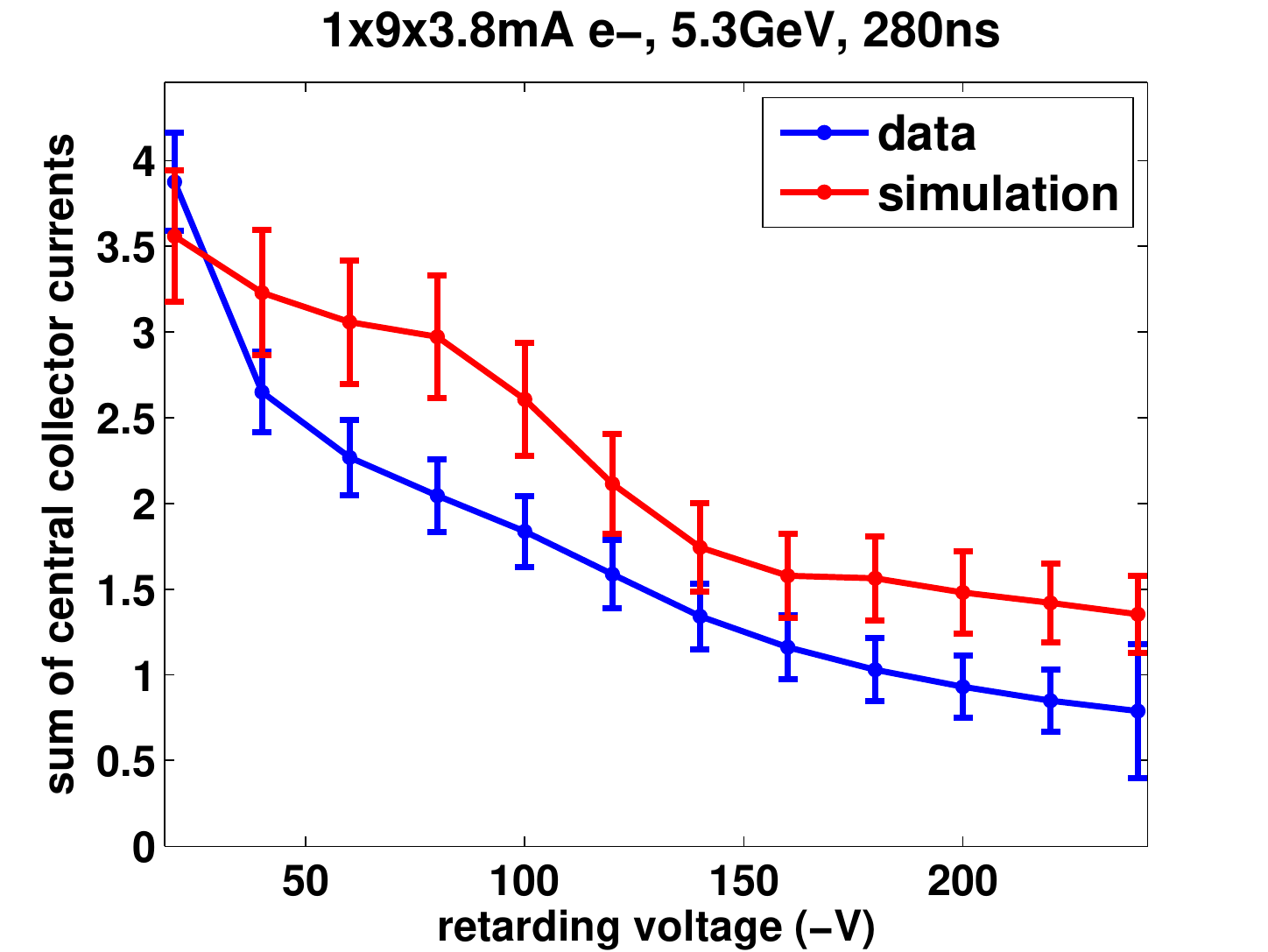}
\includegraphics[width=.32\linewidth]{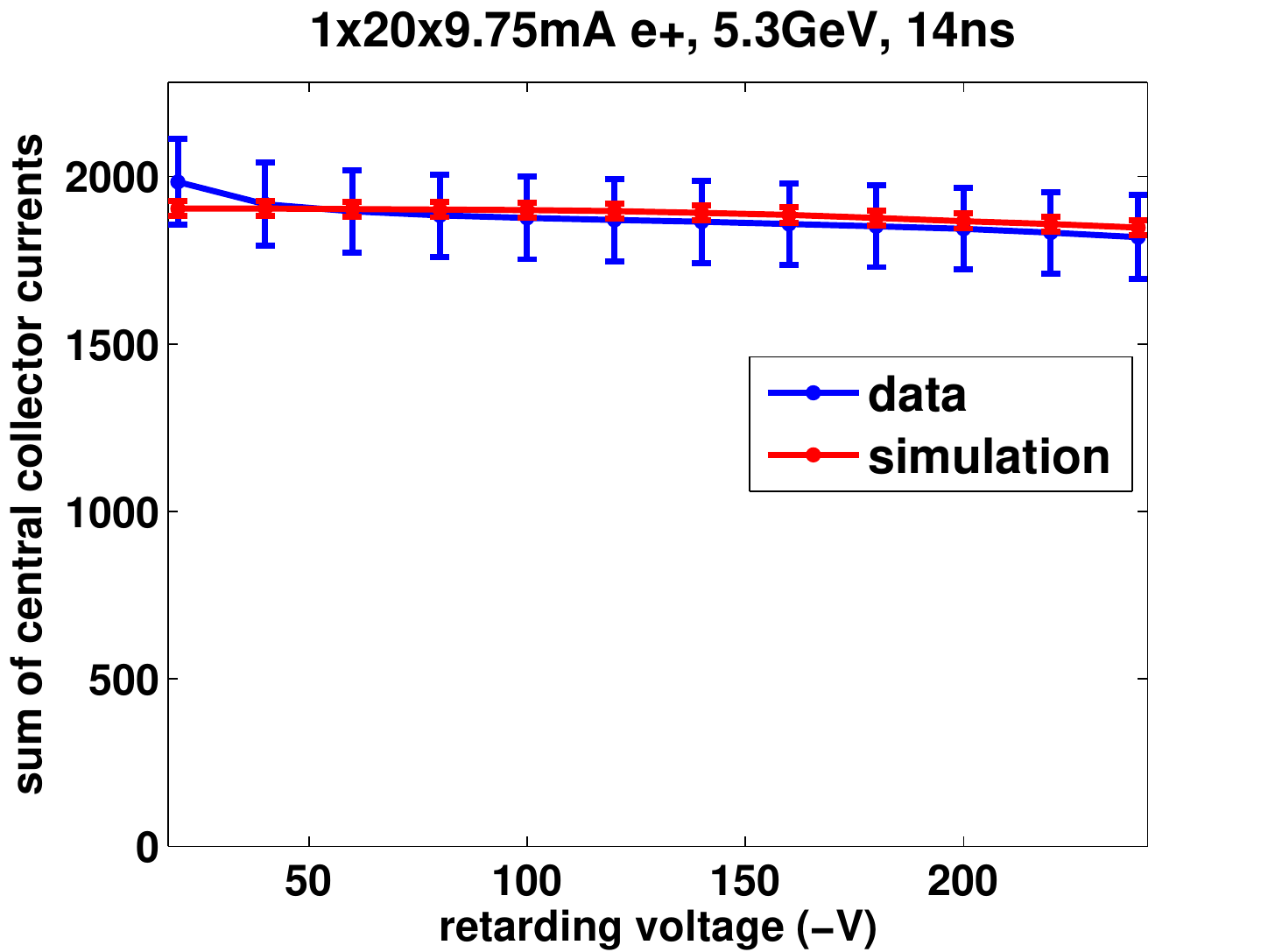}
\includegraphics[width=.32\linewidth]{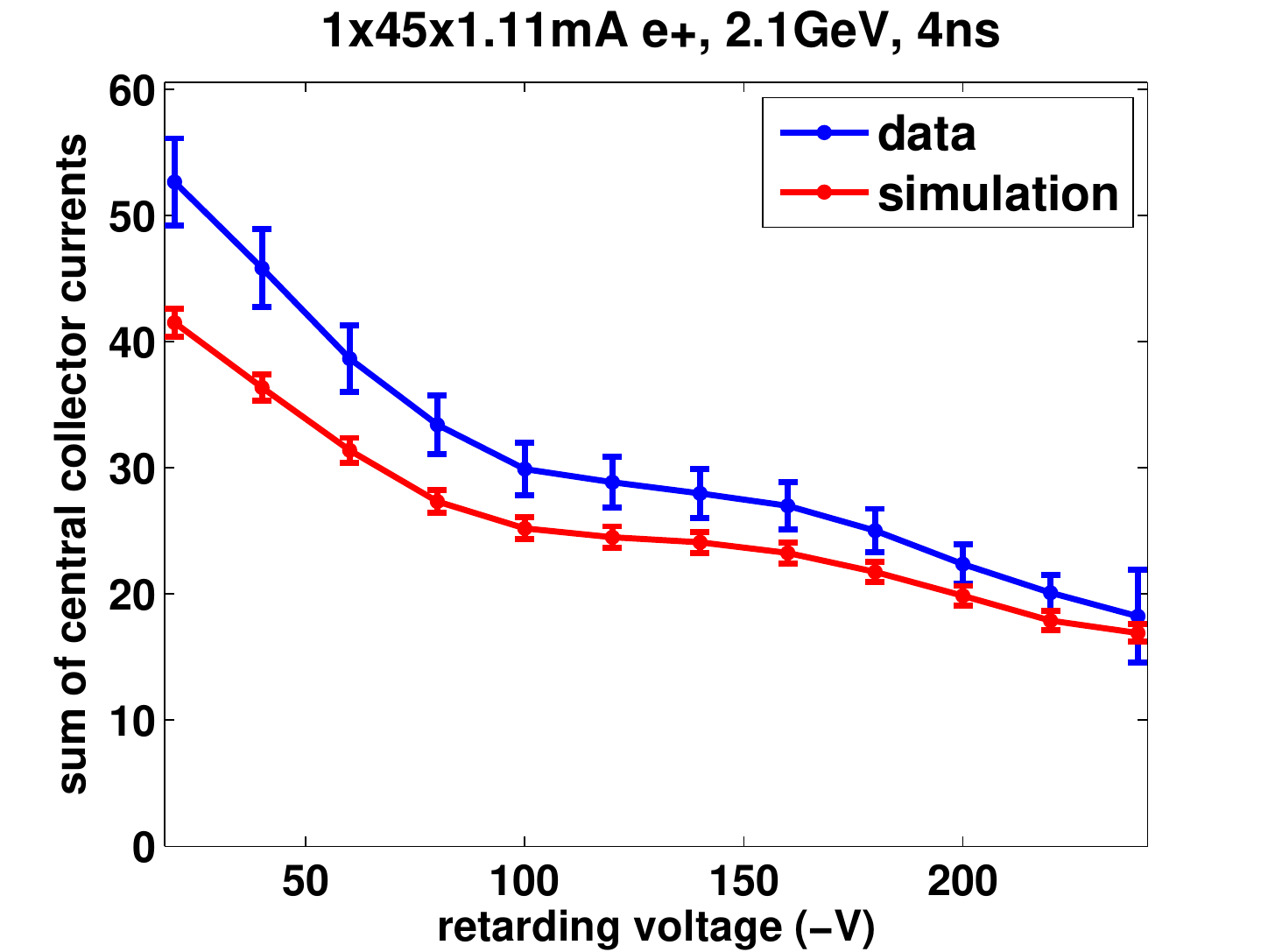} \\
%\hline \hline
\end{tabular}
\caption[Comparison of Q15W TiN data and simulation, using best fit parameters.]{\label{fig:rfa_results_TiN} Comparison of Q15W TiN RFA data and simulation, using best fit parameters (Table~\ref{tab:param_result}).  The RFA is ``thin" style (Table~\ref{tab:rfa_styles}).  The plots show the signal in the central three collectors vs retarding voltage.}
%The plots show the signal across the 9 RFA collectors at three different retarding voltages.}
%\vskip 30mm
\end{minipage}
\end{figure*}

\begin{figure*}
\begin{minipage}{.98\textwidth}
\centering
\begin{tabular}{cc}
\includegraphics[width=.32\linewidth]{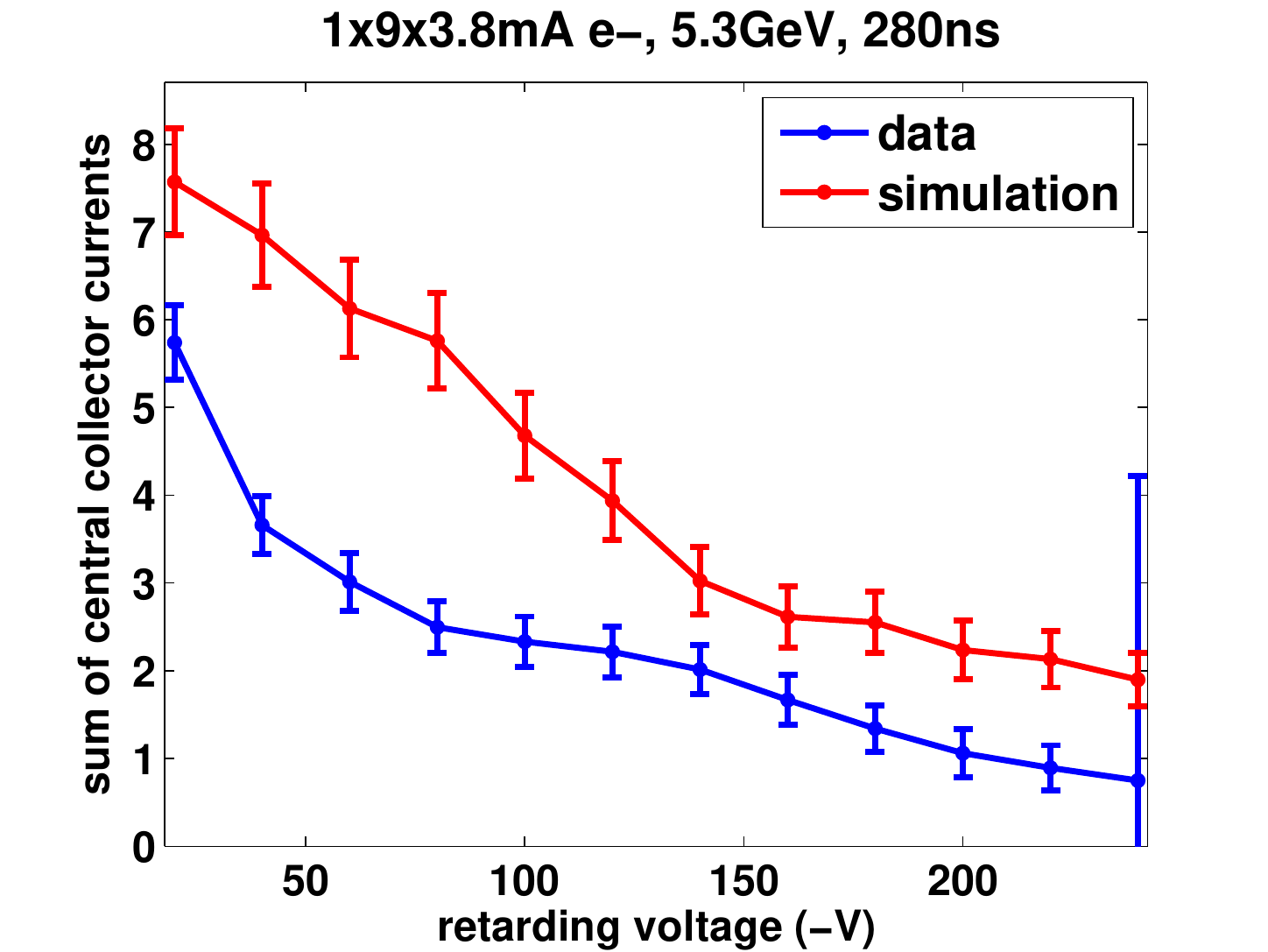}
\includegraphics[width=.32\linewidth]{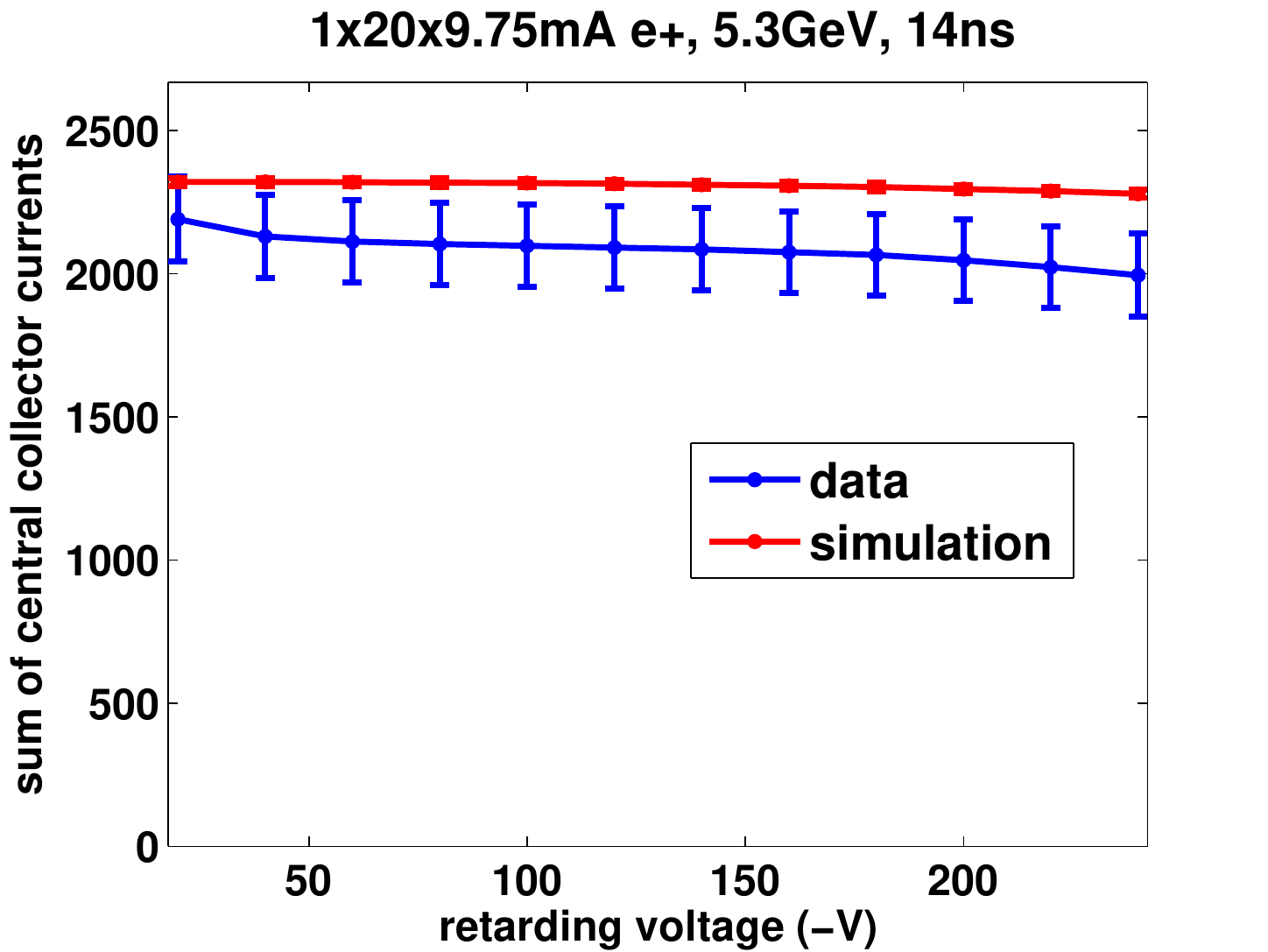}
\includegraphics[width=.32\linewidth]{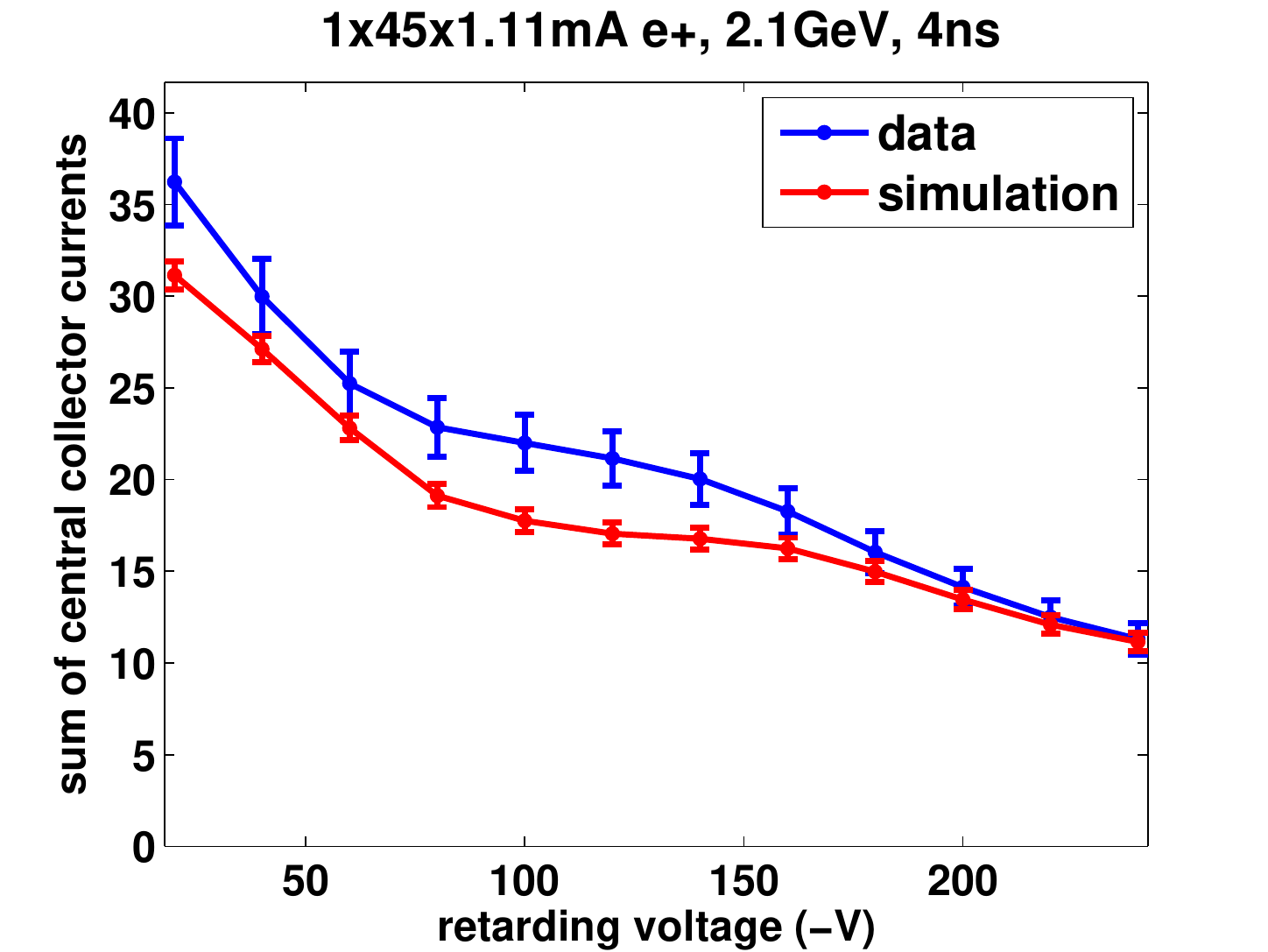} \\
%\hline \hline
\end{tabular}
\caption[Comparison of Q15E aC data and simulation, using best fit parameters.]{\label{fig:rfa_results_aC} Comparison of Q15E aC RFA data and simulation, using best fit parameters (Table~\ref{tab:param_result}).  The RFA is ``thin" style (Table~\ref{tab:rfa_styles}).  The plots show the signal in the central three collectors vs retarding voltage.}
%The plots show the signal across the 9 RFA collectors at three different retarding voltages.}
%\vskip 30mm
\end{minipage}
\end{figure*}

\begin{figure*}
\begin{minipage}{.98\textwidth}
\centering
\begin{tabular}{cc}
\includegraphics[width=.32\linewidth]{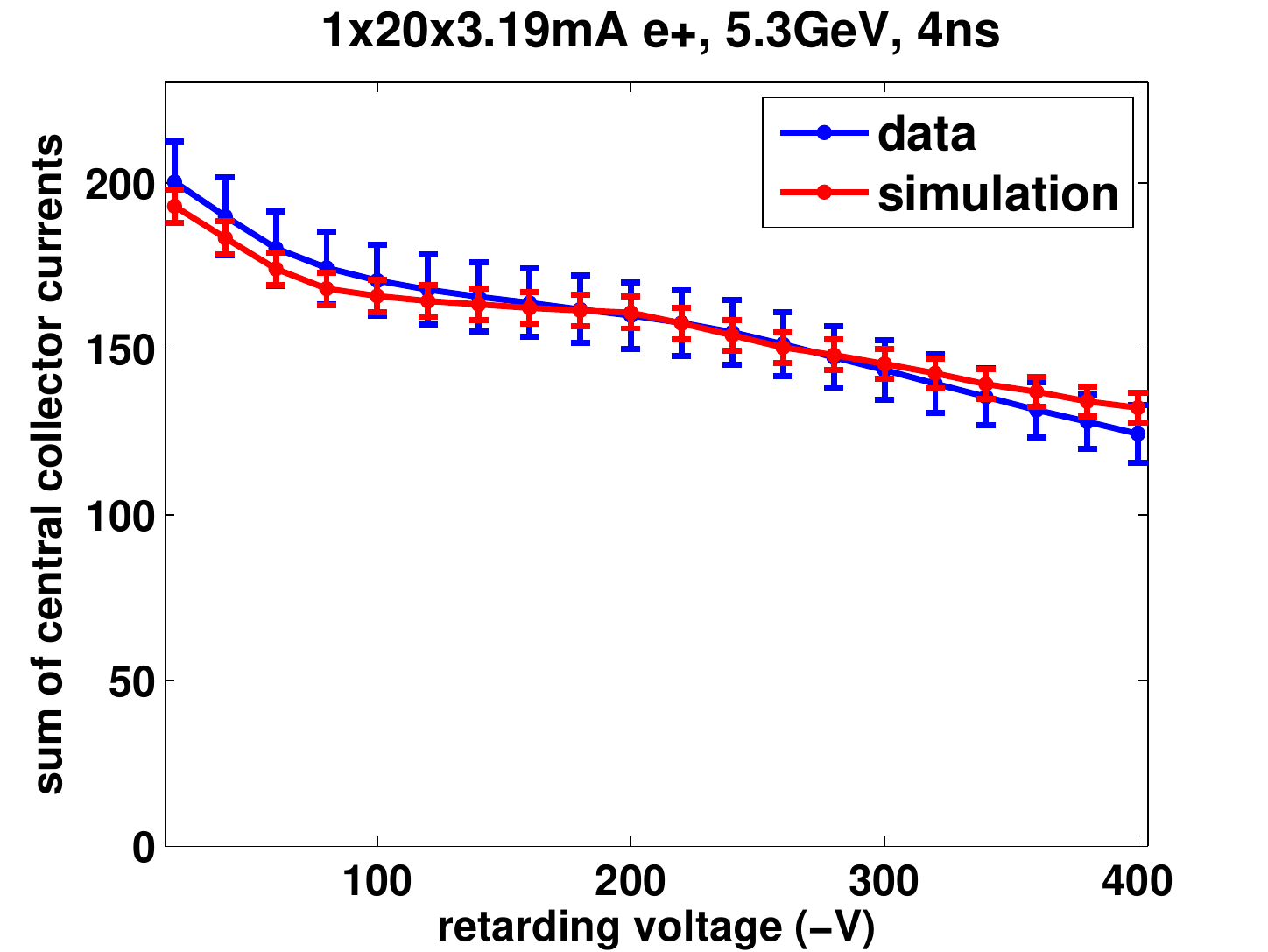}
\includegraphics[width=.32\linewidth]{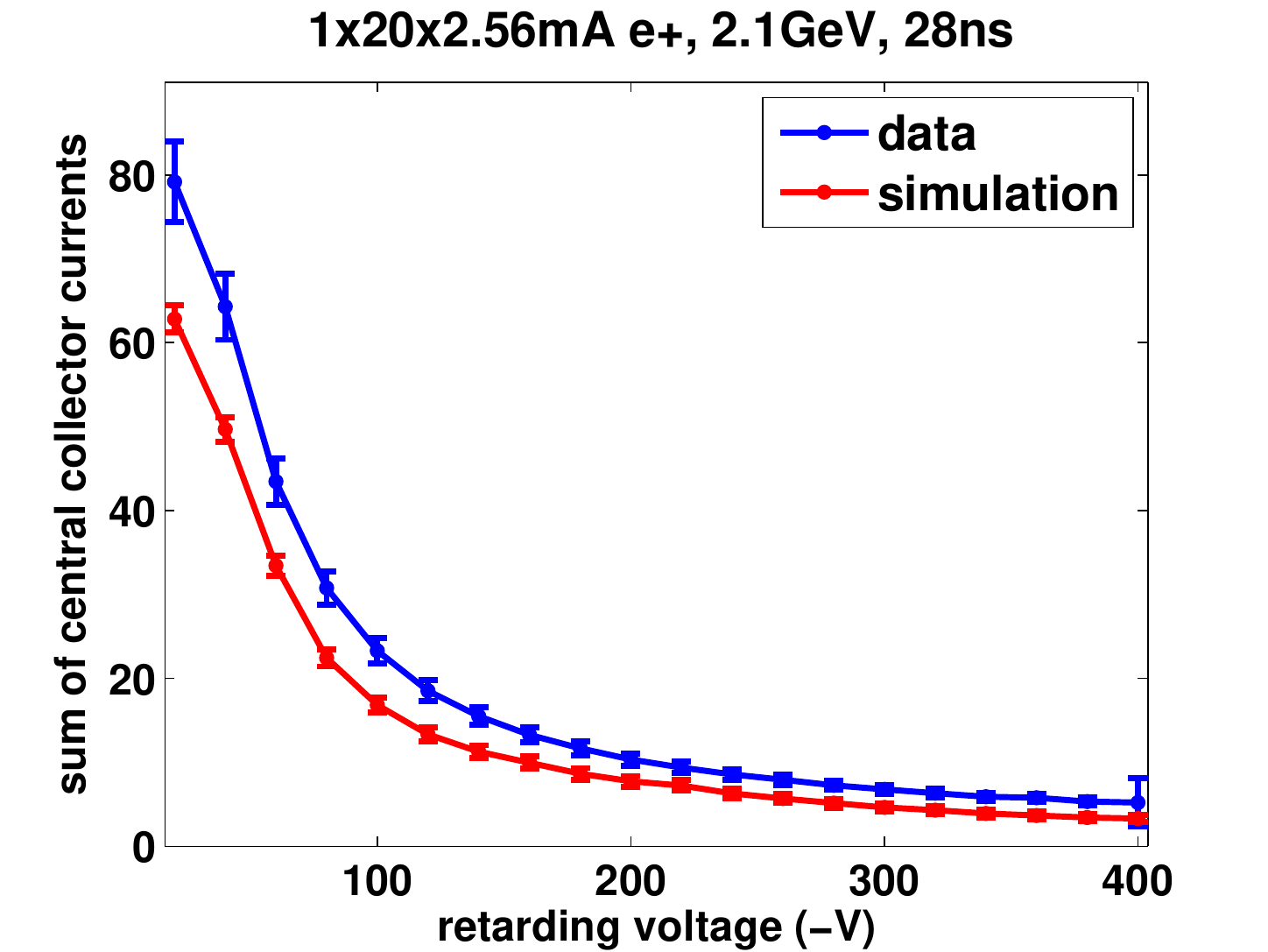}
\includegraphics[width=.32\linewidth]{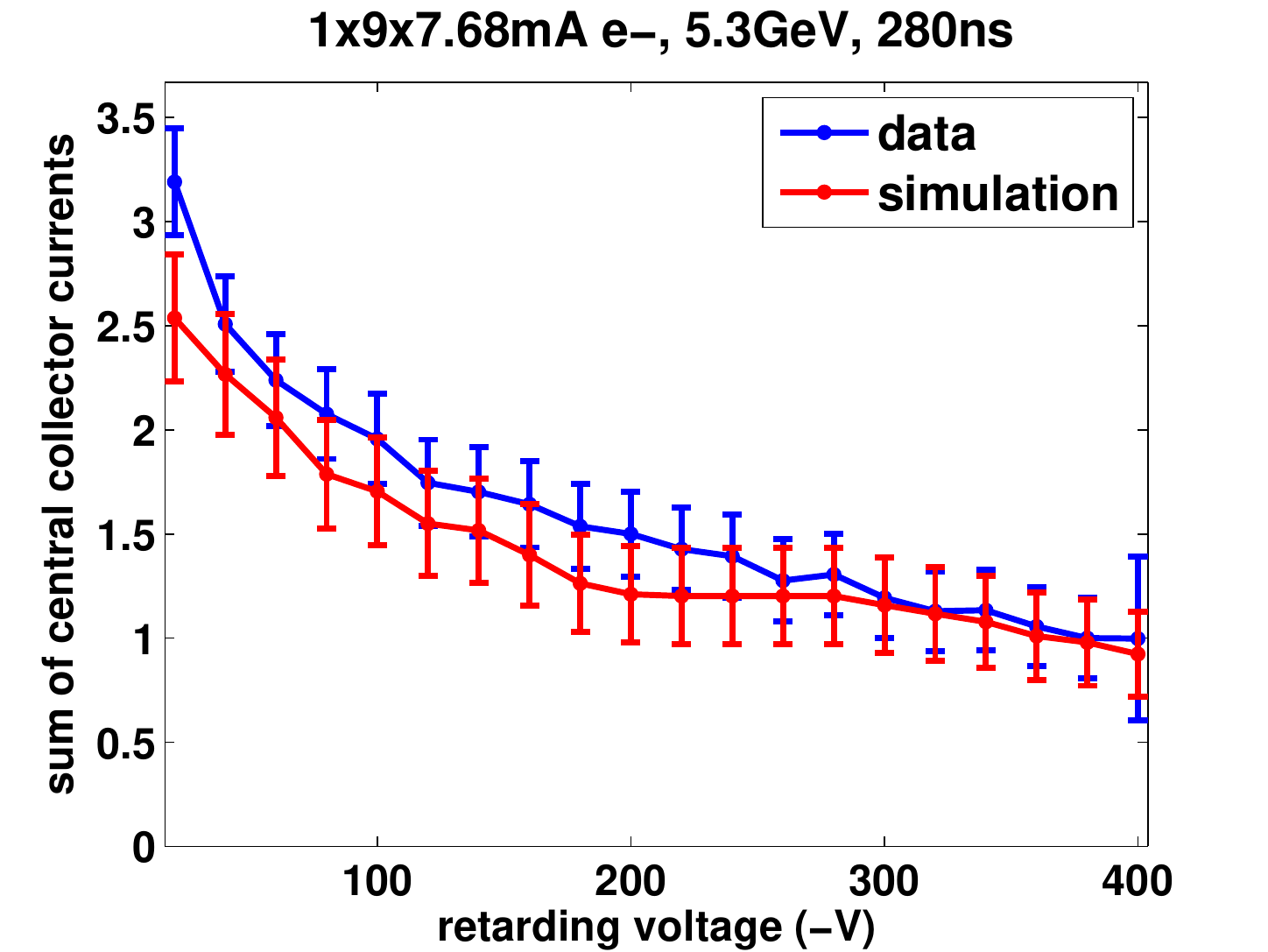} \\
%\hline \hline
\end{tabular}
\caption[Comparison of Q15W DLC data and simulation, using best fit parameters.]{\label{fig:rfa_results_DLC} Comparison of Q15W DLC RFA data and simulation, using best fit parameters (Table~\ref{tab:param_result}).  The RFA is ``Insertable II" style (Table~\ref{tab:rfa_styles}).  The plots show the signal in the central three collectors vs retarding voltage.}
%The plots show the signal across the 9 RFA collectors at three different retarding voltages.}
%\vskip 30mm
\end{minipage}
\end{figure*}

\begin{figure*}
\begin{minipage}{.98\textwidth}
\centering
\begin{tabular}{cc}
\includegraphics[width=.32\linewidth]{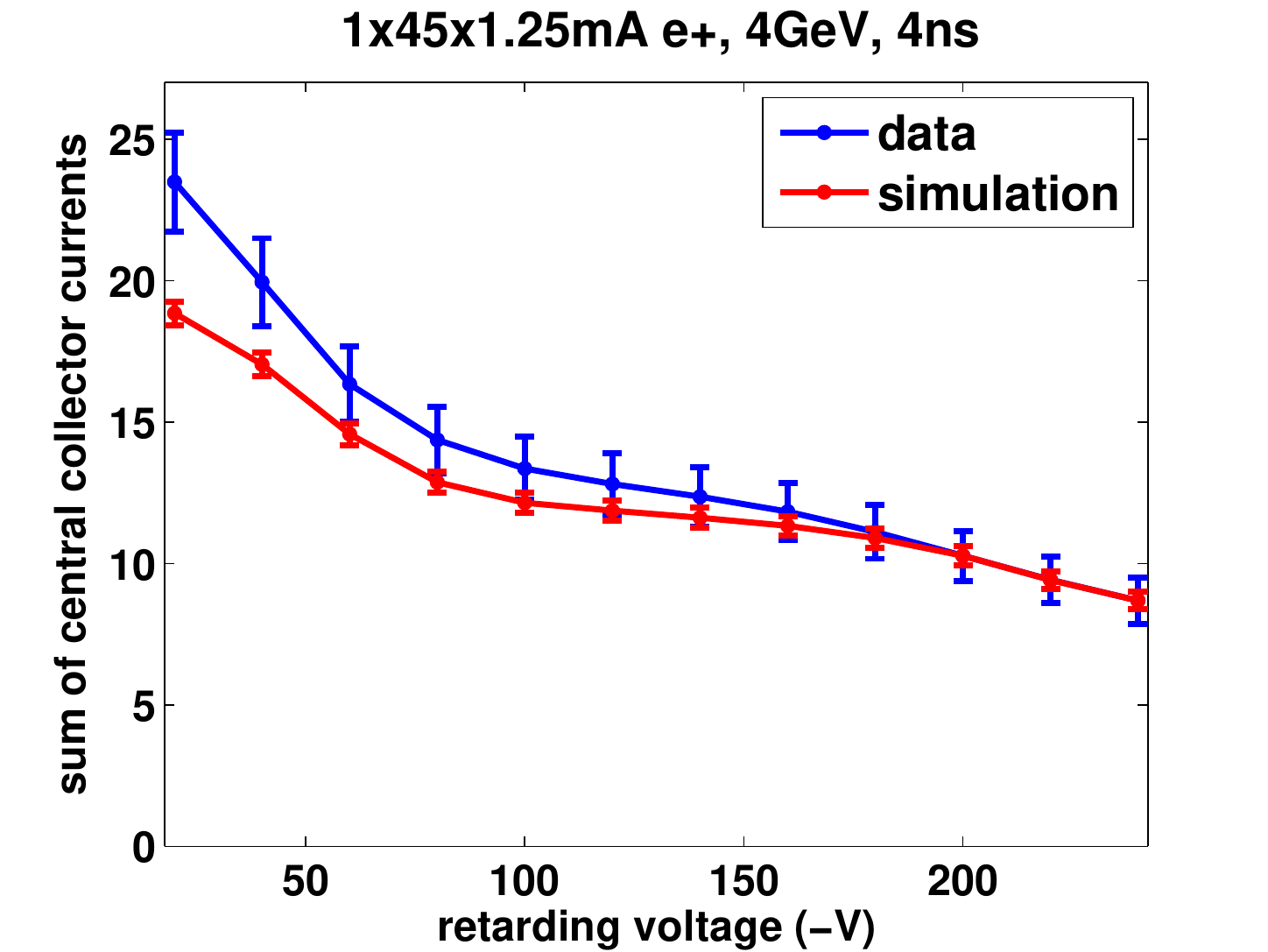}
\includegraphics[width=.32\linewidth]{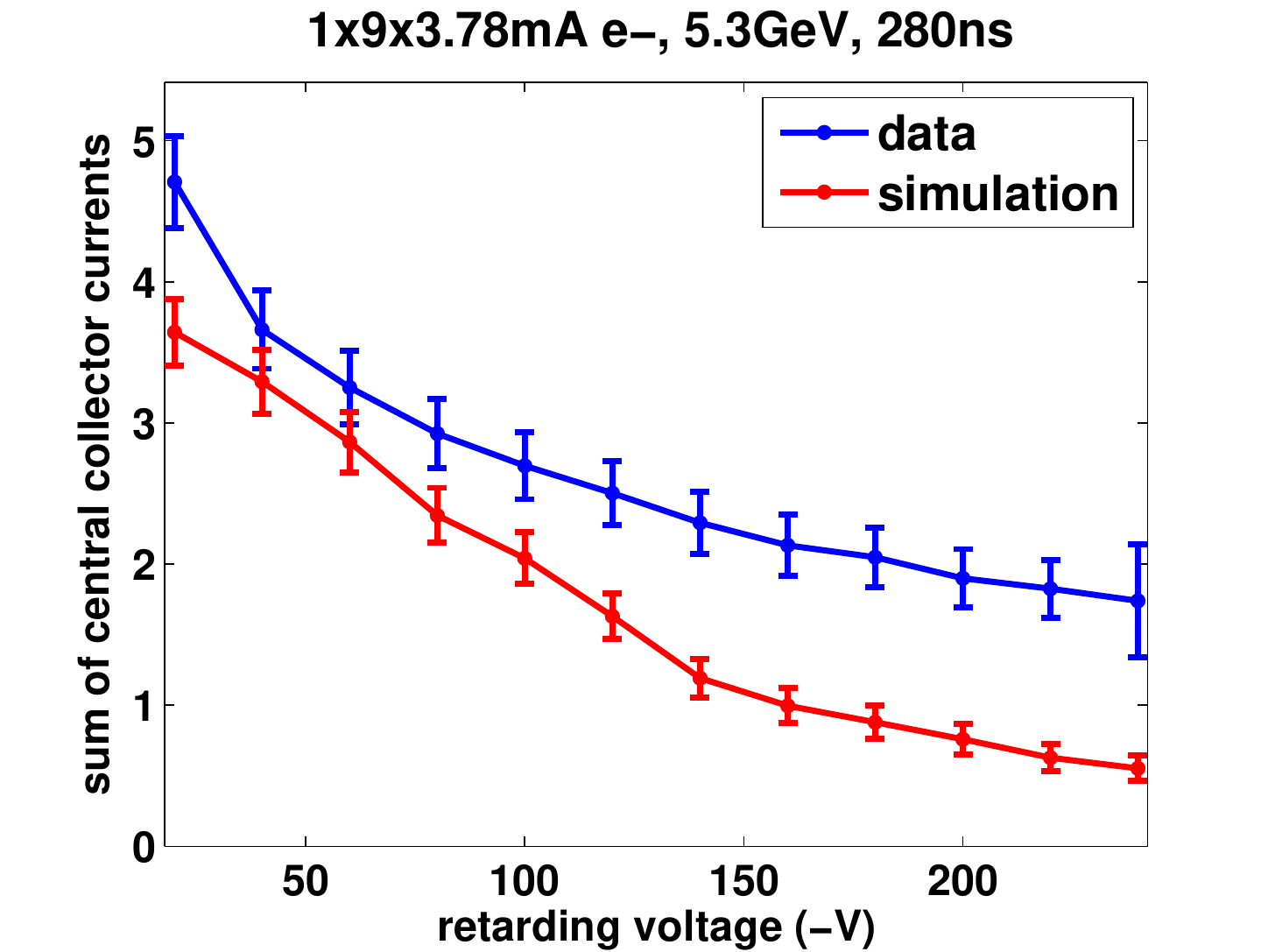}
\includegraphics[width=.32\linewidth]{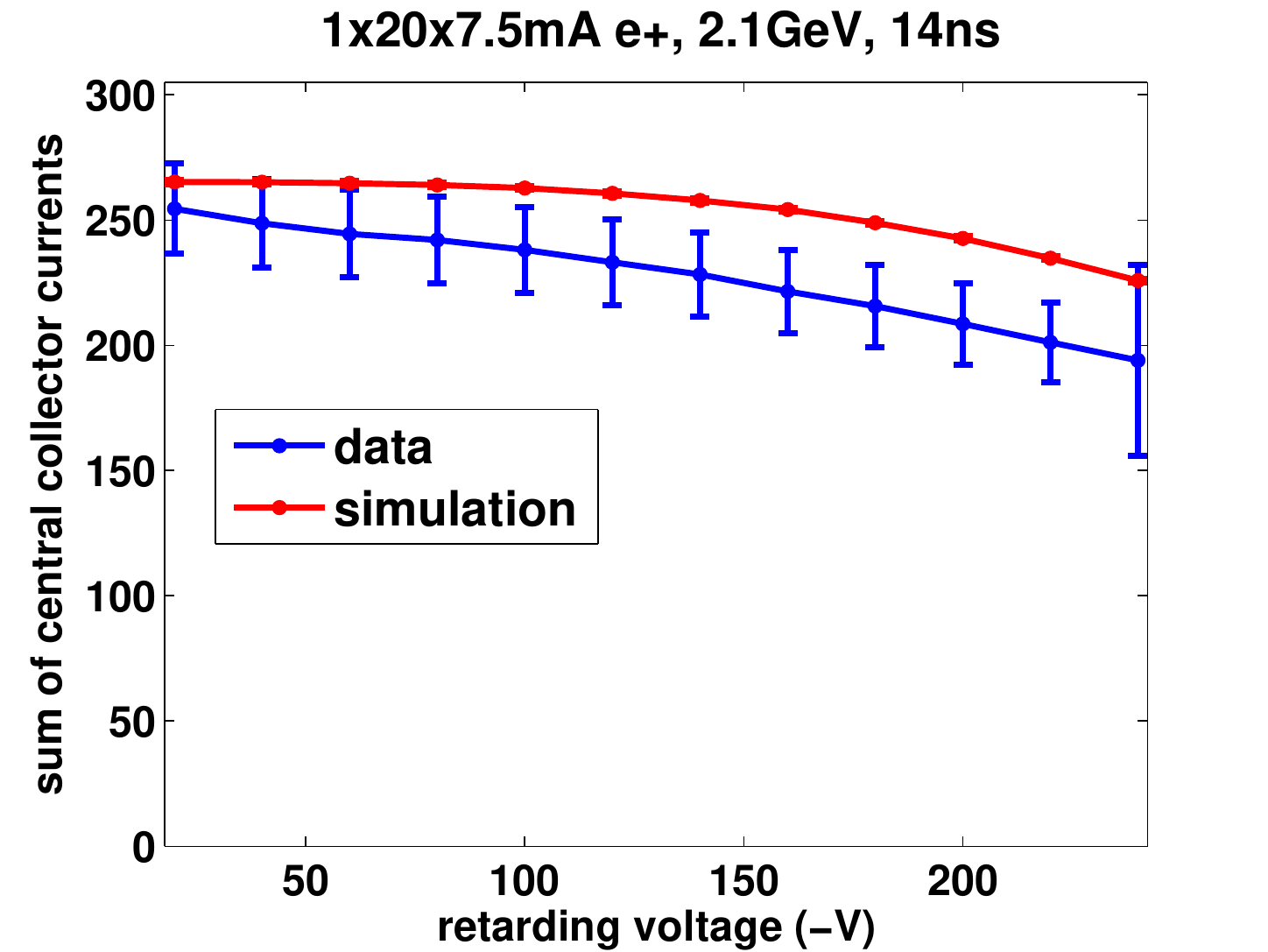} \\
%\hline \hline
\end{tabular}
\caption[Comparison of Q14E Cu data and simulation, using best fit parameters.]{\label{fig:rfa_results_Cu} Comparison of Q14E Cu RFA data and simulation, using best fit parameters (Table~\ref{tab:param_result}).  The RFA is ``Insertable I" style  (Table~\ref{tab:rfa_styles}).  The plots show the signal in the central three collectors vs retarding voltage.}
%The plots show the signal across the 9 RFA collectors at three different retarding voltages.}
%\vskip 30mm
\end{minipage}
\end{figure*}

\begin{figure*}
\begin{minipage}{.98\textwidth}
\centering
\begin{tabular}{cc}
\includegraphics[width=.32\linewidth]{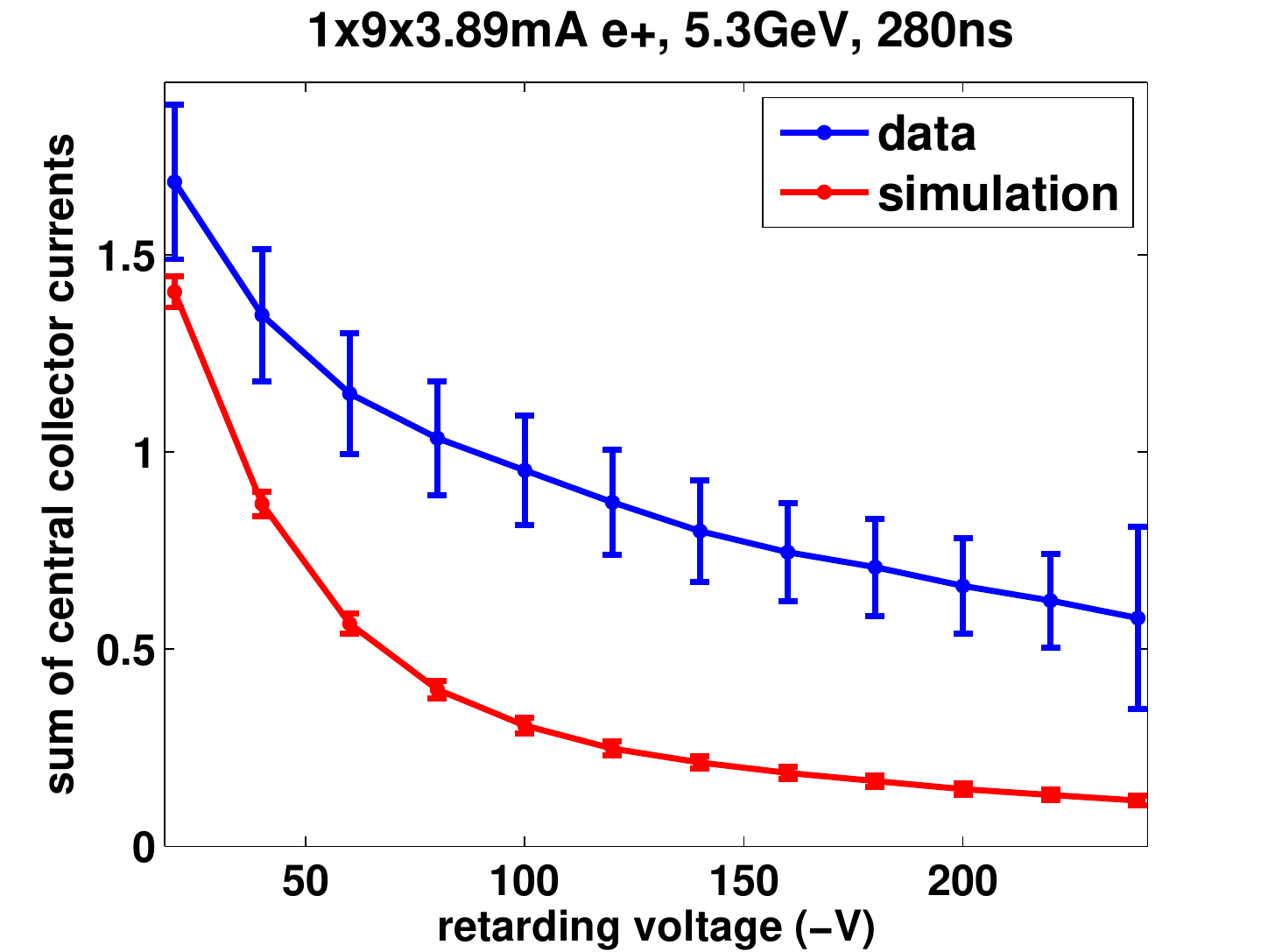}
\includegraphics[width=.32\linewidth]{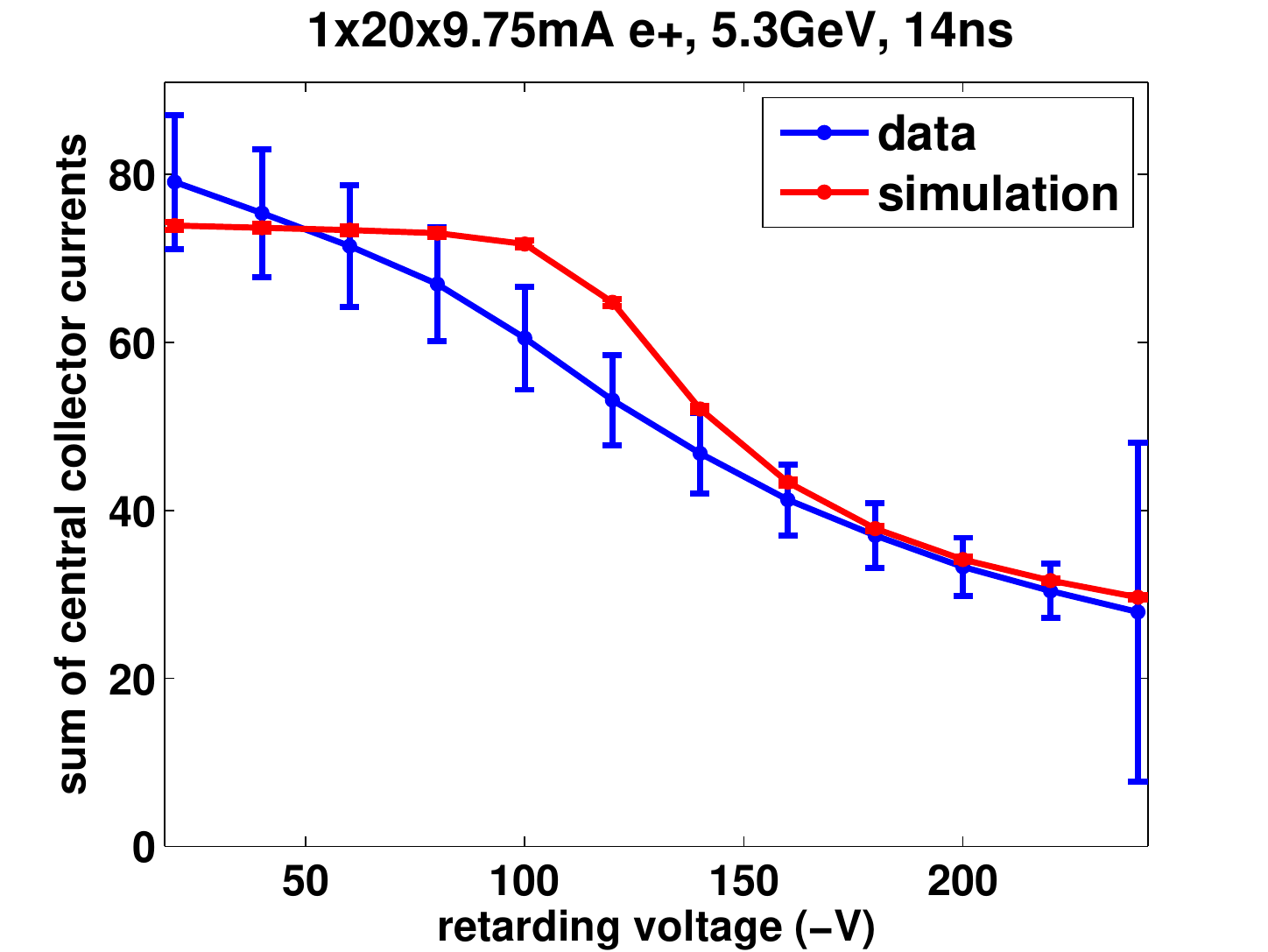}
\includegraphics[width=.32\linewidth]{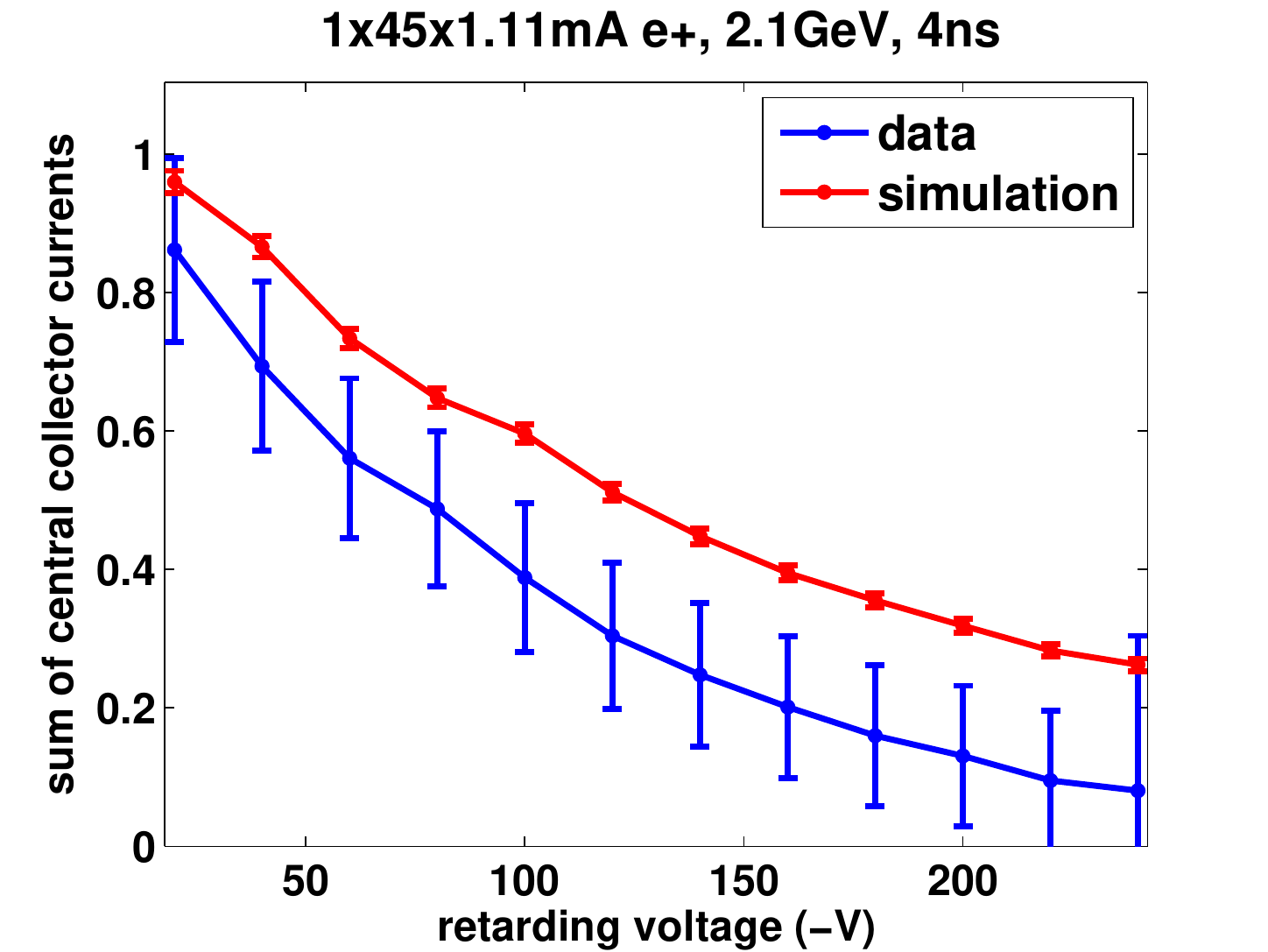} \\
%\hline \hline
\end{tabular}
\caption[Comparison of L3 NEG data and simulation, using best fit parameters.]{\label{fig:rfa_results_NEG} Comparison of L3 NEG RFA data and simulation, using best fit parameters (Table~\ref{tab:param_result}).  The RFA is ``APS" style (Table~\ref{tab:rfa_styles}).  The plots show the signal in the collector vs retarding voltage.}
%The plots show the signal across the 9 RFA collectors at three different retarding voltages.}
%\vskip 30mm
\end{minipage}
\end{figure*}

\end{document}